\documentclass[a4,usenatbib,useAMS,fleqn]{mn2e}

\usepackage{amsmath}
\usepackage{graphicx}
\usepackage{epstopdf}
\usepackage{morefloats}
\usepackage{dblfloatfix}
\usepackage{url}
\usepackage{times}
\usepackage{caption}
\usepackage{subcaption}
\usepackage{fixltx2e}
\usepackage{journals}
\usepackage{xfrac}

\usepackage{mathrsfs}

\newcommand{\Reff}{{R$_{e}$~}}

\newcommand{\galaxyone}{{p9-127B~}}

\newcommand{\galaxythree}{{p4-127A~}}
\newcommand{\galaxyfour}{{p9-127A~}}

\newcommand{\galaxyeight}{{p9-61A~}}

\newcommand{\galaxythirteen}{{p4-19B~}}
\newcommand{\galaxyfourteen}{{p4-19C~}}
\newcommand{\galaxyfifteen}{{p9-19E~}}

\newcommand{\galaxyseventeen}{{p9-19D~}}
\newcommand{\galaxyeighteen}{{p9-19B~}}

\newcommand{\asec}{{$^{\prime\prime}$}}

\newcommand{\tx}[1]{\textrm{#1}}

\newcommand{\Rekpc}{$R_{\tx{e}}~$}

\newcommand{\N}[1]{N$_{#1}$}
\newcommand{\firefly}{{\sc FIREFLY}}
\newcommand{\logageL}{$\langle \log {\rm age} \rangle_{\rm L}$~}
\newcommand{\logageM}{$\langle \log {\rm age} \rangle_{\rm M}$~}
\newcommand{\ageL}{$\langle {\rm age} \rangle_{\rm L}$~}
\newcommand{\ageM}{$\langle {\rm age} \rangle_{\rm M}$~}

\newcommand{\metalL}{$\langle {\rm [Z/H]} \rangle_{\rm L}$~}
\newcommand{\metalM}{$\langle {\rm [Z/H]} \rangle_{\rm M}$~}

\makeatletter

\renewcommand\section{\@startsection{section}{1}{\z@}%
                                  {-3.5ex \@plus -1ex \@minus -.2ex}%
                                  {2.3ex \@plus.2ex}%
                                  {\flushleft\normalfont\large\bfseries}}

\renewcommand\subsection{\@startsection{subsection}{1}{\z@}%
                                  {-3.5ex \@plus -1ex \@minus -.2ex}%
                                  {2.3ex \@plus.2ex}%
                                  {\flushleft\normalfont\bfseries}}
\makeatother

\title[P-MaNGA: Full spectral fitting]{P-MaNGA: Full spectral fitting and stellar population maps from prototype observations}

\author[David M. Wilkinson {\it et al.}]{
	\parbox[h]{\textwidth}{\raggedright David~M.~Wilkinson%
	$^{1}$\thanks{}, \mbox{Claudia~Maraston,$^{1}$} \mbox{Daniel~Thomas,$^{1}$} \mbox{Lodovico~Coccato,$^{1,}$$^2$}
	 \mbox{Rita~Tojeiro,$^{3}$} \mbox{Michele~Cappellari,$^{4}$}
	 \mbox{Francesco~Belfiore,$^{5,6}$} \mbox{Matthew~Bershady,$^{7}$} \mbox{Mike~Blanton,$^{8}$} \mbox{Kevin~Bundy,$^{9}$}
	\mbox{Sabrina~Cales,$^{10}$}  \mbox{Brian~Cherinka,$^{11}$} \mbox{Niv~Drory,$^{12}$}
	 \mbox{Eric~Emsellem,$^{2,13}$} \mbox{Hai~Fu,$^{14}$} \mbox{David~Law,$^{15}$} \mbox{Cheng~Li,$^{16}$}
	 \mbox{Roberto~Maiolino,$^{5,6}$} \mbox{Karen~Masters,$^{1}$}
	 \mbox{Christy~Tremonti,$^{7}$} \mbox{David~Wake,$^{7}$} \mbox{Enci~Wang,$^{16}$}
	 \mbox{Anne-Marie~Weijmans,$^{3}$} \mbox{Ting~Xiao,$^{16}$}
	 \mbox{Renbin~Yan,$^{17}$} \mbox{Kai~Zhang,$^{16,17}$}
	 \mbox{Dmitry~Bizyaev,$^{18}$} \mbox{Jonathan~Brinkmann,$^{18}$} \mbox{Karen~Kinemuchi,$^{18}$}
	 \mbox{Elena~Malanushenko,$^{18}$} \mbox{Viktor~Malanushenko,$^{18}$}
	 \mbox{Daniel~Oravetz,$^{18}$} \mbox{Kaike~Pan$^{18}$ and}
	 \mbox{Audrey~Simmons$^{18}$}
	}\vspace*{14pt}\\
	Affiliations are listed at the end of the paper}
		
\pagerange{\pageref{firstpage}--\pageref{lastpage}} \pubyear{2014}

\begin{document}
\label{firstpage}

\maketitle
\begin{abstract}
\noindent MaNGA (Mapping Nearby Galaxies at Apache Point Observatory) is a 6-year SDSS-IV survey that will obtain resolved spectroscopy from 3600~\AA~to 10300~\AA~for a representative sample of over 10,000 nearby galaxies. In this paper, we derive spatially resolved stellar population properties and radial gradients by performing full spectral fitting of observed galaxy spectra from P-MaNGA, a prototype of the MaNGA instrument. These data include spectra for eighteen galaxies, covering a large range of morphological type. We derive age, metallicity, dust and stellar mass maps, and their radial gradients, using high spectral-resolution stellar population models, and assess the impact of varying the stellar library input to the models.  We introduce a method to determine dust extinction which is able to give smooth stellar mass maps even in cases of high and spatially non-uniform dust attenuation. 

With the spectral fitting we produce detailed maps of stellar population properties which allow us to identify galactic features among this diverse sample such as spiral structure, smooth radial profiles with little azimuthal structure in spheroidal galaxies, and spatially distinct galaxy sub-components. 
In agreement with the literature, we find the gradients for galaxies identified as early-type to be on average flat in age, and negative (-- 0.15 dex / \Rekpc) in metallicity, whereas the gradients for late-type galaxies are on average negative in age (-- 0.39 dex / \Rekpc) and flat in metallicity.
We demonstrate how different levels of data quality change the precision with which radial gradients can be measured. We show how this analysis, extended to the large numbers of MaNGA galaxies, will have the potential to shed light on galaxy structure and evolution.\\

\vspace*{10pt}
\end{abstract}
\footnotetext[1]{E-mail: david.wilkinson@port.ac.uk}

\renewcommand{\thefootnote}{\textsuperscript{\arabic{footnote}}}

\graphicspath{{\string plots_submitted_png/}}
\epstopdfsetup{outdir=./}
%\epstopdfsetup{outdir=/Users/david/Dropbox/supervisor_meetings/manga_paper/plots/}

\section{Introduction}\label{introduction}

Large spectroscopic surveys in the past 15 years (e.g. SDSS \citep{2000AJ....120.1579Y}, GAMA \citep{2011MNRAS.413..971D}, 2dFGRS \citep{1999MNRAS.308..459F}) have proved tremendously successful in deriving the physical properties of galaxies with high accuracy. Data across a range of redshift slices have been able to provide significant insight into the evolution of galaxy properties. Pairing these data with evolutionary population synthesis models (e.g. \cite{2003MNRAS.344.1000B}, \cite{1998MNRAS.300..872M}, \cite{1997A&A...326..950F}, \cite{1999ApJS..123....3L}, \cite{2005MNRAS.357..945G}, \cite{2003MNRAS.339..897T}, \cite{1996ApJS..106..307V}, \cite{2010MNRAS.404.1639V}, \cite{2009ApJ...699..486C}, \cite{2011MNRAS.412.2183T} and \cite{2011MNRAS.418.2785M}) 
via a full spectral fitting method 
(e.g. PPXF by \cite{2004PASP..116..138C}, VESPA by \cite{2007MNRAS.381.1252T}, PCA by \cite{2012MNRAS.421..314C}, STARLIGHT \cite{2005MNRAS.358..363C}, STECKMAP by \cite{2006MNRAS.365...46O})
, or using selected absorption lines (\cite{2005ApJ...621..673T}, \cite{2010MNRAS.404.1775T}) allows the derivation of galaxy properties, such as star formation history, age of formation, metallicity and chemical content, dust content, star formation rate and stellar mass.

These large galaxy surveys - in order to collect large samples - could only perform single aperture observations per galaxy, typically across their central regions. Therefore galaxy evolution properties as a function of internal structures, such as stellar population profiles across the core, disk or halo, cannot be measured, and the global properties are very sensitive to the aperture position and coverage (\cite{2013A&A...553L...7I}). Physical processes such as radial migration, higher level structure such as dust structure and clumped star-formation, and signatures of merging and interactions cannot be easily constrained by single-aperture observations (\cite{2012MNRAS.420..197G}, \cite{2012ApJ...753..114W}).

In order to surpass these limitations, many modern galaxy surveys use integral field spectroscopy (IFS) (SAURON \citep{2001MNRAS.326...23B}, ATLAS$^{\rm 3D}$ \citep{2011MNRAS.413..813C}, CALIFA \citep{2012A&A...538A...8S}, SAMI \citep{2012MNRAS.421..872C}, VENGA \citep{2013AJ....145..138B}, DiskMass \citep{2010ApJ...716..198B}. This is a method by which multiple spectra of the same galaxy covering many 2D positions in the sky are obtained. This allows detailed internal information of each observed galaxy to be obtained.

Many of the previous IFU surveys have been limited in the number of galaxies they have observed, meaning that detailed statistics of properties of a range of galaxy types is limited. MaNGA  (Mapping Nearby Galaxies at Apache Point Observatory, \cite{2015ApJ...798....7B}) is designed to complement these small-sample surveys by providing spectral information for a large statistical sample of 10,000 galaxies across a range of galaxy types and morphologies, to be observed over a period of six years. MaNGA is one of the four surveys of the SDSS-IV (Sloan Digital Sky Survey IV) project. A unique feature of MaNGA with respect to other IFU surveys is its extended spectral coverage, from 3600 \AA~to~$1\mu$. As is well known, the wider the spectral range the easier it is to break degeneracies such as the age/metallicity/dust degeneracies (e.g. \cite{2012MNRAS.422.3285P}). Moreover, this spectral range includes absorptions features such as Sodium and Calcium triplet, that are important diagnostics of the Initial Mass Function (IMF, e.g. \cite{2002ApJ...579L..13S}; \cite{2012ApJ...760...71C}; \cite{2013MNRAS.429L..15F}; \cite{2014MNRAS.438.1483S}). 

MaNGA aims to answer fundamental questions about the relative importance of stellar accretion, mergers and disk instabilities to mass assembly, the formation of bulges, the regulation of star formation, and the growth of galactic disks. In order to be able to answer these questions, the MaNGA main samples are selected to achieve a uniform spatial coverage that reaches either 1.5 \Reff\ (``Primary+,'' roughly \sfrac{2}{3} of the final sample) or 2.5 \Reff (``Secondary,'' roughly \sfrac{1}{3} of the final sample), see Wake et al.~(in preparation) for details. The launch of the MaNGA survey was July 1st, 2014, and the program will run for 6 years, utilizing half of the dark time available in SDSS-IV. Details on the instrumentation including the design, testing, and assembly of the fiber-IFUs are given in \cite{2015AJ....149...77D}. Wake et al. (in preparation) presents the sample design, optimization, and final selection of the survey. The software and data framework as well as the reduction pipeline are described in Law et al. (in preparation). A description of the commissioning, the quality of survey observations, and further operational details will be given in Yan et al. (in preparation).

In order to test the performances of the survey in advance, preliminary `MaNGA-type' galaxy data have been obtained with
a MaNGA prototype instrument called P-MaNGA (described in detail in section \ref{data}). The P-MaNGA dataset contains 3D spectra (x position, y position, and wavelength) for 18 galaxies observed with a variety of IFU sizes and observational conditions. The P-MaNGA data were obtained through a generous donation of observing time by the SDSS-III Collaboration \citep{2011AJ....142...72E}. These data are described in detail in \cite{2015ApJ...798....7B}. P-MaNGA observations are used to understand and quantify the capabilities of MaNGA and help predicting the results we should expect from the full dataset. 

The physical properties of galaxies are obtained by analysing spectra with stellar population models. 
In this paper we perform full spectral fitting of population models on the P-MaNGA data. We use a new full spectral fitting code, \firefly~(Wilkinson \& Maraston (in preparation)) and experiment with the full suite of high-resolution stellar population models by \cite{2011MNRAS.418.2785M}, which in particular provide models for various input spectral libraries. We generate 2D stellar population maps and 1D radial profiles and analyse the results.

The paper is organised as follows. Section~\ref{data} describes the P-MaNGA observations, while Section~\ref{analysis tools} the tools of analyses used in this paper, namely the stellar population models and the fitting code. The description of our new treatment of dust reddening is also placed there, as well as the methodology to measure radial gradients. The analysis of individual stellar population maps and gradients is in Section~\ref{results}. General results are in Section~\ref{sampleresults} where we include descriptions of tests we performed to assess the robustness of our work, and Section~\ref{analcomp} presents the comparison with the other two early science P-MaNGA articles. Conclusions and discussion follow and close the paper.

\section{P-MaNGA observations}\label{data}

MaNGA is both an instrument suite and a survey that uses the Sloan 2.5m telescope \citep{2006AJ....131.2332G}. By September 2014, identical sets of MaNGA hardware were installed in six SDSS cartridges, allowing six individual plug-plates to be observed on a given night.  Each MaNGA cartridge (or ``cart'' for short) consists of 17 science integral field units (IFUs) ranging in size from 19 fibers (12\farcs5 diameter) to 127 fibers (32\farcs5 diameter) that patrol a 3$^\circ$ diameter field of view.  Twelve 7-fiber ``mini-bundles'' are used for standard star flux calibration and 92 individual fibers for sky subtraction.  

All 1423 fibers have a 2\farcs0 core diameter and a 2\farcs5 outer diameter and---when the cartridge is mounted to the telescope---feed the highly sensitive BOSS spectrographs (\cite{2013AJ....146...32S}) which provides continuous wavelength coverage from 3600 \AA\ to 10,300 \AA\ at a spectral resolution R $\sim 2000$ (R $\approx$ 1600 at 4000 \AA, and R $\approx$ 2300 at 8500 \AA, ) with a total system throughput of $\sim$25\%.  We use subsets of this wavelength range in the spectral fitting analysis of this paper, due to the limitations of the empirical simple stellar population models that are used, as discussed in Section \ref{sec:models}, which will be expanded in future work. Details on the MaNGA hardware are given by \cite{2015AJ....149...77D}. Instrument commissioning began in January 2014.

In this work, we use data obtained using the MaNGA engineering prototype instrument (P-MaNGA) in January 2013.  P-MaNGA was designed to explore a variety of instrument design options, observing strategies, and data reduction algorithms. These P-MaNGA data offer a valuable look at MaNGA's potential but differ substantially from the MaNGA survey data in several ways. We summarise these differences in Table \ref{pmanga_manga} and describe them below.

First, P-MaNGA used only 560 total fibers distributed across just one of the two BOSS spectrographs.  470 of these fibers had a 2\farcs0 diameter core, while 90 had either a 3\farcs0 or 5\farcs0 arcsec core and are not used for the present analysis.  410 of the 2-arcsec fibers were bundled into 8 IFUs with three sizes: 19 fibers (\N{19}), 61 fibers (\N{61}), and 127 fibers (\N{127}).  The P-MaNGA IFU complement was dramatically different from the MaNGA survey instrument, with $5 \times $\N{19} (instead of just two), $1 \times $\N{61} (instead of 4), $2 \times $\N{127} (instead of five), and no 37-fiber or 91-fiber IFUs. The masses and sizes of the P-MaNGA galaxies observed are therefore not representative of the final MaNGA sample.

\begin{table}
\centering
{\footnotesize
\begin{tabular}{ c | c | c}
\hline\hline
Survey & P-MaNGA & MaNGA \\ \hline\hline
\multicolumn{3}{l}{Target galaxy properties} \\ \hline\hline
Sample size & 18 & $\sim$ 10 000 \\
$z$  & 0.01 --- 0.06 & 0.01 --- 0.15\\
$M_i$ / mag & --18.5 --- --22.7 & $\sim$ --17 --- --23 \\ 
\Rekpc & 0.7 -- 3.4 & 1.5 (Primary) and \\
&& 2.5 (secondary) \\ \hline
\multicolumn{3}{l}{Instrument design} \\ \hline\hline
Flux standard type & Single 2\asec fiber & Mini-bundle \\
Fibers used per cart & 560 & 1423  \\ \hline
& 5 x \N{19} (12\asec),  & 2 x \N{19} (12\asec), \\
& & 4 x \N{37} (17\asec), \\
IFU & 1 x \N{61} (22\asec), &  4 x \N{61} (22\asec), \\
distribution &&  2 x \N{91} (27\asec), \\
& 2 x \N{127} (32\asec). & 5 x \N{127} (32\asec),  \\
&& 12 x \N{7} mini-bundles. \\ \hline\hline
\end{tabular}
\caption{Summary of differences between P-MaNGA (this paper) and the full MaNGA survey.}
\label{pmanga_manga}}
\end{table}

P-MaNGA observations were obtained for 3 galaxy fields using plates 6650, 6651, and 6652 (see summary in Table \ref{tab:sample}).  In each case, observations were obtained in sets of three 20-minute exposures dithered by roughly a fiber radius along the vertices of an equilateral triangle to provide uniform coverage across each IFU. These three fields were observed to varying depths, and in varying conditions as required by the P-MaNGA engineering tasks that they were designed for.  Although plate 6650 (Field 9) was observed to a depth comparable to what will be regularly achieved during MaNGA operations, plates 6651 (Field 11) and 6652 (Field 4) are both significantly shallower than MaNGA survey data, and plate 6651 was intentionally observed at high airmass resulting in particularly poor image quality.

Some of the P-MaNGA targets were drawn from early versions of the MaNGA sample design, but in many cases P-MaNGA targets were chosen for specific reasons. In each of the three plates, one \N{127} IFU was allocated to a galaxy already observed by the CALIFA survey \citep[Calar Alto Large Integral Field Area, ][]{2012A&A...538A...8S} for comparison purposes, even if it would not otherwise satisfy the MaNGA selection cuts.  Additionally, the non-optimal IFU complement of the P-MaNGA instrument required some targets to be selected manually.  Altogether 18 P-MaNGA galaxies were observed (see Table \ref{tab:sample}).

The raw data was reduced using a prototype of the MaNGA Data Reduction Pipeline (DRP), which is described in detail by Law et al. (in prep).  In brief, individual fiber flux and inverse variance spectra were extracted using a row-by-row algorithm, wavelength calibrated using a series of Neon-Mercury-Cadmium arc lines, and flat-fielded using internal quartz calibration lamps.  Sky-subtraction of the IFU fiber spectra was performed by constructing a cubic basis spline model of the sky background flux as seen by the 41 individual fibers placed on blank regions of sky, and subtracting off the resulting composite spectrum shifted to the native wavelength solution of each IFU fiber. 

Flux calibration of the P-MaNGA data is performed by fitting \cite{1979ApJS...40....1K} (and revisions) model stellar spectra to the spectra of calibration standard stars covered with single fibers at each of the three dither positions. Such methods are shown from SDSS DR6 \citep{2008ApJS..175..297A} to yield median calibration errors over all plates of less than 2\%. This method is preferable to observing flux-standard stars at different times from the galaxy observations, as even on the same night systematic errors (e.g. due to changing airmass) are larger than the calibration error quoted above. The flux calibration vectors derived from these single-fiber spectra were found to vary by $\sim$ 10\% from exposure to exposure, depending on the amount of light lost from the fiber due to atmospheric seeing and astrometric misalignments.  The flux calibration error is due to the positional error of the fibers, which can be different from plate to plate. However, error is a slowly varying function of wavelength which is absorbed by the dust method described in Section \ref{measuringdust}, thereby minimising the impact on the spectral fitting. While this uncertainty is acceptable for the present science purposes, the flux calibration uncertainty of the single fibers ultimately drove the design decision of the MaNGA survey to instead use 7-fiber IFU ``mini-bundles'' for each calibration standard star, which results in a photometric accuracy better than 2.5\% between Halpha and Hbeta, and better than 7\% between [NII] 6584 and [OII] 3726, 3729 for MaNGA survey data (see Yan et al., in prep).

Flux calibrated spectra from the blue and red cameras were combined together across the dichroic break using an inverse-variance weighted basis spline function.  Astrometric solutions were derived for each individual fiber spectrum that incorporate information about the IFU bundle metrology (i.e., fiber location within an IFU), dithering, and atmospheric chromatic differential refraction, among other effects.  Fiber spectra from all exposures for a given galaxy were then combined together into a single datacube (and corresponding inverse variance array) using these astrometric solutions and a nearest-neighbor sampling algorithm similar to that used by the CALIFA survey.  For the P-MaNGA datacubes, a spaxel size of 0\farcs5 was chosen.  The typical effective spatial resolution in the reconstructed datacubes can be described by a Gaussian with ${\rm FWHM} \approx 2$\farcs5.  When binning the datacubes, we scale the resulting error vectors to approximately account for wavelength and spatial covariance in the P-MaNGA error cubes. The spectral resolution is given by ${\rm FWHM} = 2.35 \sigma$, which is a function of wavelength as described in \cite{2013AJ....146...32S}. In this analysis we use P-MaNGA datacubes, which have been processed through the prototype DRP and a prototype data analysis pipeline (PDAP) being developed at the Institute of Cosmology and Gravitation of the University of Portsmouth\footnote{For information on the PDAP contact Daniel Thomas, daniel.thomas@port.ac.uk}. The relevant details are provided below.

\subsection{The prototype data analysis pipeline (PDAP)}\label{pdap}

\begin{figure*}
\begin{center}
\includegraphics[width=\linewidth]{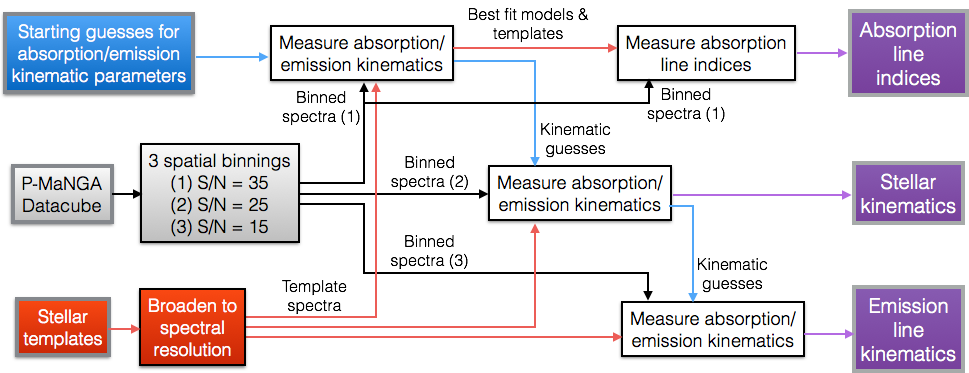}
\caption{Summary of the processes of the Prototype Data Analysis Pipeline (PDAP), used to recover the physical parameters used as input in this work in addition to the P-MaNGA datacubes. P-MaNGA datacubes are binned in three spatial schemes according to the signal-to-noise requirements of the properties extracted from the kinematic fits. Each of the processes are described in section \protect\ref{pdap}. Each kinematic fit also outputs improved starting kinematic guesses for the next spatial bin.}
\label{pdap_flow}
\end{center}
\end{figure*}

The PDAP analyses the reduced datacubes produced by the DRP to measure
galaxy's physical parameters, such as kinematics and emission line
fluxes. This is summarised as a flow diagram in Figure \ref{pdap_flow} and explained in the following paragraphs.

The first analysis block within the PDAP is to spatially bin adjacent
spectra to increase the signal-to-noise ($S/N$) ratio. This is done by
using the Voronoi Binning algorithm\footnote{Available from purl.org/cappellari/software} \citep{2003MNRAS.342..345C}.
We implement optimal weighting to determine the $S/N$ of each bin as described in section 2.1 of \cite{2003MNRAS.342..345C}, in which the weighting given to each pixel is proportional to its $S /(N^2)$. Because the various galaxy's
physical parameters require different minimum $S/N$ to be measured
with a desired accuracy, the PDAP performs the spatial binning three
times, producing three sets of spectra binned with different $S/N$
criteria according to the scientific needs.  These three sets of spectra
are processed independently in subsequent analysis blocks, to
extract the various galaxy parameters.

The $S/N$ values quoted in the following sections are computed by calculating the median signal to noise per pixel over the wavelength range $5560~<~\lambda/\mathrm{\AA}~<~6942$, the wavelength range of the SDSS $r$-band \citep{1996AJ....111.1748F}, which straddles the dichroic break of the SDSS spectrograph \citep{2013AJ....146...32S}, in order to better relate to the signal-to-noise measurements in SDSS photometry. However, the binned $S/N$ values in the P-MaNGA data do not include error covariance. Hence the $S/N$ thresholds in the following sections are empirically chosen to give reliable extraction of each of the corresponding properties.

The first spatial binning requires a minimum $S/N=35$ per bin and is
used to measure absorption line indices. This requirement is based the analysis of the errors of absorption indices as derived by \cite{2012MNRAS.421.1908J}. The second spatial binning requires a minimum of $S/N=25$ and is used to measure stellar kinematics. The requirement is derived from estimating the errors on kinematic parameters as in \cite{2004PASP..116..138C}. The third spatial binning with a minimum $S/N=15$ is used to measure the properties of the
emission lines (kinematics, fluxes, equivalent widths, reddening, and
Balmer decrement). This requirement is based on errors computed in \cite{2013MNRAS.431.1383T}.

The second analysis block within the PDAP fits the spatially
binned spectra with a series of stellar templates and Gaussian
functions. The fit is executed 3 times, one for each of the 3 groups
of spatially binned spectra produced in the first analysis block.

The processes in the PDAP are applied to the full wavelength range of P-MaNGA data, however for the spectral fitting analysis with \firefly we use a large portion of this wavelength range due to the limitations of the current empirical simple stellar population models used, as discussed in \ref{sec:models}. In future work, we will expand this analysis to the full wavelength range.

The spectral fitting routine uses an implementation of the Penalized
Pixel Fitting (hereafter, PPXF$^{3}$ \citep{2004PASP..116..138C}) and Gas AND
Absorption Line Fitting (\cite{2006MNRAS.366.1151S}), hereafter GANDALF, and it is performed into
two steps.  In the first step, the galaxy's ionized-gas lines are
masked, and the stellar continuum is fitted with a series of simple stellar population (SSP)
models from \cite{2011MNRAS.418.2785M}, which offer a large variety of calculations for various stellar libraries. 
In particular, the MILES-based models, which extend to $\sim~7000~\mathrm{\AA}$ are joined with the fully theoretical M11 models based on the MARCS-stellar spectra which, by having the very high spectral resolution of 0.1~$\AA$, makes it possible to match the required high spectral resolution of MILES ($2.54~\mathrm{\AA}$~(\cite{2011A&A...531A.109B}, \cite{2011A&A...531A.165P}, \cite{2011A&A...532A..95F}))~also in the near-IR (see Section~\ref{sec:models}).\footnote{These extended models as the others from M11 are available at: www.maraston.eu/M11.}
In this first step, stellar kinematics are measured. 

In the second step, the entire (unmasked) spectrum is fitted with a series of
SSP models and Gaussian functions, but the stellar kinematics are kept
fixed from the measurements obtained in the first step. Gaussian
functions are linked into three groups, so that each group of lines
have the same velocity and velocity dispersion. The three groups are
i) Balmer lines, ii) NaI absorption lines (5890--5896 \AA), iii) all
the other ionized-gas emission lines. Each gas line is broadened with
a Gaussian function to match the instrumental FWHM at the wavelength
that the specific line is detected at. Multiplicative and additive
polynomials, or a dust attenuation law (following the prescriptions
in \cite{2006MNRAS.366.1151S}) can be added to the fit procedure in these two
steps.

The first execution of the spectral fitting routine (both steps as described above) is
performed onto the galaxy spectra of the first spatial binning. In
this execution, 6th order multiplicative polynomials are used to fit to the continuum. The
best fitting model of the emission lines is then removed from the
galaxy binned spectra, and the strength of absorption line
indices are measured. The best fitting stellar template and the best
fitting stellar template convolved by the best fitting stellar Line Of Sight Velocity Dispersion (LOSVD)
are then used to correct absorption line indices for internal
kinematic broadening. The best fitting stellar and gas kinematics are
used as starting guesses for the next execution of the spectral
fitting routine.

The second execution (both steps as above) is
performed onto the galaxy spectra of the second spatial binning. Only
stellar templates that have been selected in the first execution are
used. In this execution, 6th order multiplicative polynomial and 3rd
order additive polynomials are used for the fit. The best fitting
stellar and gas kinematics will be used as starting guesses for the
next execution. The best fitting stellar kinematics obtained in this
step will be saved in the final output table, and will characterize
the galaxy stellar kinematics.

The third execution of the spectral fitting routine (both steps as above) is
performed to the galaxy spectra of the third spatial binning. Only
stellar templates that have been selected in the first execution are
used. In this execution, 6th order multiplicative polynomial and 3rd
order additive polynomials are used in the first step, which is a fit using PPXF.
Multiplicative polynomials, or a reddening extinction law if required,
are used in the second step, which is fit using an adapted version \citep{2013MNRAS.431.1383T} of the code 
GANDALF \citep{2006MNRAS.366.1151S}. The best fitting gas
kinematics (velocity and velocity dispersion) of each fitted emission
line will be averaged together using the line flux as weight to
characterize the mean gas kinematics of the galaxy. Gas-subtracted
galaxy spectra are de-redshifted using the stellar kinematics measured
in this third execution.

In this work, we used the PDAP to bin the galaxy spectra to reach a
minimum $S/N=5$, lower than the thresholds used previously since we use full spectral fitting rather than fits to emission or absorption lines. We chose $S/N=5$ as it seems adequate to provide well-constrained fits, based on tests done on mock galaxy spectra in Wilkinson \& Maraston (in preparation). We then fitted the binned spectra following the prescription of the third execution of the spectral fitting routine, namely to remove gas lines and to set spectra to rest-frame. The gas-free, rest-framed spectra are then analyzed, as described in Section 4.

\begin{table*}
\centering
{\footnotesize 
\begin{tabular}{lccccccccccccl}
\hline\hline
Nickname&mangaID& Bundle &RA &Dec &$z$ &$\log$ &$M_i$ &$(g-r)$ &$R_e$  & $R_{\rm IFU}$   & Morphological \\
 & & &\multicolumn{2}{c}{(J2000 degrees)} & &$M_{*}$& mag & mag & '' &(\Reff) &  type$^\diamond$ \\ \hline
\hline
\multicolumn{12}{c}{Group $\alpha$: MaNGA-like conditions and setup, has been selected for MaNGA sample.$\dagger$} \\
\hline
p9-127B & 12-131835 & ma008 & 143.7764 & +21.6277 & 0.013 & 9.1 & -18.5 & 0.51 & 6.8  & 2.4  & LTG \\
p9-19B &12-131893  & ma005 & 142.7900 & +22.7465 & 0.051 & 10.6 & -22.2 & 0.86 & 4.0  & 1.6   & ETG  \\
p9-19D &12-131577 & ma007 & 142.7878 & +20.9168 & 0.034 & 10.3 & -21.2 & 0.79 & 3.2  & 2.0   & ETG\\
p9-19E &12-131821 & ma001 & 145.1260 & +21.2538 & 0.024 & 9.7 & -20.0 & 0.73 & 2.9  & 2.2  & ETG \\
\hline
\multicolumn{12}{c}{Group $\beta$: MaNGA-like conditions and setup, not selected by MaNGA.} \\
\hline
p9-127A & 12-188794 &ma003      & 143.7400 & +21.7053 & 0.013 & 10.7 & -21.3 & 0.70 & 23.7  & 0.7   & LTG  \\ %CALIFA
p9-61A & 12-188807  & ma002   & 144.2993 & +21.6692 & 0.019 & 10.1 & -20.5 & 1.3? & 9.3  & 1.2   &  ETG \\
\hline
\multicolumn{12}{c}{Group $\gamma$: Poorer conditions and/or setup than MaNGA, has been selected for MaNGA sample.$\dagger$} \\
\hline
p4-19A &12-113557 & ma004 & 165.0504 & +36.3873 & 0.027 & 9.4 & -19.5 & 0.49 & 2.5 &  2.5   & LTG \\
p4-19B &12-113506 & ma005& 162.4946 & +36.4150 & 0.023 & 9.5 & -19.5 & 0.53 & 4.6 &  1.4    & LTG \\
p4-19C &12-109657 & ma006& 164.0237 & +36.9600 & 0.022 & 9.5 & -19.4 & 0.72 & 4.8 & 1.3    & ETG \\ \hline
p11-61A* &12-93688 & ma002     & 208.0482 & +13.9999 & 0.024 & 10.1 & -20.7 & 0.77 & 7.3 &  1.5   & LTG \\
p11-19B* &12-93551 & ma005 & 207.2918 & +13.3470 & 0.024 & 9.2 & -19.4 & 0.42 & 2.4 &  2.6    & LTG \\ 
\hline
\multicolumn{12}{c}{Group $\delta$: Poorer conditions and/or setup than MaNGA, not selected by MaNGA.} \\
\hline
p4-127A &12-109661 & ma003 & 163.9803 & +36.8615 & 0.022 & 10.7 & -22.1 & 0.84 & 10.3   &1.6    & LTG \\ %CALIFA
p4-127B &12-109682 & ma008    & 163.2461 & +37.6134 & 0.042 & 11.0 & -22.7 & 0.77 & 13.9  & 1.2  & LTG \\
p4-61A &12-113576 & ma002   & 164.4442 & +36.2827 & 0.030 & 9.7 & -20.0  & 0.71 & 3.4  & 3.4   & LTG \\ \hline
p11-127A* &12-196354 & ma003   & 207.8786 & +14.0922 & 0.024 & 10.9 & -22.4 & 0.85 & 17.0  & 1.0    & LTG \\ %CALIFA
p11-127B* &12-196727 & ma008 &  209.2309 & +14.1423 & 0.016 & 9.4 & -19.4 & 0.42 & 9.6  & 1.7   & LTG \\
p11-19A* &12-196773 & ma004& 207.5811 & +14.1407 & 0.062 & 9.5 & -21.1 & 0.85 & 1.1  & 2.8    & LTG \\
p11-19C* &12-93538 & ma006 & 206.7061 & +14.4005 & 0.021 & 10.0 & -20.9 & 0.54 & 2.8  & 2.2    & LTG \\ \hline
\hline
\end{tabular}
\caption{Details of P-MaNGA galaxy sample. Galaxy nicknames are structures as p (prototype), Field number (see Table \protect\ref{plate_table}) -- IFU bundle size, iterative name A/B/C. Properties are from the SDSS Data Release 8 database \protect\citep{2011ApJS..193...29A}. Stellar masses are from the MPA-JHU catalogue based on \protect\cite{2003MNRAS.341...33K}, in the Data Release 8 photometric pipeline, scaled to 3 arcsecond fiber magnitudes. These are compared to our mass estimates and other mass estimates from broad-band fitting in section \protect\ref{masscomparison}. Starred galaxies (*) in groups $\gamma$ and $\delta$ correspond to galaxies taken under very poor observational conditions and setup (plate 11, see Table \protect\ref{plate_table}), whereas other galaxies in these groups are taken with a setup that is closer to, but still poorer than, a MaNGA-like setup (plate 4, see Table \protect\ref{plate_table}).\\
$\diamond$ Morphologies of early-type (elliptical / lenticular) or late-type (spiral / irregular) are obtained through direct visual inspection of the corresponding SDSS images by the main author.\\
$\dagger$ P-MaNGA galaxies in these groups meet the selection criteria for either the MaNGA primary or secondary samples as described in \protect\cite{2015ApJ...798....7B}, and will be observed in MaNGA.
\label{tab:sample}}
} \end{table*}

\begin{table}
\centering
{\footnotesize
\begin{tabular}{c | c | l}
Field & Groups & Conditions and setup \\ \hline
9 & $\alpha$/$\beta$ & Exposure 3.0 hr, seeing 1\farcs7 \\
4 & $\gamma$/$\delta$ & Exposure 2.0 hr, seeing 1\farcs3 \\
11 & $\gamma$/$\delta$ & Exposure 1.0 hr, seeing 2\farcs0, (airmass $\sim$1.5) \\
\end{tabular}
\caption{Summary of observational conditions and setup\label{plate_table}}}
\end{table}

\section{Analysis tools}\label{analysis tools}

\subsection{Models}\label{sec:models}

In this work we use the stellar population models of \cite{2011MNRAS.418.2785M}, hereafter M11, in order to match observed spectral energy distributions (SEDs) to physical properties of the stellar populations observed. 
We use the spectra of the base models, i.e. simple stellar population (SSPs). SSPs are coeval populations of stars with a given age, metallicity (in terms of [Z/H]), and initial mass function (IMF). 

The M11 models assume the same energetics - stellar tracks and fuel consumptions in post-Main Sequence - as those by \cite{2005MNRAS.362..799M}, and in particular include the contribution of the Thermally-Pulsating Asymptotic Giant Branch (TP-AGB) phase (see below).
The main difference is the assumed input stellar spectra, which allows the calculations of model spectra at higher spectral resolution. M11 explored all available optical empirical stellar libraries, namely MILES \citep{2006MNRAS.371..703S}, STELIB \citep{2003A&A...402..433L}, and ELODIE \citep{2007astro.ph..3658P} and also a theoretical stellar library from the MARCS \citep{2008A&A...486..951G} model atmosphere calculations. The latter allows the calculation of high-resolution models also in the near-IR. The resulting stellar population parameter coverage is summarized in Table~\ref{tab:modelparams}.\footnote{See \url{www.maraston.eu/M11/README_M11.txt} for details.}  

A few model features are worth noticing. First, empirical libraries do not contain enough M dwarf stars to cover the full temperature and gravity extension of the theoretical Main Sequence from stellar tracks. M11 filled the empirical libraries with theoretical dwarf spectra from the MARCS library, smoothed to the appropriate resolution. Second, none of these optical empirical libraries contain Carbon or Oxygen-rich stars required to characterise the TP-AGB phase. M11 use the same low-resolution empirical spectra from \cite{2000A&AS..146..217L} that were used in \cite{2005MNRAS.362..799M} and re-bin them via interpolation in wavelength in order to match the wavelength binning of the given library. Even if the actual resolution of those spectra remain low, the effect of the TP-AGB component on the continuum and broad features is visually identifiable in the spectra.

For the full spectral fitting analysis we use the empirically-based models because they extend down to low ages (see Table \ref{tab:modelparams} and M11). This means that in practice we do not exploit the full wavelength extension of the P-MaNGA data, which we will pursue at a later stage when empirical SSP models with young stars with suitable resolution in the near-IR will be available.

An interesting finding of M11 were variations in the optical model spectra due to the input stellar library (see Figure~12 in M11). These potentially impact the galaxy properties that are obtained when different models are applied to data. For example, M11 fit the M67 globular cluster spectrum with all models finding that MILES-based models released an old (9 Gyr), half-solar metallicity best-fit solution for this cluster, while STELIB-based models gave the parameters determined from the Colour Magnitude Diagram (CMD), namely 3 Gyr and solar metallicity.  These findings motivate us to quantify these subtle model effects on the derived properties of the P-MaNGA data (see Section~\ref{librarycomparison}). Here it should be noted that when we make direct comparisons between models we use the parameter coverage contained in all three of the models. This corresponds to a wavelength range of 3900 -- 6800 \AA, ages varying between 30 Myr and 15 Gyr, and metallicities of 0.5, 1.0, and 2.0 times solar metallicity. We also use two different initial mass functions (IMFs) for each of these libraries, Salpeter \citep{1955ApJ...121..161S} and Kroupa \citep{2001MNRAS.322..231K}. In a separate project we plan to use the data to constrain more exotic IMFs.

\begin{table*}
\centering
{\footnotesize 
\begin{tabular}{| l | cccc |}
\hline\hline
Model & Wavelength coverage & Age coverage & Metallicity \\
library & (min -- max) / \AA & (min -- max) / Myr & coverage (Z/H) \\ \hline \hline
MILES & 3500 -- 7429 & 6.5 -- 15000 & [0.0001, 0.001, 0.01, 0.02, 0.04] \\
STELIB & 3201 -- 7900 & 30 -- 15000 & [0.01,0.02,0.04] \\ 
ELODIE & 3900 -- 6800 & 3 -- 15000 & [0.0001, 0.01, 0.02, 0.04] \\ \hline
\end{tabular}
\caption{Parameter coverage of the stellar population models as a function of input stellar library, available at \protect\url{www.maraston.eu/M11/README_M11.txt}\label{tab:modelparams}}
} \end{table*}

\subsection{Fitting Method}\label{fittingmethod}

To obtain stellar population properties from the data, we use a new full spectral fitting code, which we named \firefly. This was written with the purposes of mapping out inherent spectral degeneracies, working well at low S/N, comparing input stellar libraries, and making as few assumptions about the star formation histories derived as possible by modelling each star formation history as a linear combination of bursts (SSPs). The code is discussed in detail with tests to mock galaxies and example applications to SDSS galaxies in Wilkinson \& Maraston (in preparation), and is summarised in this paper in the following paragraphs.

\firefly~is a chi-squared minimisation fitting code that for a given input observed SED, returns a set (typically on the order $10^2$ -- $10^3$) of model fits, each of which is a linear combination of SSPs. 
This set of fits is obtained through an iterative algorithm that is coded as follows. 

We initially fit all base SSPs on the observed spectrum and save their chi-squared value.\footnote{Calculated as $\chi^{2} = \sum\limits_{\lambda} \frac{(O(\lambda)-M(\lambda))^{2}}{E(\lambda)^{2}}$, where $O$ is the observed SED, $M$ is the model SED, and $E$ is the error spectrum, all as a function of wavelength $\lambda$.} 

We then begin improving these fits, by checking if their chi-squared values are improved by adding a different SSP component with equal luminosity as the first one. For example, a 1 Gyr, solar metallicity SSP fit ($Fit_{1} = M_{1 Gyr, Z_\odot}$) may be improved by adding a 2 Gyr, solar metallicity component, thus creating a linear combination of SSPs as a fit to the data as $Fit_{2} = \frac{1}{2}~M_{1 Gyr, Z_\odot} + \frac{1}{2}~M_{2 Gyr, Z_\odot}$. Each of the possible combinations of two SSPs are checked for improvement on the one-SSP fits, and a selection of the best set of fits from both one- and two-SSP fits is saved for the next iteration. This means we are somewhat limited in the precision of weightings we can achieve with individual fits, since the smallest possible multiple of weighting on a given SSP will be given by 1 / (number of iterations). However, since we construct many (on the order of 1000) possible combinations, we can achieve a higher level of precision by combining these fits together by their probability, see below.

This process then iterates to the next step, where we check whether any of 2-SSP fits can be improved with a further SSP contribution. Thus following the convention in the example above, we may find an improved fit with $Fit_{3} = \frac{1}{3}~M_{1 Gyr, Z_\odot} + \frac{1}{3}~M_{2 Gyr, Z_\odot} + \frac{1}{3}~M_{0.5 Gyr, 0.5 Z_\odot}$. Thus we continue iteratively checking for additional SSP components until adding further SSPs no longer improves the fit. SSP components that already exist in the previous iteration of the fit can still be added to create new fits, thus allowing solutions with different weightings of SSPs.

To avoid overfitting and allow for convergence of the number of SSPs used in each linear combination, each iteration must improve on the previous fit by a statistically significant amount. To enforce this we employ the Bayesian Information Criterion (\cite{2007MNRAS.377L..74L}), $\textrm{BIC} = \chi^2 + k~ln~n$ at each iteration, where $k$ is the number of fitting parameters used (in our case, the number of SSPs added in combination to make a fit), and $n$ is the number of observations (in our case, the number of flux points used in the fit). Each iteration must reduce the BIC to be included in the output set of fits. This therefore adds a penalty term to the chi-squared value that scales linearly to the number of SSPs used in each linear combination. For the majority of \firefly~fits of P-MaNGA data, fits converge at around the fifth iteration.

Once a set of fits are obtained, we calculate each of their stellar masses (including the subtraction of stellar mass losses and the contribution by stellar remnants as described in \cite{1998MNRAS.300..872M}; \cite{2005MNRAS.362..799M}), light-weighted and mass-weighted ages and metallicities, and the chi-squared values. We then calculate likelihoods for each of these fits by using the chi-squared cumulative probability distribution as follows:

\begin{equation}\label{chi_prob}
 \  P ( X > \chi^2_o ) = \int_{\chi^2_o}^{\inf} \frac{1}{2^{(k/2)}\Gamma(\frac{k}{2})}X^{(k/2)-1} e^{-X/2} \mathrm{d}X,
\end{equation}

where $\Gamma$ is the Gamma function, $\chi^2_o$ is the critical value of chi-squared we use to find the cumulative probabilities. $k$ is the degrees of freedom, which can be expressed as $k = N-\nu-1$, where $N$ is the number of (independent) observations and $\nu$ is the number of fitting parameters. Hence for our method, $N$ is the number of data flux points that we are fitting model fluxes to (of order 3000 for P-MaNGA observations), and $\nu$ is the number of SSPs used in linear combination to obtain our fits. In cases where errors are badly estimated (either over- or under-estimated), resulting in a poorly constrained likelihood distribution, we assume that the error, and therefore the chi-squared values, is offset by a constant multiplicative factor in order to bring the lowest chi-squared value equal to the degrees of freedom, and re-calculate the likelihoods of the fits accordingly.

We then use these probabilities to fit a gaussian profile\footnote{Since we achieve enough fits with each run of \firefly, these properties are generally very well fit by Gaussian profiles.} to each of the marginal distributions of the parameters in order to estimate their averages and standard deviations. The standard deviations are used as estimates for the errors on each of the properties. We note that the error estimate using this method will include the spectral degeneracies between the stellar population properties folded into the random errors. Generally however, higher random errors taken over a large range of different stellar populations will correspond to higher error estimates. Separating out the statistical errors on the spectra from the spectral degeneracies requires careful simulation, which will be analysed in future work. The exception to this method is the dust calculation, hence degeneracies with dust are not folded into these error estimates. The method for calculating dust is discussed in the next subsection.

We note that the errors derived in this paper are statistical plus degeneracy errors only. They do not include systematic errors from the modelling, fitting code or covariance between pixels. For a detailed discussion of the treatment of covariance for the SAMI IFU survey, we direct the reader to \cite{2015MNRAS.446.1551S}. However, we do quantify the size of the first two of these systematic effects in sections \ref{librarycomparison} and \ref{califacomp}. For the effect of covariance between pixels we note that MaNGA data will have more accurate estimation of spectral errors, allowing accurate propagation through to the final datacubes from which we can construct covariance measures between pixels, but this is beyond the timescale of this paper. Hence in this paper we assume that each 2\farcs0 fiber is an independent observation, which with dithering will increase in area slightly, to approximately 4 square arcseconds. Thus each 0\farcs5 spaxel the deconstructed datacube, in dependent on 15 other spaxels. Therefore when binning datacubes in this paper, for example in elliptical bins for the radial profiles, we calculate standard areas assuming that the number of independent observations is equal to the number of grid points analysed, divided by 16. We note that always applying this 1/16 factor to the variance in each radial bin means that the overall error on a single radial bin will be an overestimate, since all 16 pixels from a given fiber may not be included in each radial bin. However, this means that the covariance between bins is therefore included in the radial profiles and thus the gradient estimations. The codependence of neighbouring pixels will be analysed in more detail in future work.

\firefly~is able to work with either emission-cleaned spectra or implement masking over emission features as appropriate. In this paper we use emission-cleaned spectra, as described in section \ref{data}. GANDALF, when fitting to emission lines, also has the added benefit of releasing accurate measurements of the Doppler broadening of the SEDs due to the velocity dispersion of the stars in the object, based on pPXF fits (\cite{2004PASP..116..138C}). This broadening effect essentially `downgrades' the resolution to which we can identify absorption features in that SED, often below that of the instrumental resolution of the observation, and very often below that of the model spectra that we have used. Therefore one needs to downgrade the model SSP resolution to match this effective resolution. In all of our analyses this is done based on the measurement of this broadening from GANDALF with pPXF.

A particular advantage of the code when applied to P-MaNGA is the ability to map out degeneracies in age, metallicity and dust for each individual spectrum. These effects are also explored in detail for whole-galaxy spectra in Wilkinson \& Maraston (in preparation). The P-MaNGA data is binned spatially into Voronoi bins (see \cite{2003MNRAS.342..345C}) in order to give a target S/N per spatial bin of 5. This gives a 2D map of 1D spectra (in other words, a 3D datacube) which are then fit individually with \firefly~for each model combination.

\subsection{Measuring dust}\label{measuringdust}

A feature of the code that P-MaNGA has allowed us to hone is the measurement of dust extinction. The flux calibration issues discussed in Section \ref{data} require us to develop a more sophisticated approach to dealing with the large-scale wavelength feature of a dust attenuation curve, the measurement of which would otherwise become degenerate with the similarly large-scale feature of an unknown flux calibration curve. Visual inspection of the SDSS images corresponding to the P-MaNGA objects allows us to provide a sanity check on the results.

\begin{figure}
\begin{center}
\begin{subfigure}{\linewidth}
	\hspace{1cm}
	\includegraphics[width=0.7\linewidth]{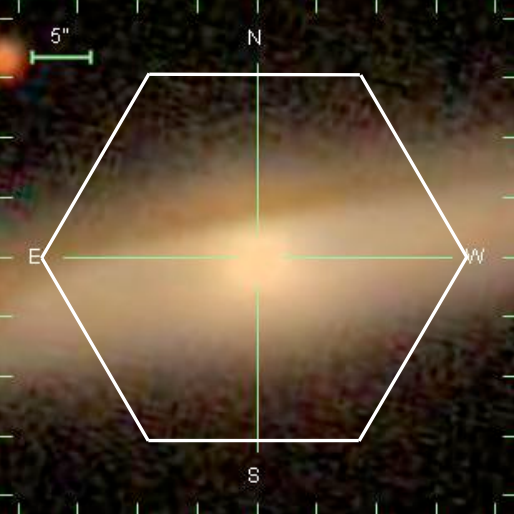}
	\caption{SDSS imaging data with the P-MaNGA footprint in white.}
\end{subfigure}
\begin{subfigure}{\linewidth}
	\includegraphics[width=\linewidth]{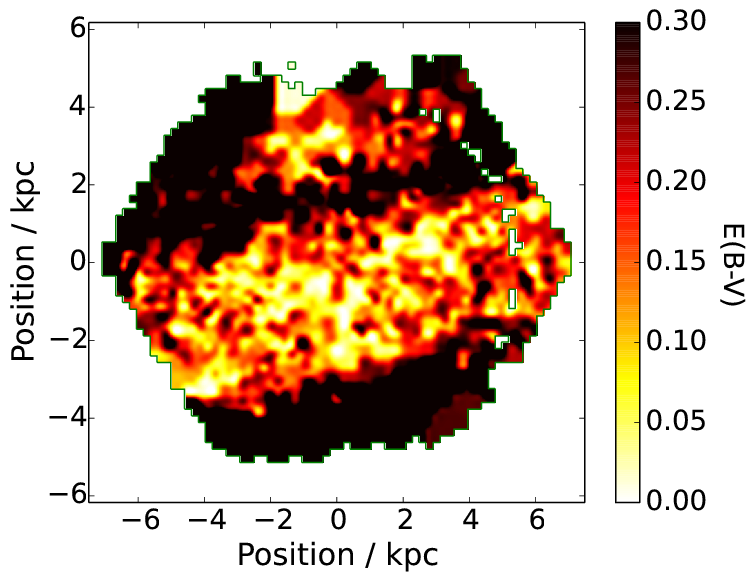}
	\caption{Dust values obtained when folding dust in as an additional parameter to the fitting code.}
\end{subfigure}
\begin{subfigure}{\linewidth}
	\includegraphics[width=\linewidth]{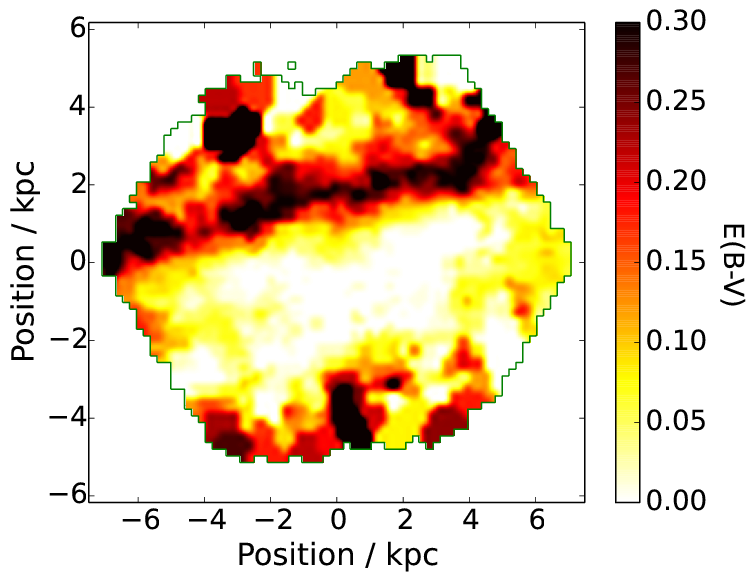}
	\caption{Dust values obtained when using a filter to measure the dust attenuation curves before fitting the stellar population properties.}
\end{subfigure}
\caption{The effect of different methods for dealing with dust attenuation in fitting codes with the example galaxy \galaxythree (in table \protect\ref{tab:sample}), analyzed using MILES-based M11 models with its full parameter range.}
\label{fig:baddust}
\end{center}
\end{figure}

Consider the galaxy in figure \ref{fig:baddust}a, where a dust lane clearly covers the top of the P-MaNGA footprint.

One could attempt to fold in dust extinction into the fitting code by treating it as an additional free parameter to the models. However in practice the dust-age-metallicity degeneracy together with flux calibration issues become so strong as to conspire to produce unphysical solutions. This is shown in figure \ref{fig:baddust}b, where the E(B-V) values correspond to the value of the single parameter Calzetti law \citep{2000ApJ...533..682C} to derive attenuation curves\footnote{In this paper we shall use this form of the Calzetti law, with R$_V$ = 4.05, as default in all of our derivations.}. The E(B-V) values have a high degree of scatter, are high in general, and the band of higher values that could be identified as a dust lane is clumpy and difficult to interpret. The result is clearly not an optimal representation of this galaxy.

\begin{figure*}
%\begin{center}
%\includegraphics[width=8.0cm]{}
\begin{subfigure}[t]{0.31\linewidth}
	\includegraphics[width=\linewidth]{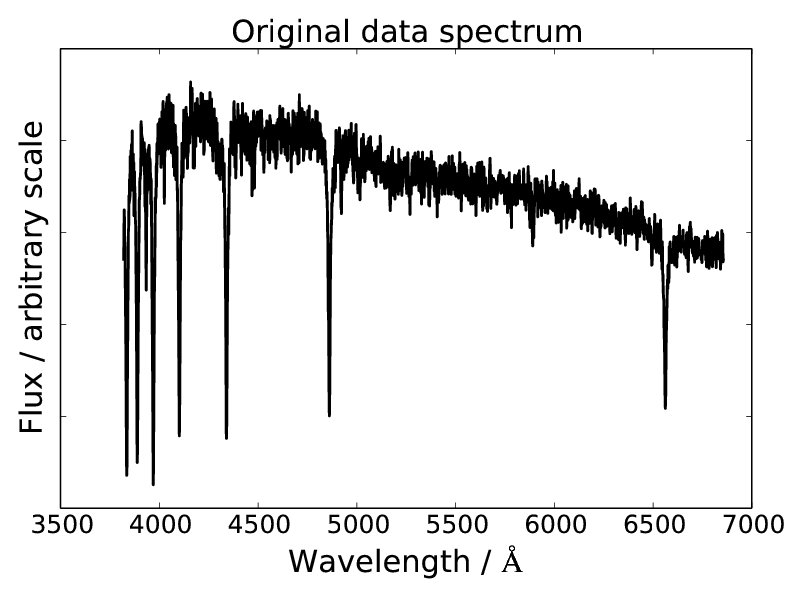}
	\caption{}
\end{subfigure}\hfill
\begin{subfigure}[t]{0.31\linewidth}
	\includegraphics[width=\linewidth]{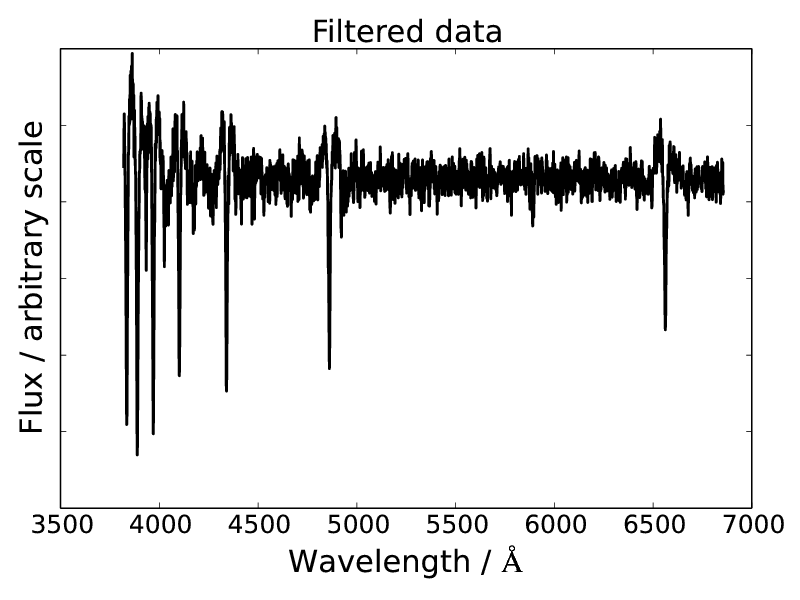}
	\caption{}
\end{subfigure}\hfill
\begin{subfigure}[t]{0.31\linewidth}
	\includegraphics[width=\linewidth]{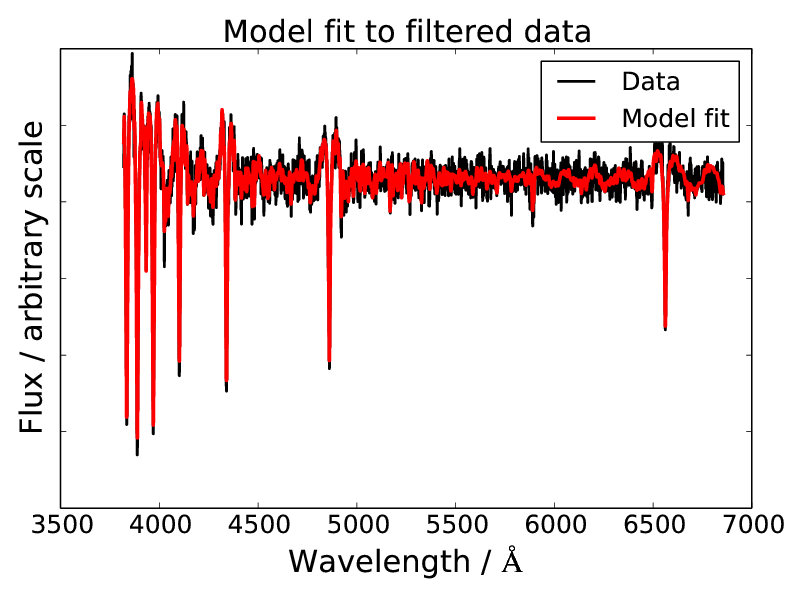}
	\caption{}
\end{subfigure}\hfill
\begin{subfigure}[t]{0.31\linewidth}
	\includegraphics[width=\linewidth]{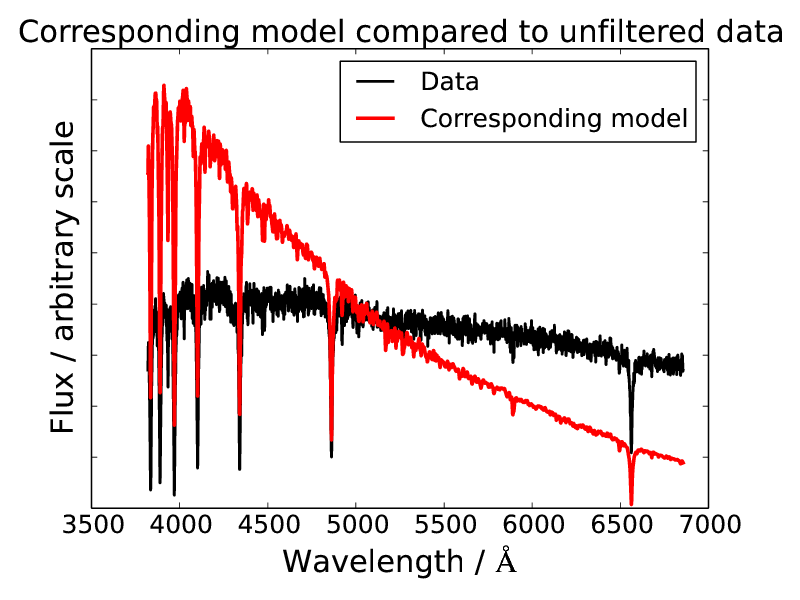}
	\caption{}
\end{subfigure}\hfill
\begin{subfigure}[t]{0.31\linewidth}
	\includegraphics[width=\linewidth]{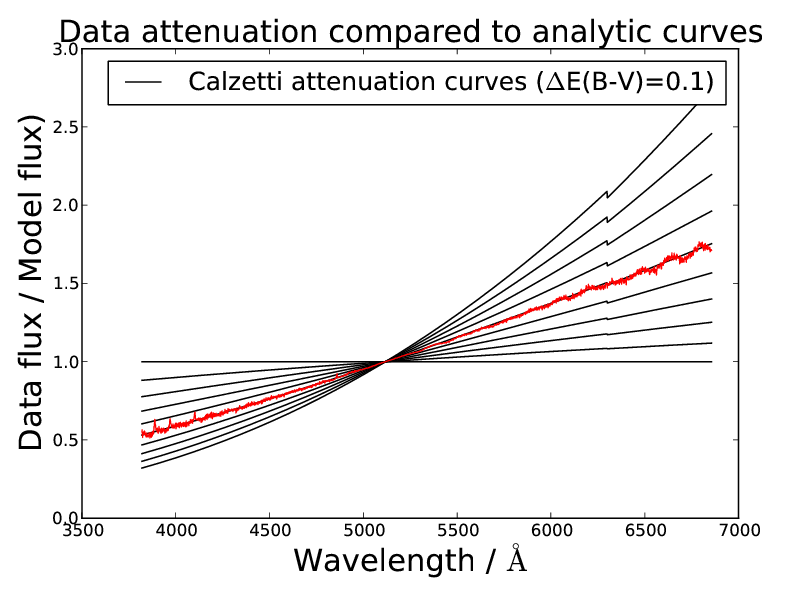}
	\caption{}
\end{subfigure}\hfill
\begin{subfigure}[t]{0.31\linewidth}
	\includegraphics[width=\linewidth]{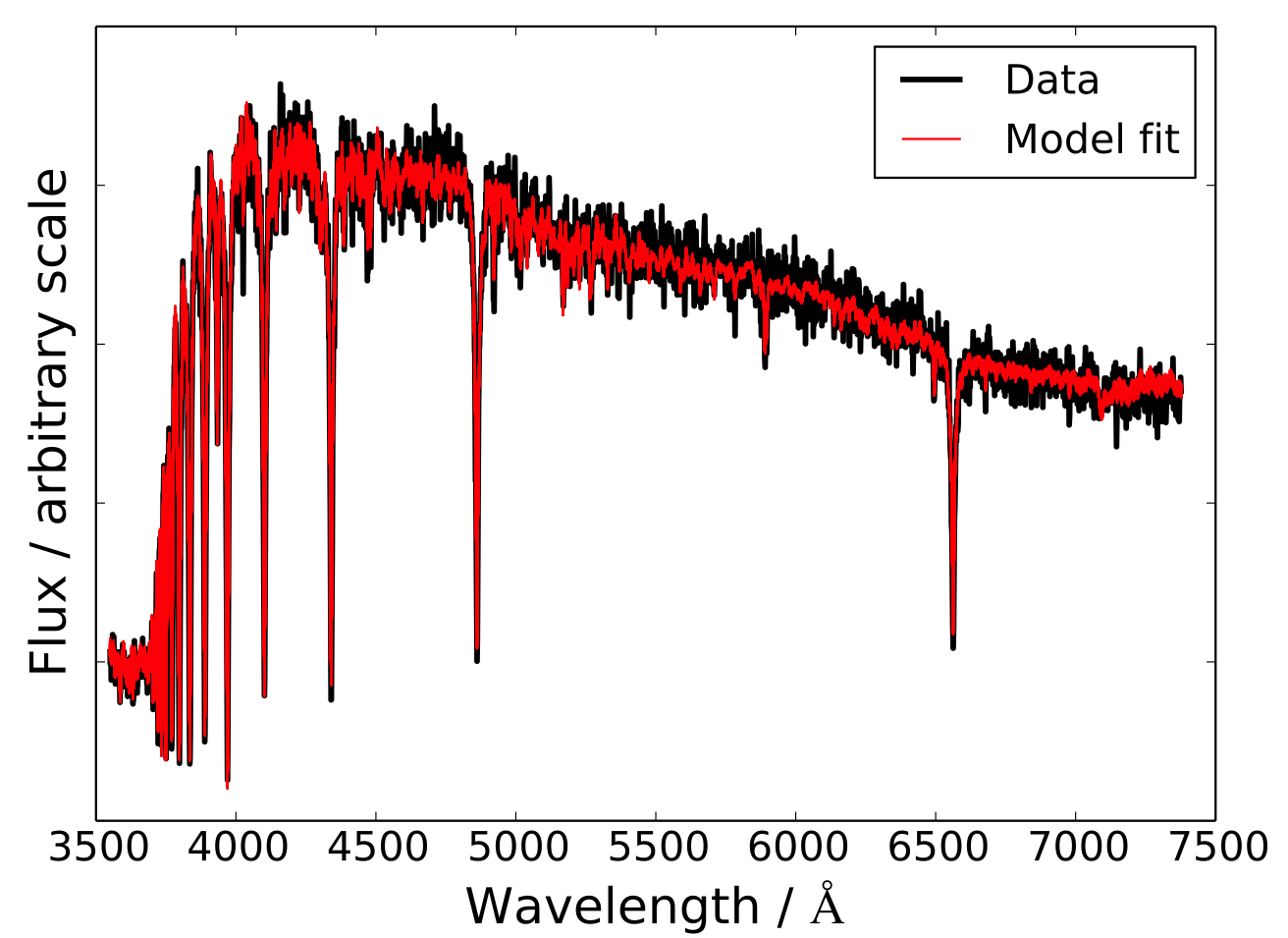}
	\caption{}
\end{subfigure}\hfill
\caption{Example of fitting a spectrum for dust using a high-pass filter. Starting top-left and moving horizontally, the panels describe each stage of the filter and fitting process. Panel a)  shows the input model spectrum corresponding to a 30 Myr, [Z/H]~=~0 SSP with a signal-to-noise applied at each flux point of 50. Panel b) shows that data processed through a high-pass filter that removes large-scale features such as continuum shape and dust extinction. Panel c) shows the fit to the filtered spectrum using models that are filtered in the same way. All models used to make this fit have E(B-V)~=~0 since we would be removing the effect of dust attenuation on the models anyway. Using the same parameters determined through the fit in c), panel d) shows the corresponding parameters in the unfiltered model case. We expect to see a systematic tilt between the data and models, corresponding to a dust attenuation curve. In panel e) we find the best value for E(B-V) by calculating the [data / model] spectrum and fitting this to a range of attenuation curves across the whole spectrum. In practice we used $\Delta$E(B-V) of 0.01, but intervals of 0.10 are shown here for clarity. The best fitting E(B-V) for this case is 0.50, exactly matching our input value of 0.50. Lastly we use the value of E(B-V) obtained to re-fit the original data using unfiltered attenuated models. Panel f) shows the final fit obtained across the whole wavelength range.}
%\end{center}
\label{fig:hpf}
\end{figure*}

The problem with this approach is that the large-scale shape of a spectrum is the product of both the stellar population components, the flux calibration factor, and also dust attenuation. On the other hand, the flux values of a spectrum at small-scales is a result of only the stellar population components and are not sensitive to dust or flux calibration. Typically fits to galaxy SEDs are driven by the continuum, due to it having better signal-to-noise and less noise correlation compared to individual features. Therefore these fits are very susceptible to dust, which can become particularly problematic. If there are (even small) offsets in the continuum shape (i.e. if the continuum is flux-calibrated incorrectly or badly modelled). It can be difficult to separate out the degeneracy between dust and other stellar population properties, since both affect the overall continuum shape.

Our solution is to use an analytical function across all wavelengths to rectify the continuum before deriving the stellar population parameters. We use a functional form called high-pass filter (HPF). A HPF removes the large-scale features of a spectrum, such as continuum shape and dust extinction, through the use of a window function applied to the Fourier transform of the spectra as follows:

\begin{equation}\label{hpf}
\	\mathrm{Flux}_{\lambda}^{\mathrm{output}} = \mathrm{Flux}_{\lambda}^{\mathrm{input}} \otimes \mathrm{W}_{\lambda},
\end{equation}

where this is the convolution in real space, and the window function W$_\lambda = \mathscr{F}^{-1}$W$_{k}$ describes which modes are removed by the Fourier filter, where $k$ are the modes of the spectra. Roughly these modes can be thought of as corresponding to features of the spectrum with wavelength size given by $\sim$ number of wavelength points in the spectrum divided by $k$. Since we remove the long modes of the spectrum, we want a window function that only allows small modes. By testing the results with mock galaxies made from simple stellar populations we find that the following step function gives good results, with the derived properties being stable to small changes in the function:

\begin{equation}\label{windowfunction}
\	\mathrm{W}_k = \begin{cases} 0 & k \le k_{crit} \\  1 & \text{otherwise,} \end{cases}
\end{equation}

where we set $k_{crit} = 40$ in this paper. 

For example, a spectrum with regular wavelength intervals of 1~\AA~between 3000~\AA~and 9000~\AA~will be completely represented by 6000 modes. A filter that excludes the first 20 modes of a spectrum will therefore ignore all features in the spectrum that are larger than 6000 / 20 = 300~\AA. The choice of modes excluded by the window function is chosen to remove potentially badly modelled or badly calibrated modes. 

The step-by-step process of obtaining a dust attenuation curve is described in figure \ref{fig:hpf}. 
\begin{enumerate}
\item In figure \ref{fig:hpf}a we show an example spectrum. This corresponds to a 30 Myr SSP from the MILES-based M11 models with a known E(B-V) of 0.5, applied using the Calzetti law. We apply a Gaussian displacement on each point corresponding to signal to noise of 50 for clarity, although this process works as well down to S/N of 5, which is the lowest value for the P-MaNGA data outputs that we will be using in this paper. This spectrum is chosen to represent what the stellar populations of a portion of the dust lane in the galaxy in figure \ref{fig:baddust}a might contain, using a known set of properties such that we can test their recovery. We apply random errors at each flux point assuming a signal-to-noise of 50. 
\item We apply equation \ref{hpf} to the data, giving the output shown in figure \ref{fig:hpf}b. This equation is also applied to all of the models, which are not attenuated by dust. 
\item These filtered models are then fit to the filtered data in Figure~\ref{fig:hpf}c. 
\item We then use the parameters measured from this fit in the unfiltered models and data (shown in figure \ref{fig:hpf}d), dividing the best model fit by the data to give a residual attenuation curve. 
\item As shown in figure \ref{fig:hpf}e, we then use the dust attenuation models of \cite{2000ApJ...533..682C} to find the best matching E(B-V) value that fits the recovered curve. The best fit model in this case returns E(B-V) = 0.5, exactly as input. Down to a signal-to-noise of 5, we can recover dust in mock galaxies to within 0.05 in E(B-V).
\item The best fitting curve is then applied to the models, which are then fit to the full unfiltered data spectrum to give the final fit in figure \ref{fig:hpf}f.\footnote{Errors on the dust extinction measured are evaluated separately to other stellar population properties since we fit this separately, and so consequently errors on age, metallicity and stellar mass do not include errors arising from their degeneracies with dust.}
\end{enumerate}

\begin{figure}
\centering
\begin{subfigure}[t]{\linewidth}
	\includegraphics[width=\linewidth]{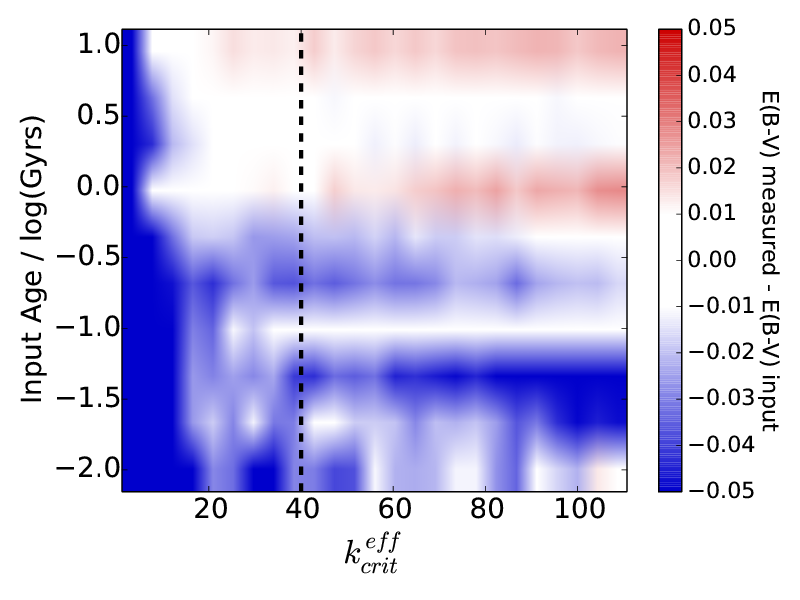}
\end{subfigure}
\begin{subfigure}[t]{\linewidth}
	\includegraphics[width=\linewidth]{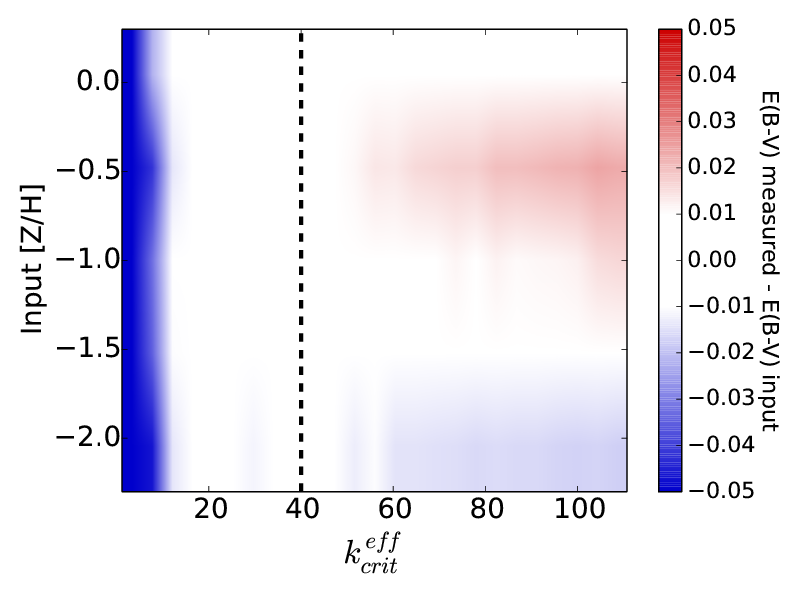}
\end{subfigure}
\begin{subfigure}[t]{\linewidth}
	\includegraphics[width=\linewidth]{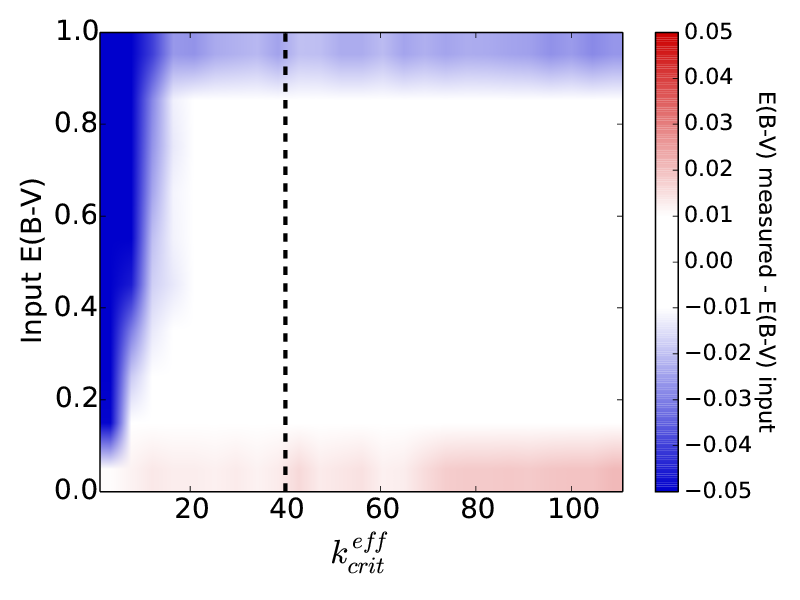}
\end{subfigure}
\caption{Tests showing the recovery of dust values (in terms of difference between E(B-V) recovered to input E(B-V)) from mock galaxy spectra from MILES-based models as a function of the number of large-scale modes excluded in the fit. The number of modes excluded is described by an effective $k_{crit}$ parameter (in equation \protect\ref{windowfunction}). The three plots show recovery of dust as a function of the following input properties: age, metallicity and extinction respectively. In each case, the values plotted are marginalised over the other two input properties. The black dashed line marks the value of $k_{crit}$ used in this paper for the analysis of P-MaNGA galaxies. Each mock galaxy spectra is constructed by perturbing each flux point of an input model SSP by a Gaussian probability distribution on each point, assuming a flat signal to noise of 10 across the whole wavelength range. A signal-to-noise of ten was chosen as it is typical of the outer regions of most P-MaNGA galaxies.}
\label{ktest}
\end{figure}

This number of large-scale modes excluded in the fit, selected by $k_{crit}$, has been tested empirically on a range of mock galaxy spectra based on SSPs. Figure \ref{ktest} shows the dust recovery, in terms of extinction in B-V, as a function of input age, metallicity and extinction. Each of these properties are marginalised over the other two, meaning that, for example, the bottom plot shows the recovery of dust as a function of $k_{crit}$ and dust input summed over all possible values of SSP age and metallicity input.

The plots show some dependance on dust recovered with age, visible as red or blue structure in the top plot, coming from age-dust degeneracy. However the E(B-V) values are recovered generally within a 0.03 margin within a large range of $k_{crit}$ from that used in this paper. The metallicity plot shows that dust is recovered very well except when excluding modes $k_{crit} > 60$, where some small dependencies ($<$ 0.02 in E(B-V)) start to become apparent. The input dust plot, bottom, similarly shows that dust recovery is excellent. We some edge effects where we are not able to fit below or above the dust parameter range we allow, but otherwise dust is recovered to within 0.01 in E(B-V). 

From these plots we also conclude that our method gives stable solutions with small changes in the window function around the $k_{crit} = 40$ value used in this paper. At around $k_{crit} < 15$ we see that we almost always underestimate the amount of dust, except for low values of input dust where we cannot construct an attenuation curve to remove. Conversely, at $k_{crit} > 90$ we see increasing instability in the age plot (top) as we are removing more information from our data in terms of wavelength modes used. 

To properly compare with the P-MaNGA analysis we have downgraded the wavelength sampling of the P-MaNGA spectra to match that of the wavelength sampling of the model spectra mock spectra. In these tests we have used MILES-based models, which have been interpolated to give a constant resolution of $R = \lambda / \Delta\lambda = 4343$. The P-MaNGA spectra, with sampling of $R = 10 000$ are therefore downgraded to the model sampling. Changing the wavelength sampling by a constant factor will change the $k_{crit}$ values by that same factor accordingly. Hence the value of the number of modes excluded is an effective $k_{crit}$, written as $k_{crit}^{eff}$, that applies to the P-MaNGA wavelength sampling. 

Mock spectra were made by taking input model SSPs and perturbing each of their flux points by an amount drawn randomly from a Gaussian probability distribution on each point given an input signal-to-noise. For these tests we use a signal-to-noise of 10, since these represent the typical values for P-MaNGA data in the outer regions of most of the sample galaxies. We note that lowering signal-to-noise to 5 increases the errors by approximately factor 2, whereas increasing the signal-to-noise to 20 reduces the errors by approximately a factor 2, but show very similar patterns to the plots shown in Figure \ref{ktest} otherwise.

It is important to note that the particular window function used remains self-consistent as it is applied in exactly the same way to both the data and model SEDs\footnote{We apply the filter directly to the model SSPs, and fit combinations of these to the filtered models. This takes considerably less CPU time than applying the filter to each of the model combinations, and makes only an extremely small difference to the fluxes at each point ($<$ 0.5 percent on any given flux point, most considerably less than this). This is because the filtering process tends towards a distributive operation at the large number of flux points on any given spectrum that we analyse.}.

Using this method we derive the dust map shown in figure \ref{fig:baddust}c. This is a clear improvement, as dust values are lower  and more consistent, and especially show a band of higher dust values that matches the position of the observed dust lane in figure \ref{fig:baddust}a.

This example case shows that the functions described in equations \ref{hpf} and \ref{windowfunction} can recover dust values well despite potential flux calibration problems in at least this galaxy. We also find sensible spatially confined dust determinations for other P-MaNGA galaxies, and hence use this approach as standard in this paper. Whether this more sophisticated approach to dust attenuation is advantageous for cases where flux calibration is not thought to be a problem will be discussed in future work.

To derive radial gradients of stellar population properties, we bin the stellar population properties output by \firefly~into elliptical annuli that are scaled in along the major axis such that the bins are constant in effective radius. From these summed properties we derive surface mass profiles, mass-weighted stellar age profiles, and mass-weighted stellar metallicity profiles. Example profiles for the galaxy \galaxyeighteen are shown in Figures~\ref{maps_eighteen}.

\section{Analysis of Stellar Population Maps}\label{results}

For each P-MaNGA galaxy we derive maps of stellar mass density, light-weighted and mass-weighted stellar age and metallicity in terms of [Z/H], and dust extinction in terms of E(B-V). We use the following conventions for denoting the age and metallicity properties: metallicities are written as \metalL~and \metalM~and ages are written as \ageL~and \ageM~(or \logageL~and \logageM~for the logarithmic values) for luminosity- and mass-weighted properties respectively. 

As noted above, we provide two types of stellar population maps. The first set shown in the plots are maps that are computed by weighting each stellar population component by their total luminosity across the wavelength range fitted, and find the geometric average of this to compute `light-weighted' properties as a function of position. The second set of maps and also the radial profiles are computed by weighting each stellar population component by their stellar mass contribution, again finding the geometric average of this to compute `mass-weighted' properties. 

In this paper we show both mass- and light-weighted properties since they can complement each other well to identify certain processes more clearly. For example, recent star formation will dramatically reduce the light-weighted stellar age deduced due to the high luminosity it contributes, despite perhaps being a very small mass component. A featureless \ageM~map with a steep negative \ageL~gradient can thus be interpreted as a galaxy that has most of its stars distributed uniformly across the galaxy but with some more star formation in its outer regions. We discuss case-by-case where these differences are greatest. Conversely, small or negligible differences in light- and mass-weighted properties come from having star formation histories that are well represented by a single-burst of star formation at its formation. Moderate differences can therefore be interpreted as having extended episode(s) of star formation. We note that \firefly ~fits stellar population components using their light contributed, and thus mass-weighted properties are decoded from the light-weighted properties by calculating the mass-to-light ratio of each stellar population component.

In all radial profile plots and analysis of gradients we use mass-weighted property values since these give insight into the mass assembly history of the galaxies. Mass-weighted properties better trace the whole evolutionary history of the galaxy, rather than being highly affected by phases contributing a high proportion of light, such as recent star formation. 

In the following sections we will discuss some of the features identified in the stellar population maps by grouping galaxies by their similarity in data quality to the full MaNGA samples, both primary and secondary, and whether they would be included in the MaNGA target selection. Since in this work we do not calculate covariance between pixels due to the preliminary nature of the error estimates on the spectra, as discussed in Section \ref{fittingmethod}, we calculate the statistical standard errors on the gradients obtained assuming that each pixel is dependent on 15 other pixels that in total represent a dithered 2\farcs0 fiber observation.

After discussing each of the groups ($\alpha$ to $\delta$) from Table \ref{tab:sample} separately in this section, we provide summary plots in section \ref{sampleresults} showing the gradients obtained across the whole sample and the statistical errors obtained on the profiles derived as a function of observational condition. We then test the stability of the results obtained as a function of stellar library. Lastly, we compare with literature values for both total stellar mass and detailed IFU stellar population information.

\subsection{MaNGA-selected galaxies with high-quality data: Group \large{$\alpha$}}

\begin{figure*}
\centering
%\vspace{2cm}
\hspace{0.3cm}
\begin{subfigure}{0.30\linewidth}
	\hspace{0.4cm}
	\includegraphics[width=0.72\linewidth]{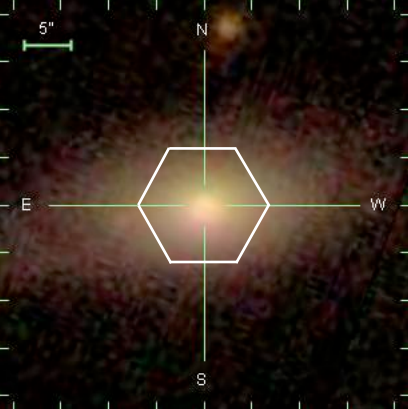}
	\caption{SDSS image with the P-MaNGA footprint.}
\end{subfigure} 
\begin{subfigure}{0.32\linewidth}
	\includegraphics[width=\linewidth]{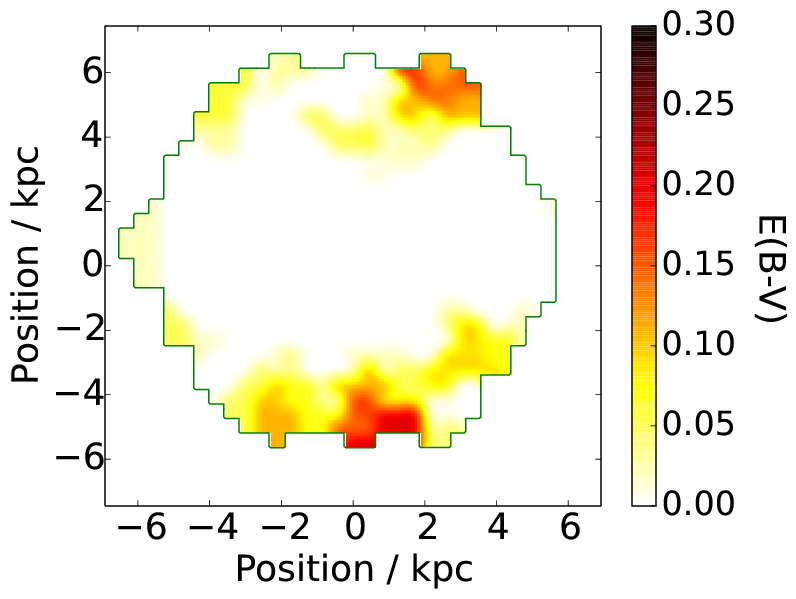}
	\caption{Dust extinction, E(B-V).}
\end{subfigure}
\begin{subfigure}{0.32\linewidth}
	\includegraphics[width=\linewidth]{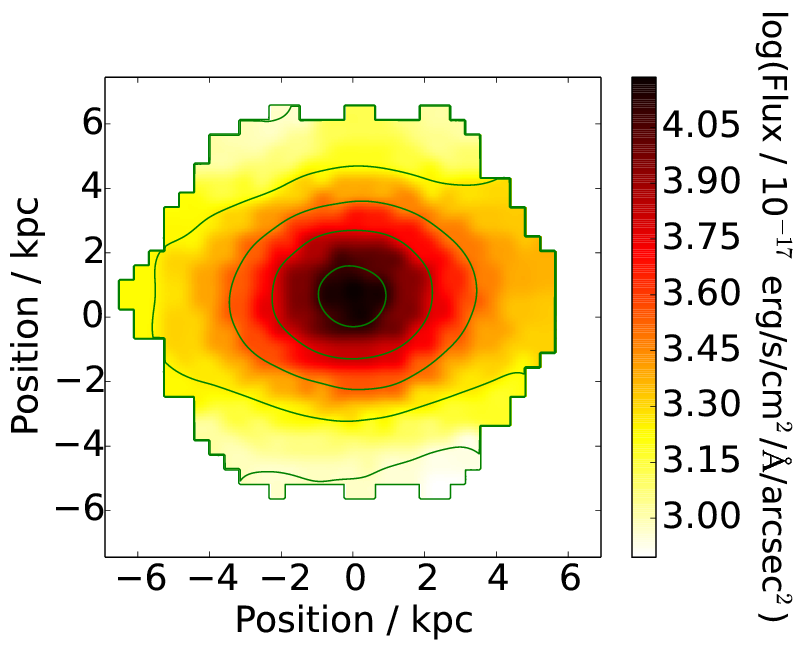}
	\caption{Flux map with isoflux contours (green).}
\end{subfigure}\hspace{0.1cm}
\begin{subfigure}{0.32\linewidth}
	\includegraphics[width=\linewidth]{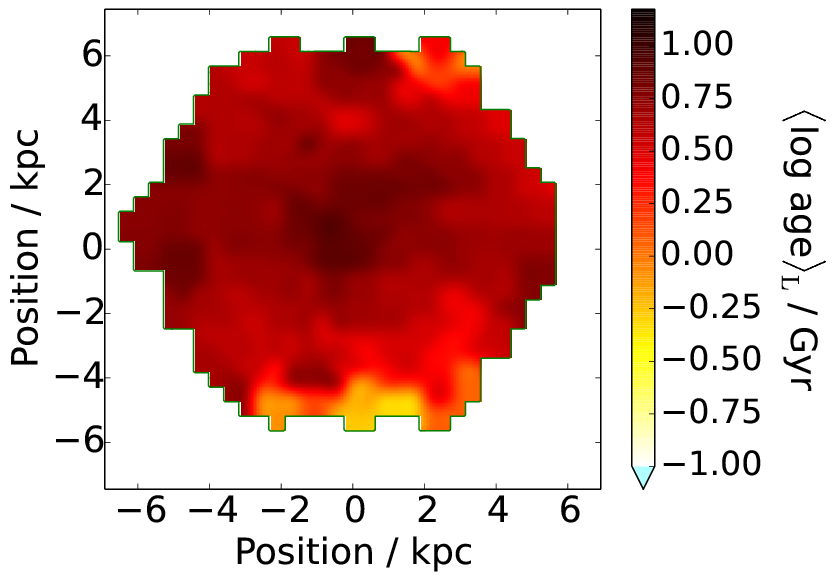}
	\caption{Luminosity-weighted stellar age.}
\end{subfigure}
\begin{subfigure}{0.32\linewidth}
	\includegraphics[width=\linewidth]{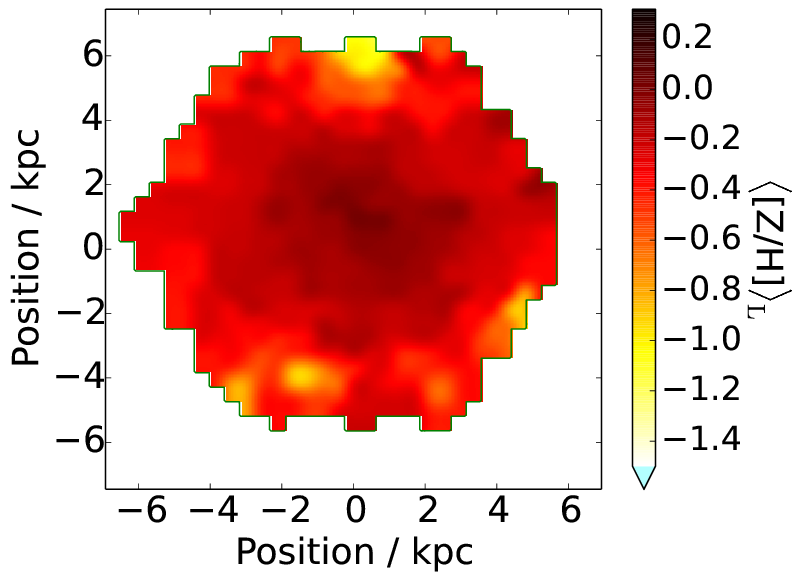}
	\caption{Luminosity-weighted metallicity.}
\end{subfigure}
\begin{subfigure}{0.32\linewidth}
	\includegraphics[width=\linewidth]{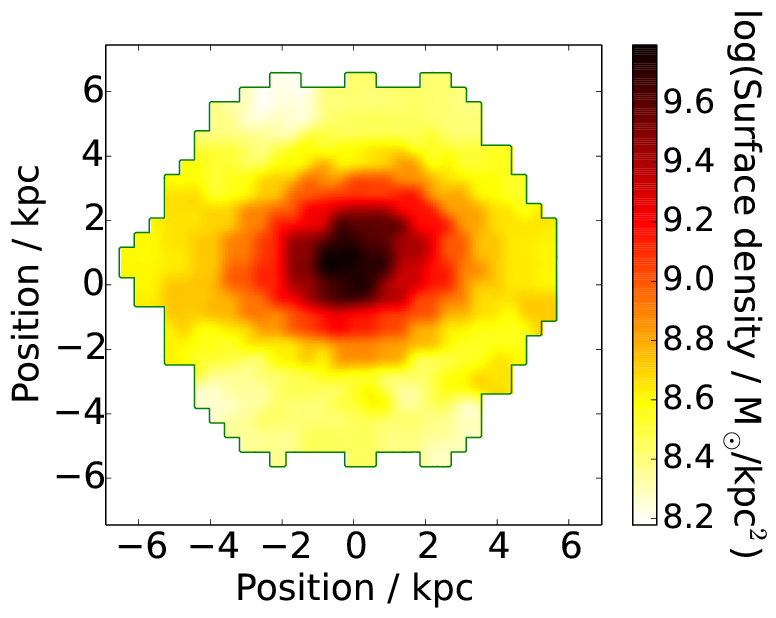}
	\caption{Stellar mass.}
\end{subfigure}
\begin{subfigure}{0.32\linewidth}
	\includegraphics[width=\linewidth]{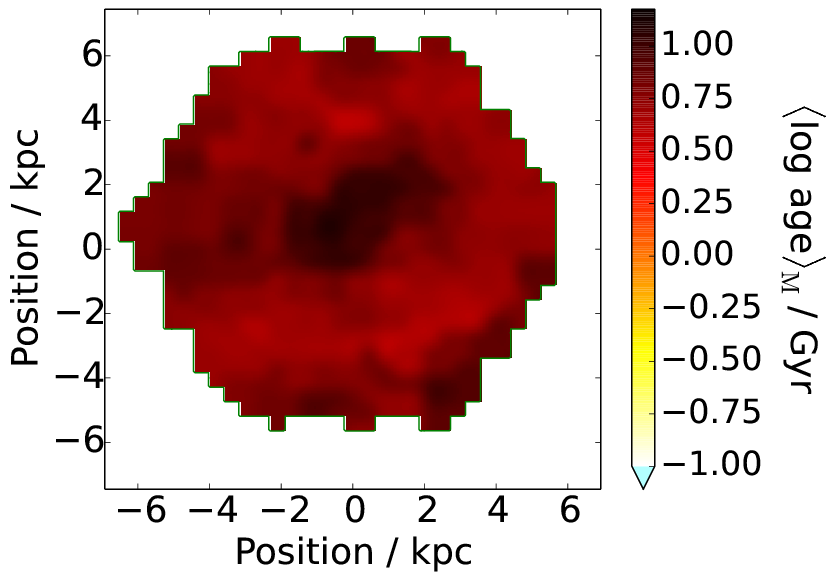}
	\caption{Mass-weighted stellar age.}
\end{subfigure}
\begin{subfigure}{0.32\linewidth}
	\includegraphics[width=\linewidth]{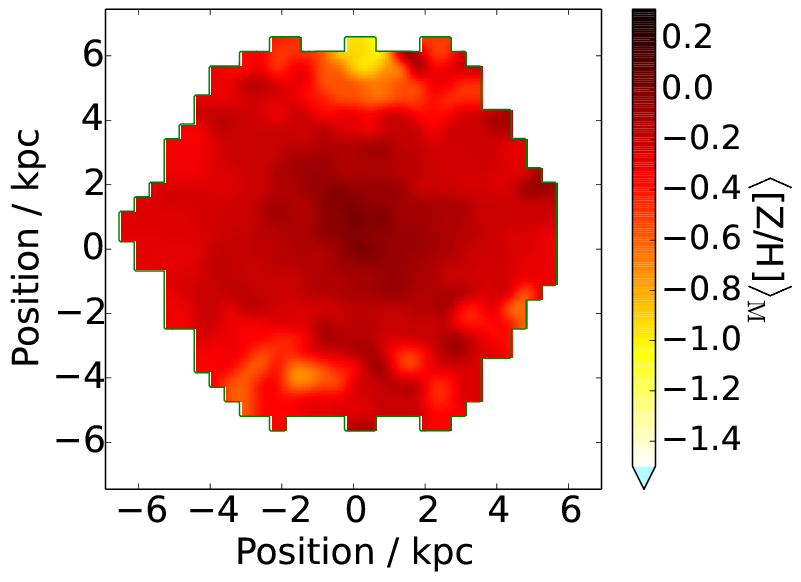}
	\caption{Mass-weighted metallicity.}
\end{subfigure}
\begin{subfigure}{0.32\linewidth}
	\includegraphics[width=\linewidth]{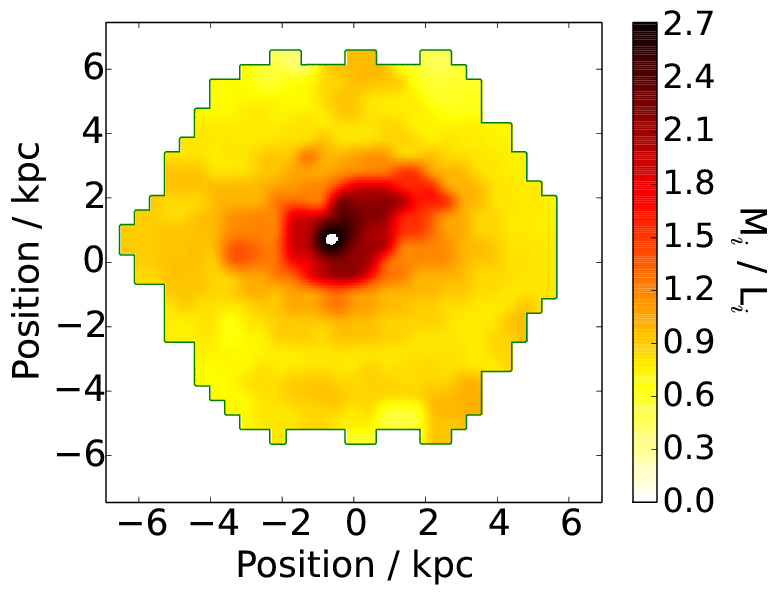}
	\caption{Stellar mass-to-light ratio in the SDSS $i$-band.}
\end{subfigure}
\begin{subfigure}{0.3\linewidth}
	\includegraphics[width=\linewidth]{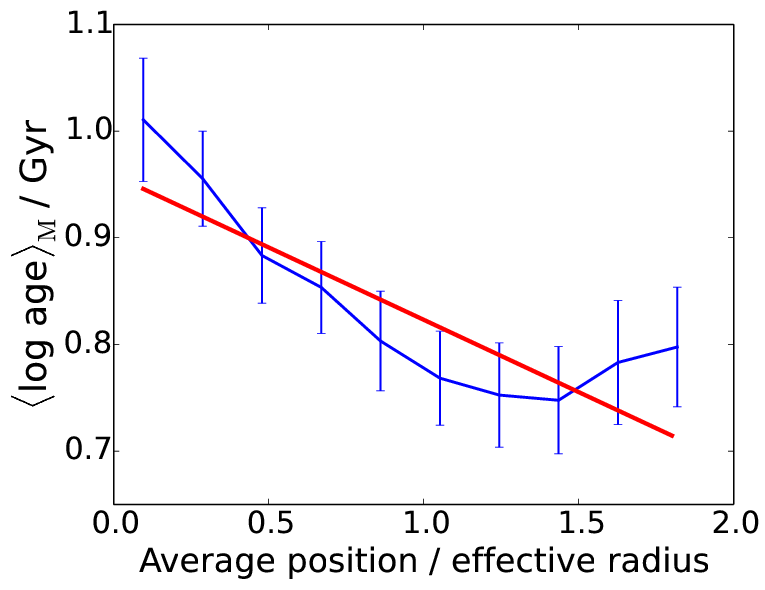}
	\caption{Radial age profile.}
\end{subfigure}\hspace{0.2cm}
\begin{subfigure}{0.3\linewidth}
	\includegraphics[width=\linewidth]{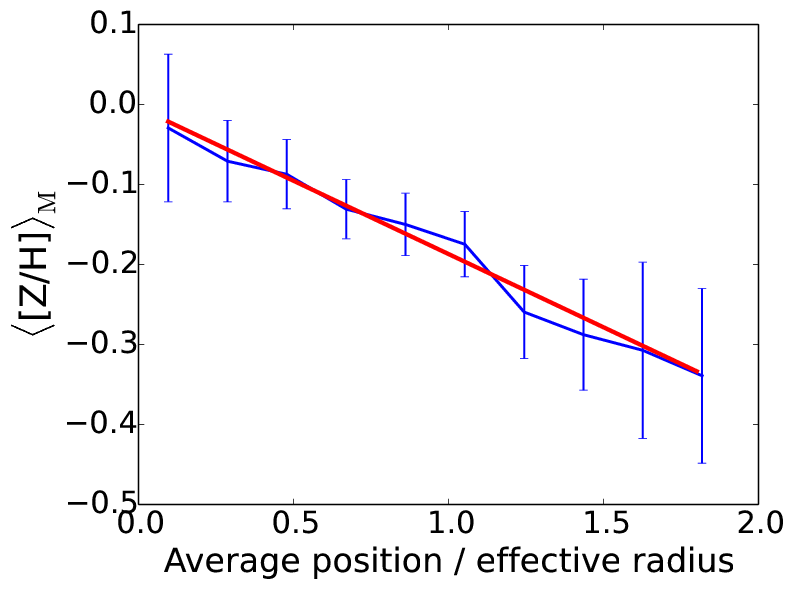}
	\caption{Radial metallicity profile.}
\end{subfigure}\hspace{0.4cm}
\begin{subfigure}{0.3\linewidth}
	\includegraphics[width=\linewidth]{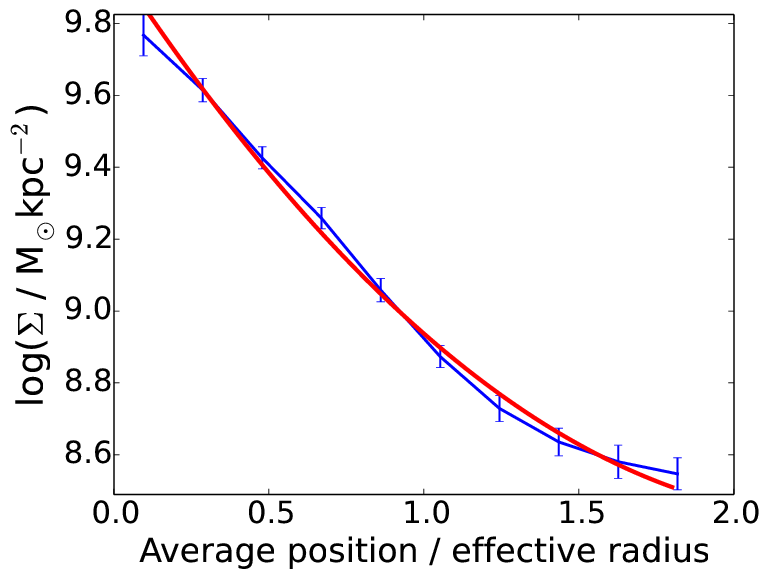}
	\caption{Stellar mass surface density gradient profile.}
\end{subfigure}\hspace{0.9cm}
\caption{{\bf Group $\alpha$, galaxy \galaxyeighteen} as in table \protect\ref{tab:sample}. Stellar population maps and profiles analyzed using MILES-based models with their full parameter range. The radial profiles are over-plotted in red with a linear fit to the data, in the age and metallicity profiles to show the derived gradient, and a quadratic fit in the stellar mass density profile in order to guide the eye. We use a pixel-to-pixel Gaussian filter to smooth the maps for ease of visualisation, but use a 0.5 sigma kernel in order to retain small-scale features.This galaxy has been observed with good observational conditions with a MaNGA-like exposure time and dithering, and is selected in the MaNGA primary sample. Therefore out of the P-MaNGA dataset, these observations are the most similar to the expected output of the MaNGA primary sample.}
\label{maps_eighteen}
\vspace{2cm}
\end{figure*}

Two galaxies from P-MaNGA - p9-19B (Figure \ref{maps_eighteen}) and p9-19D (Appendix Figure \ref{maps_seventeen}) - fall into the MaNGA primary sample selection whilst also maintaining MaNGA-like data quality with 3 hour exposures and decent (1\farcs7) seeing conditions. Both have been observed with the \N{19} IFU setup. 

These galaxies are visually identified as early-type galaxies and indeed show a high level of radial symmetry in their age, metallicity and stellar mass density maps. 

Galaxy p9-19B has a \logageL = 8 Gyr / \logageM = 10 Gyr, slightly super-solar metallicity core going down to \logageL = \logageM =  5 Gyr with sub-solar metallicity at above 1 effective radius, and is largely dustless except for at very high radii along the minor axis. Galaxy p9-19D is similar but with a weaker age and metallicity gradient.

In both of these cases we recover extremely smooth negative stellar population profiles with very low errors on each point, with the median of errors of 0.05, 0.10 dex in age, 0.05, 0.08 dex in metallicity, and 0.03, 0.07 dex in surface mass density for p9-19B and p9-19D respectively. This results in errors on their respective gradients determined to 0.03, 0.05 dex / \Rekpc in age and 0.05 -- 0.09 dex / \Rekpc in metallicity, which bodes well for MaNGA observations of early-type galaxies like these. % Low errors -> great! Discuss?gal 18

Two further galaxies from P-MaNGA fall into the MaNGA secondary sample, having wider radial coverage. The stellar population maps and profiles of these galaxies are shown and described in the Appendix.We note for both of these cases the more complex nature of the stellar populations as a function of position, coupled with the lower spatial resolution than the MaNGA primary sample, means that the uncertainties on the profiles derived are greater. Profiles are determined to a median of 0.14, 0.11 dex in age, 0.24, 0.17 dex in metallicity, and 0.10, 0.07 dex in stellar mass density, corresponding to errors on the gradients to be  0.07, 0.07 dex / \Rekpc in age and 0.14, 0.13 dex / \Rekpc in metallicity, for p9-19E and p9-127B respectively. This still allows recovery of significant gradients in all cases, showing that the MaNGA secondary sample should be able to find a good sample of clearly defined age gradients though may find difficulties for weak metallicity gradients.

\subsection{Non-MaNGA galaxies with high-quality data: Group \large{$\beta$}}

\begin{figure*}
\centering
\vspace{1.0cm}
\hspace{0.3cm}
\begin{subfigure}{0.30\linewidth}
	\hspace{0.4cm}
	\includegraphics[width=0.72\linewidth]{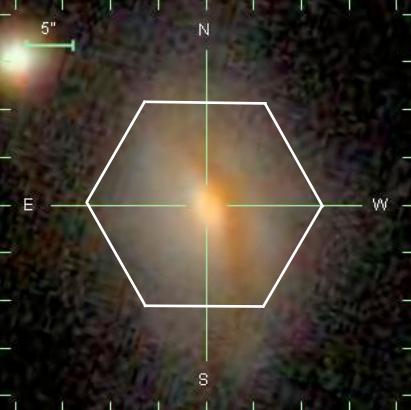}
	\caption{SDSS image with the P-MaNGA footprint.}
\end{subfigure} 
\begin{subfigure}{0.32\linewidth}
	\includegraphics[width=\linewidth]{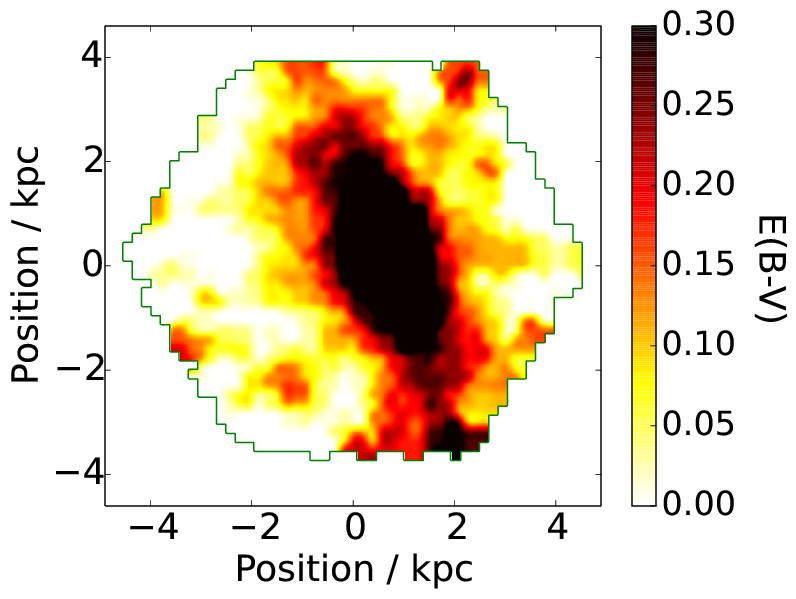}
	\caption{Dust extinction, E(B-V).}
\end{subfigure}
\begin{subfigure}{0.32\linewidth}
	\includegraphics[width=\linewidth]{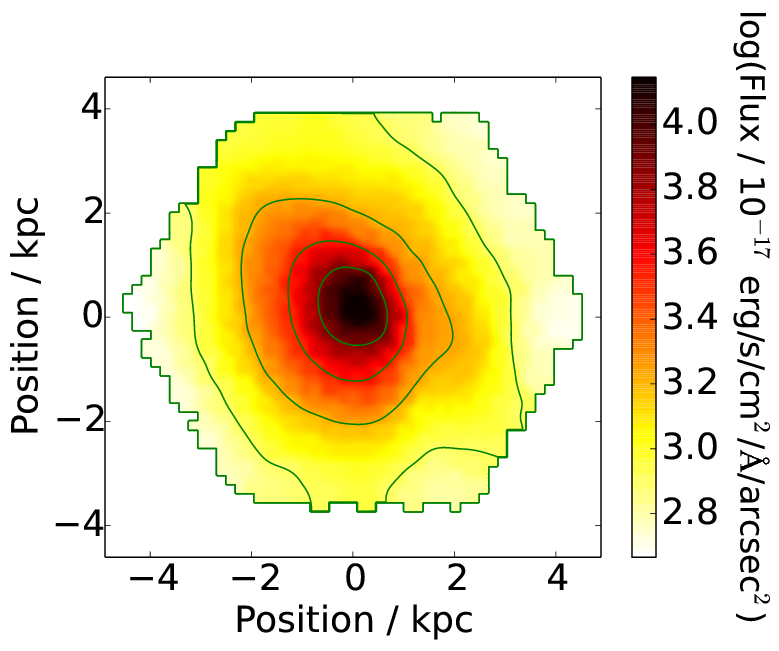}
	\caption{Flux map with isoflux contours (green).}
\end{subfigure}\hspace{0.1cm}
\begin{subfigure}{0.32\linewidth}
	\includegraphics[width=\linewidth]{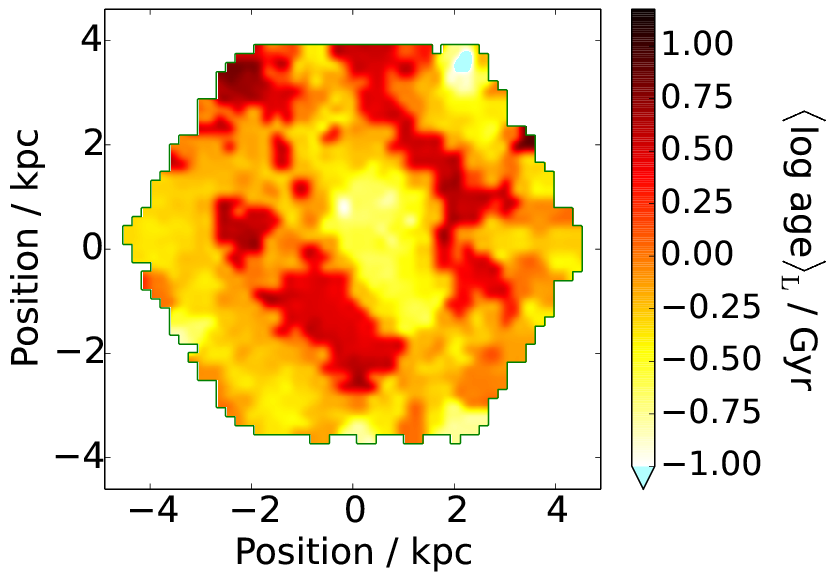}
	\caption{Luminosity-weighted stellar age.}
\end{subfigure}
\begin{subfigure}{0.32\linewidth}
	\includegraphics[width=\linewidth]{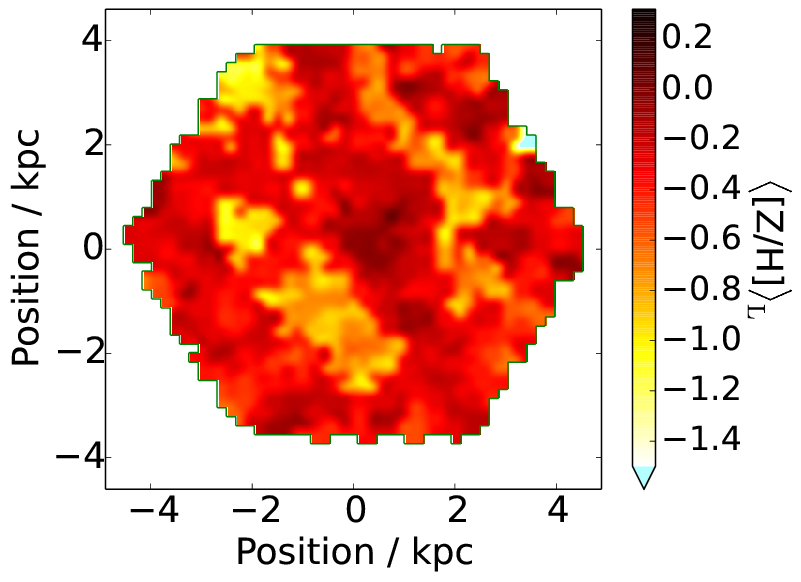}
	\caption{Luminosity-weighted metallicity.}
\end{subfigure}
\begin{subfigure}{0.32\linewidth}
	\includegraphics[width=\linewidth]{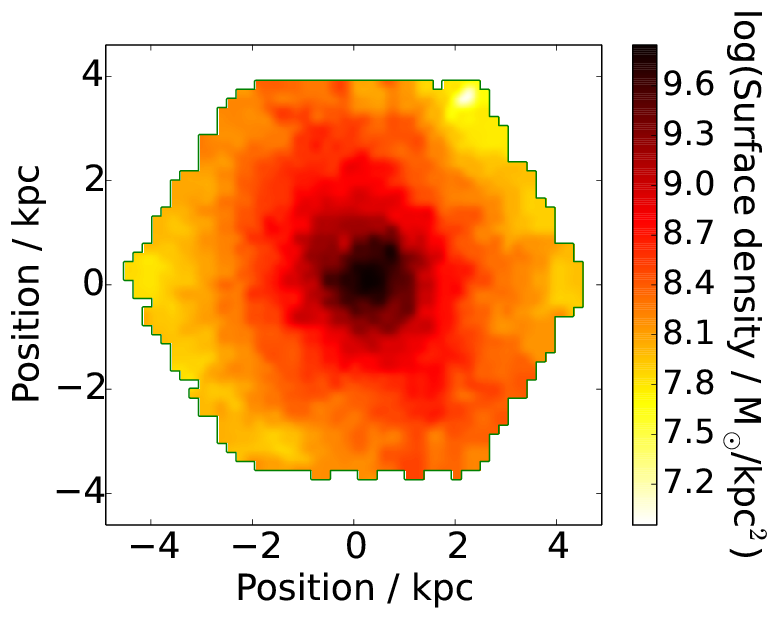}
	\caption{Stellar mass.}
\end{subfigure}
\begin{subfigure}{0.32\linewidth}
	\includegraphics[width=\linewidth]{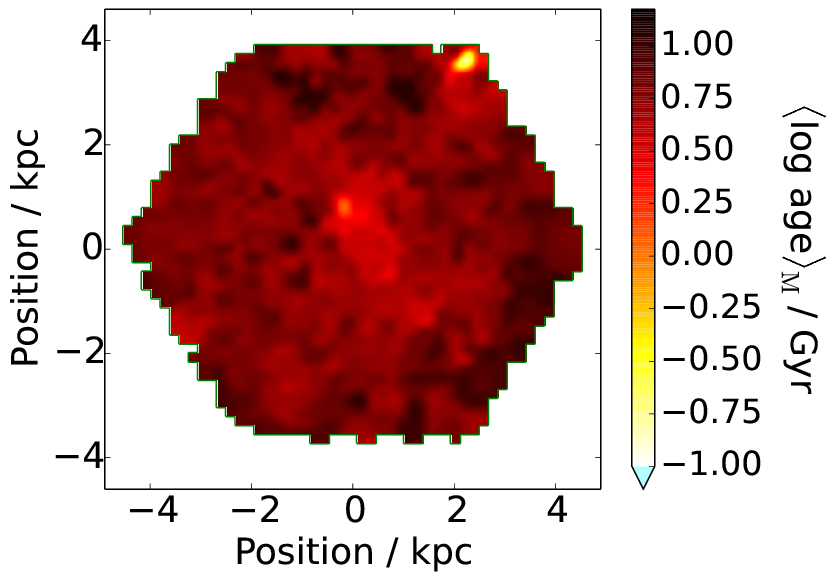}
	\caption{Mass-weighted stellar age.}
\end{subfigure}
\begin{subfigure}{0.32\linewidth}
	\includegraphics[width=\linewidth]{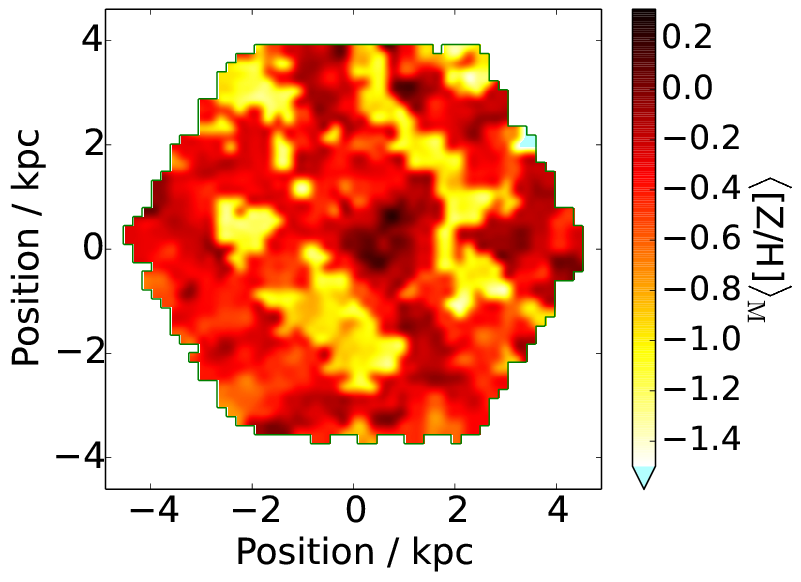}
	\caption{Mass-weighted metallicity.}
\end{subfigure}
\begin{subfigure}{0.32\linewidth}
	\includegraphics[width=\linewidth]{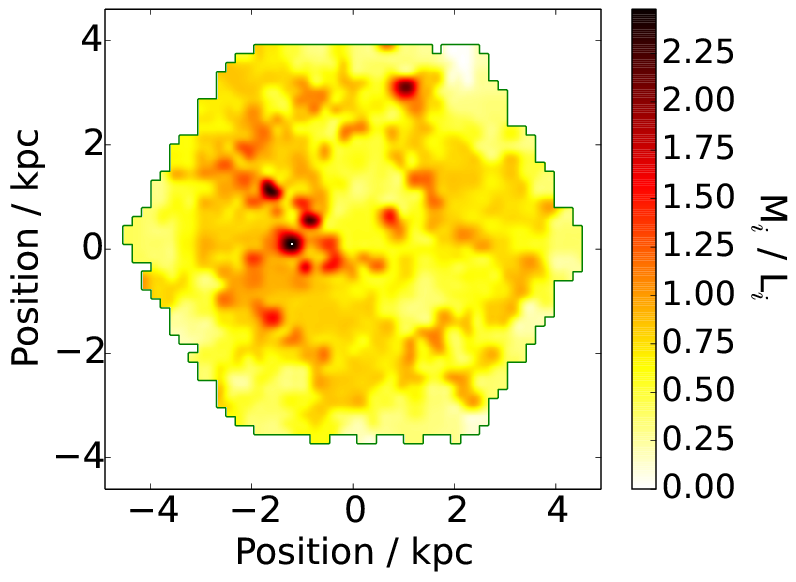}
	\caption{Stellar mass-to-light ratio in the SDSS $i$-band.}
\end{subfigure}
\begin{subfigure}{0.3\linewidth}
	\includegraphics[width=\linewidth]{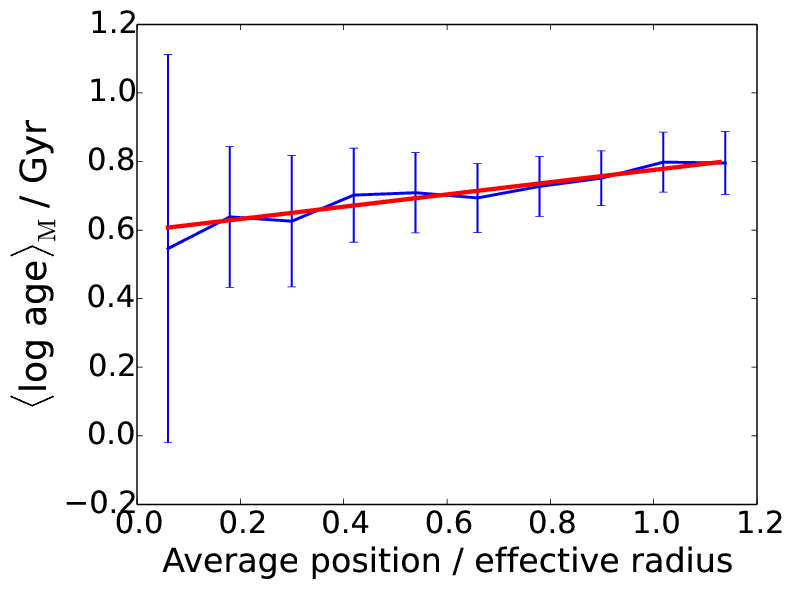}
	\caption{Radial age profile.}
\end{subfigure}\hspace{0.2cm}
\begin{subfigure}{0.3\linewidth}
	\includegraphics[width=\linewidth]{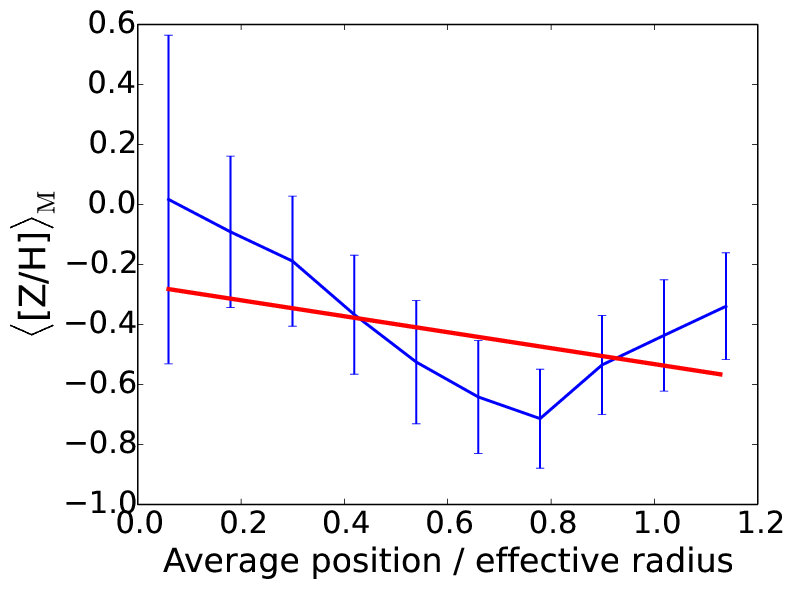}
	\caption{Radial metallicity profile.}
\end{subfigure}\hspace{0.4cm}
\begin{subfigure}{0.3\linewidth}
	\includegraphics[width=\linewidth]{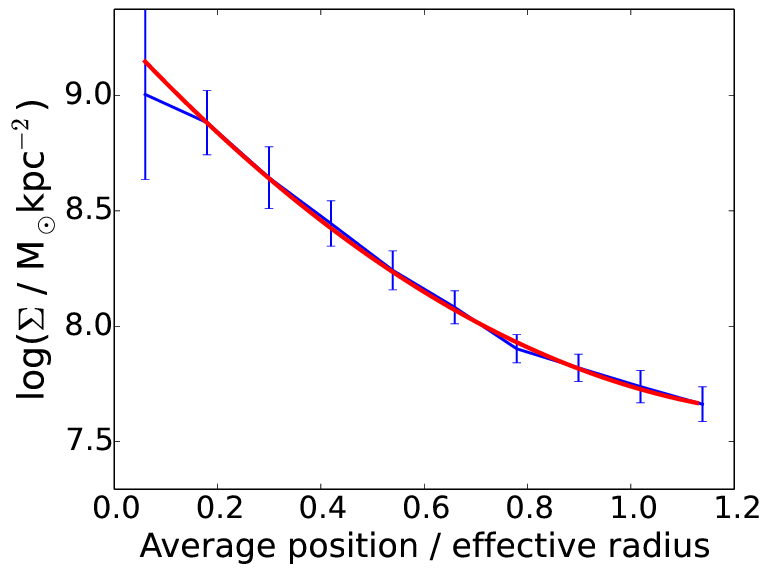}
	\caption{Stellar mass surface density gradient profile.}
\end{subfigure}\hspace{0.9cm}
\vspace{0.5cm}
\caption{{\bf Group $\beta$, galaxy \galaxyeight} as in table \protect\ref{tab:sample}. Stellar population maps and profiles analyzed using MILES-based models with their full parameter range, as described in detail in Figure \protect\ref{maps_eighteen}. Observed under the same setup as galaxy \galaxyeighteen (Figure \protect\ref{maps_eighteen}), except with a \N{61} IFU, and observed to a higher spatial resolution than would be targeted for the MaNGA sample.}
\label{maps_eight}
\end{figure*}

The last two galaxies from the high-quality subsample of P-MaNGA would not be selected in the MaNGA samples, due to their low redshift and hence low radial coverage by the MaNGA IFUs. These galaxies, p9-127A (Appendix Figure \ref{maps_four}) and p9-61A (Figure \ref{maps_eight}), are observed with \N{127} and \N{61} IFUs, cover 0.7 and 1.2 \Rekpc, and are visually identified as a face-on spiral and an early type galaxy, respectively. The high number of fibers per unit radius means that we are able to derive more detailed, but radially-limited, data than would be achievable with the full MaNGA samples.

As an example, galaxy p9-61A shows in its population maps and gradients a nearly flat metallicity profile with a ring of younger metallicities around the young central core. Its older population is shown as a 4 Gyr ring in \logageL around the core before becoming younger in its halo population. This is more difficult to see in \logageM, suggesting that the mass fraction of young stars in the halo is small. The metallicity gradient is nearly flat on average, apart from the low metallicity ring causing a dip in the profile. The dust lane stretching vertically is clearly recovered in the dust attenuation map. We can clearly see the effect of the dust in the age and metallicity maps too, with the dust regions corresponding to lower ages (particularly in \logageL, very suggestive of recent star formation or a continuous star formation history) and higher metallicities.

The high spatial detail in these two cases will not be possible to recover in MaNGA due to the higher redshift, and therefore lower spatial resolution of the target galaxies. Nonetheless the radial profiles are similar in the range of errors to the group $\alpha$ primary galaxies, with 0.04, 0.11 dex in age, 0.05, 0.19 dex in metallicity and 0.03, 008 dex in surface mass density profiles for p9-127A and p9-61A respectively. Due to the smaller radial extent probed (0.7 and 1.2 \Rekpc) this gives somewhat higher errors on the gradients recovered of 0.07, 0.13 dex / \Rekpc in age and 0.09, 0.18 dex / \Rekpc in metallicity, for p9-127A and p9-61A respectively.

\subsection{MaNGA-selected galaxies with lower-quality data: Group \large{$\gamma$}}

Five galaxies fall into the category of being selected in MaNGA, but due to testing of the P-MaNGA instrument, have lower exposure times, and/or were taken in poor observational conditions. Specifically, plate 11 and plate 4 have one and two hour exposure times respectively compared to the MaNGA exposure time of three hours, and plate 11 has the additional problem of having been observing under an airmass of 1.5, see table \ref{tab:sample}.

\begin{figure*}
\centering
\vspace{1.0cm}
\hspace{0.3cm}
\begin{subfigure}{0.30\linewidth}
	\hspace{0.4cm}
	\includegraphics[width=0.72\linewidth]{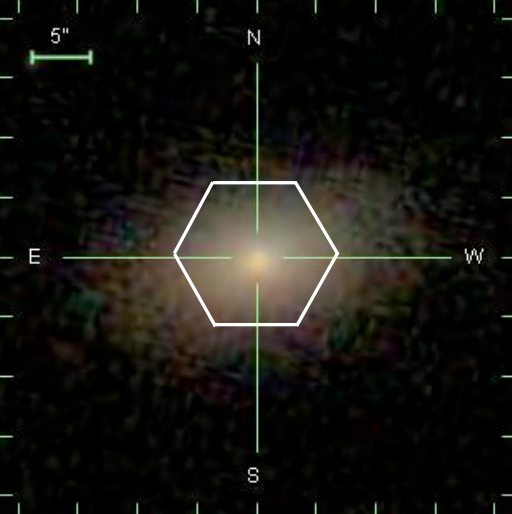}
	\caption{SDSS image with the P-MaNGA footprint.}
\end{subfigure} 
\begin{subfigure}{0.32\linewidth}
	\includegraphics[width=\linewidth]{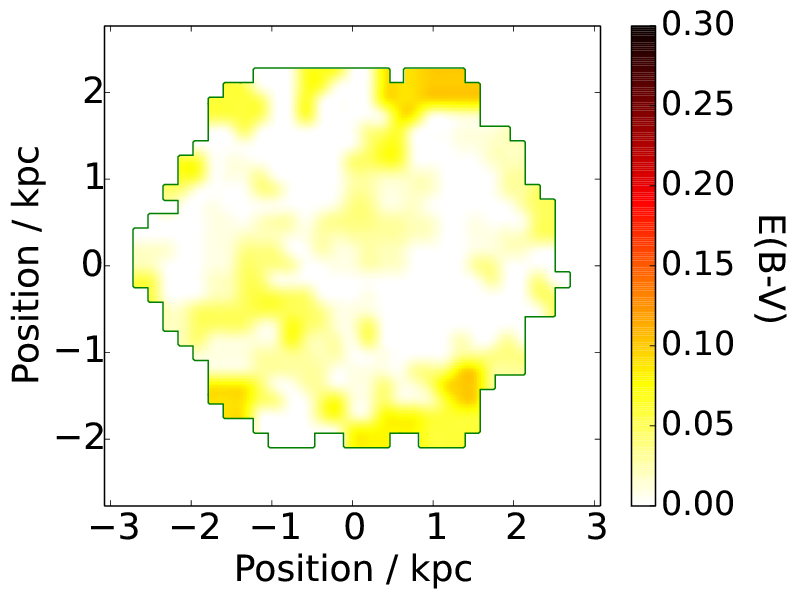}
	\caption{Dust extinction, E(B-V).}
\end{subfigure}
\begin{subfigure}{0.32\linewidth}
	\includegraphics[width=\linewidth]{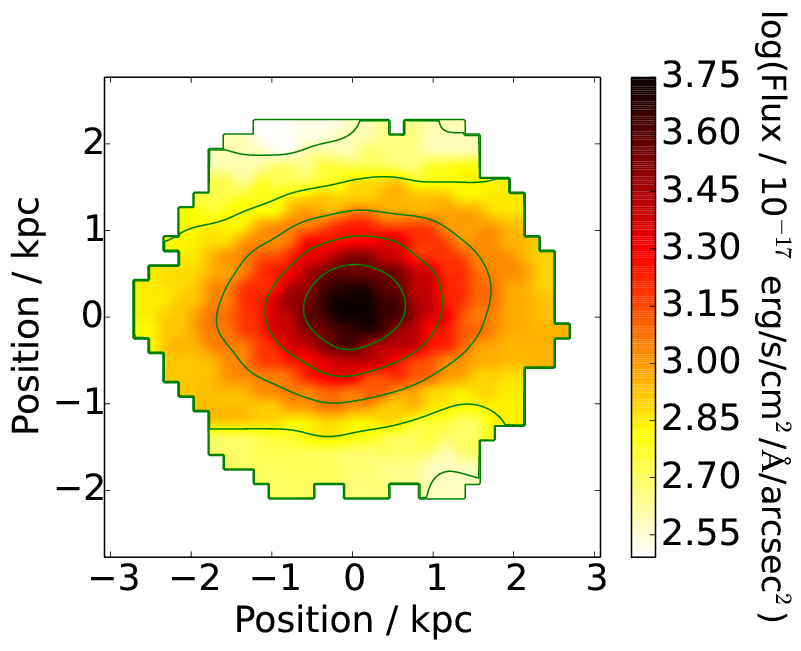}
	\caption{Flux map with isoflux contours (green).}
\end{subfigure}\hspace{0.1cm}
\begin{subfigure}{0.32\linewidth}
	\includegraphics[width=\linewidth]{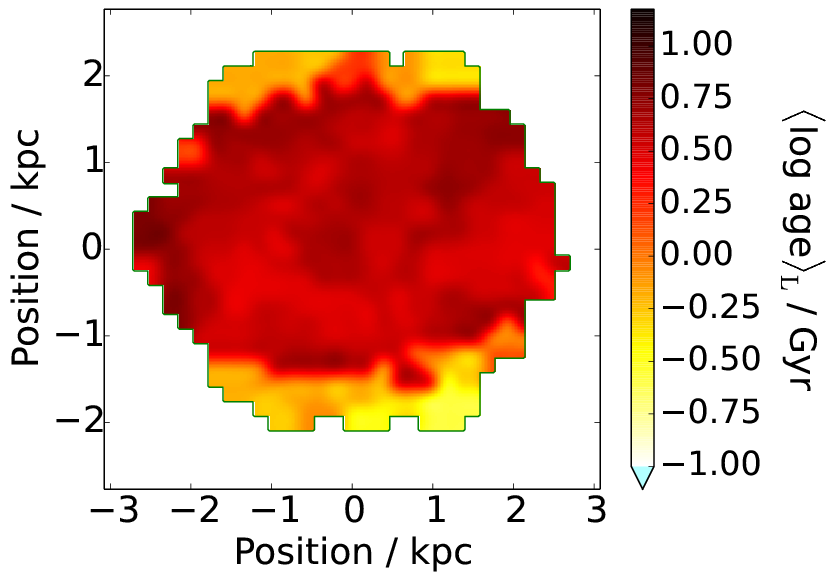}
	\caption{Luminosity-weighted stellar age.}
\end{subfigure}
\begin{subfigure}{0.32\linewidth}
	\includegraphics[width=\linewidth]{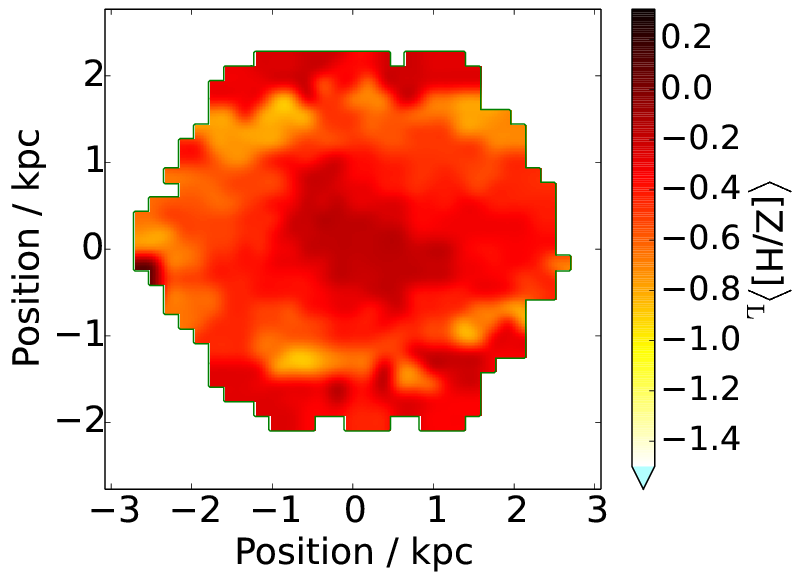}
	\caption{Luminosity-weighted metallicity.}
\end{subfigure}
\begin{subfigure}{0.32\linewidth}
	\includegraphics[width=\linewidth]{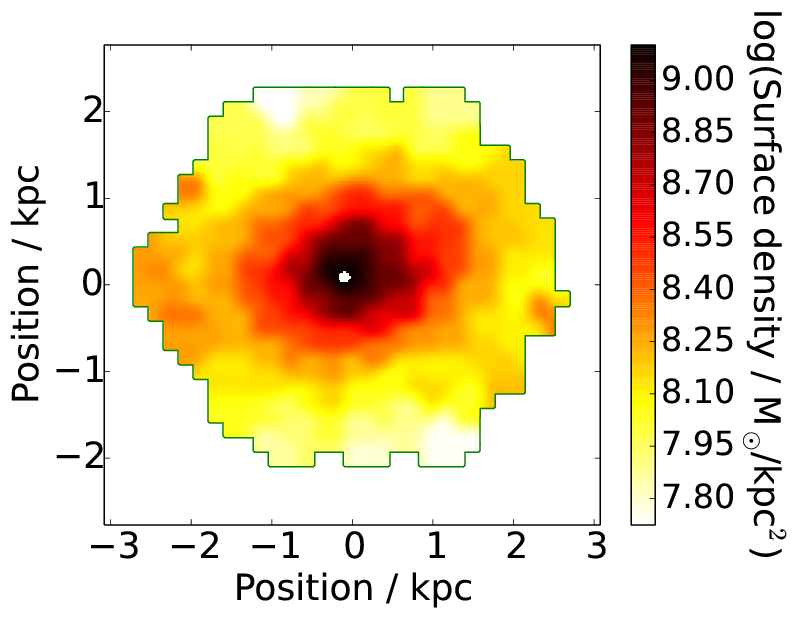}
	\caption{Stellar mass.}
\end{subfigure}
\begin{subfigure}{0.32\linewidth}
	\includegraphics[width=\linewidth]{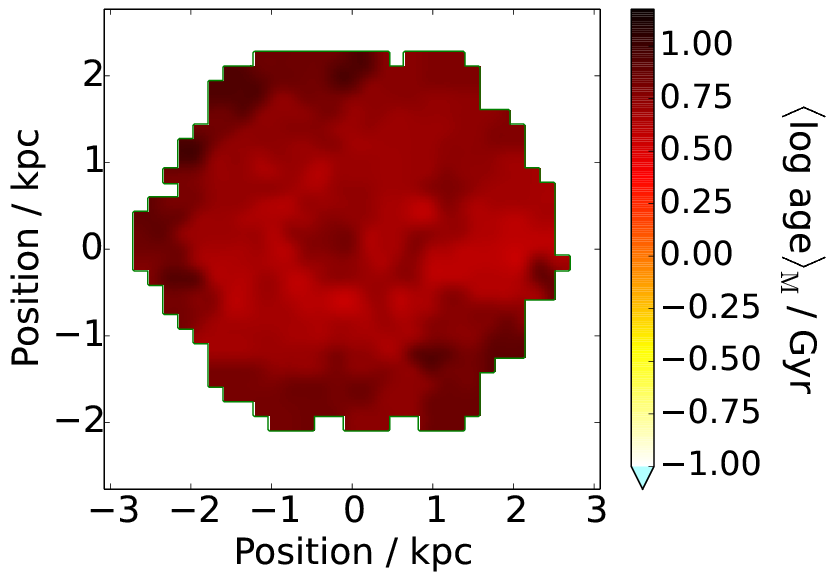}
	\caption{Mass-weighted stellar age.}
\end{subfigure}
\begin{subfigure}{0.32\linewidth}
	\includegraphics[width=\linewidth]{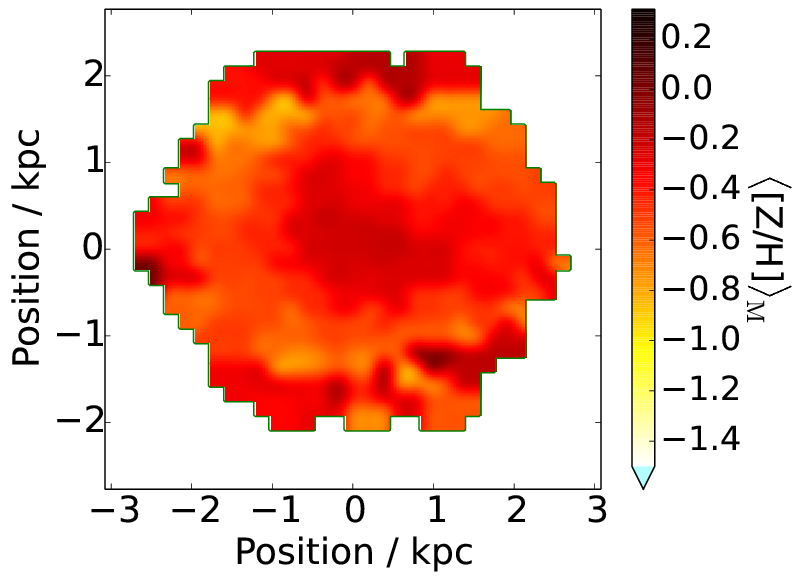}
	\caption{Mass-weighted metallicity.}
\end{subfigure}
\begin{subfigure}{0.32\linewidth}
	\includegraphics[width=\linewidth]{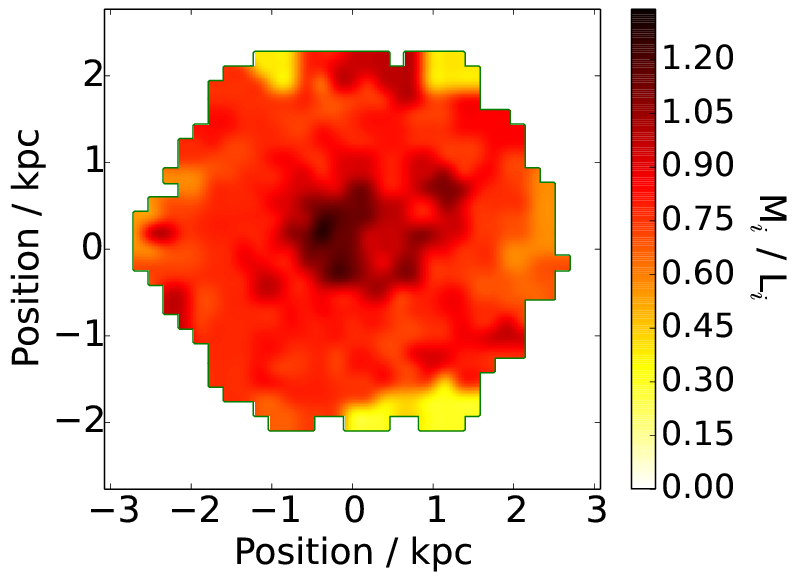}
	\caption{Stellar mass-to-light ratio in the SDSS $i$-band.}
\end{subfigure}
\begin{subfigure}{0.3\linewidth}
	\includegraphics[width=\linewidth]{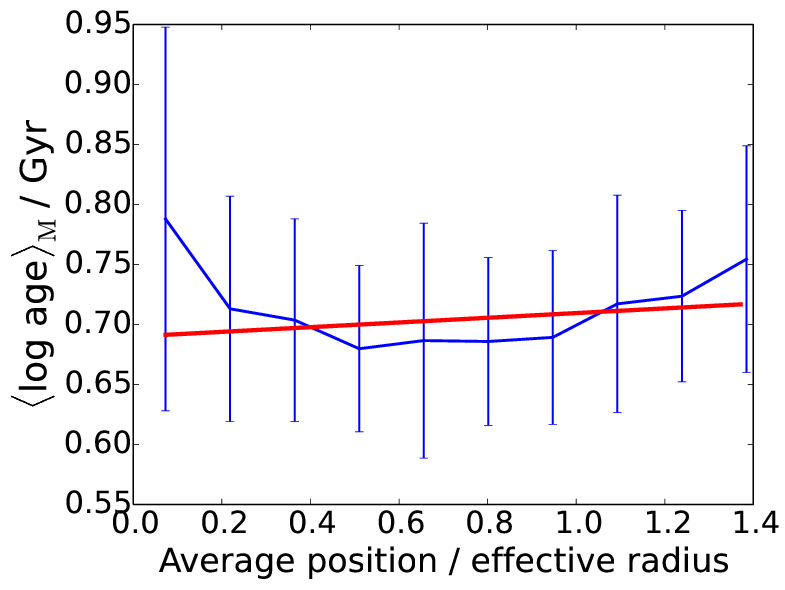}
	\caption{Radial age profile.}
\end{subfigure}\hspace{0.2cm}
\begin{subfigure}{0.3\linewidth}
	\includegraphics[width=\linewidth]{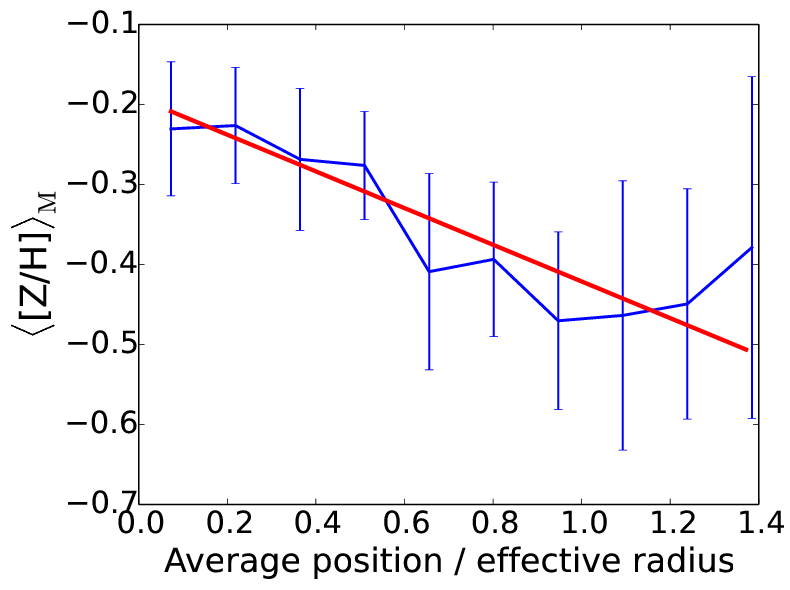}
	\caption{Radial metallicity profile.}
\end{subfigure}\hspace{0.4cm}
\begin{subfigure}{0.3\linewidth}
	\includegraphics[width=\linewidth]{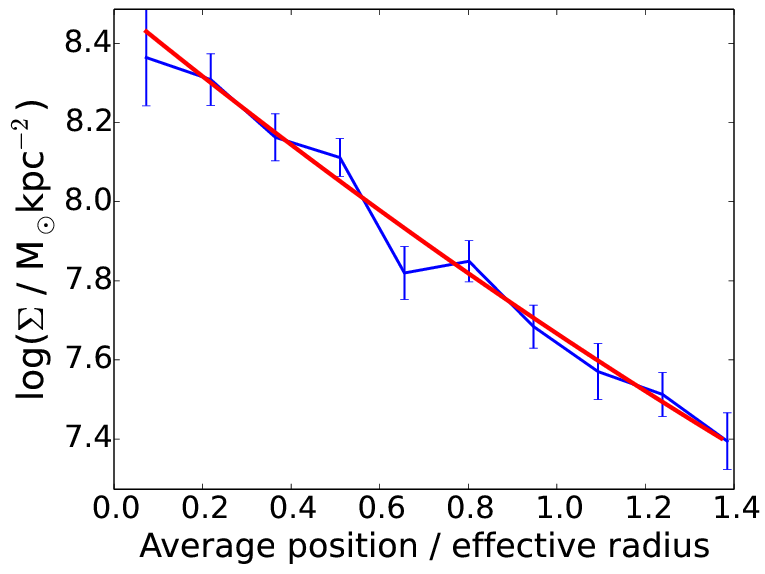}
	\caption{Stellar mass surface density gradient profile.}
\end{subfigure}\hspace{0.9cm}
\vspace{0.5cm}
\caption{{\bf Group $\gamma$, galaxy \galaxyfourteen} as in table \protect\ref{tab:sample}. Stellar population maps and profiles analyzed using MILES-based models with their full parameter range, as described in detail in Figure \protect\ref{maps_eighteen}. This galaxy has been observed under poorer conditions and a lower exposure and dithering setup than expected for MaNGA. Nonetheless, we recover a simple and smooth stellar population structure with negative age and metallicity gradients.}
\label{maps_fourteen}
\end{figure*}

We begin by looking at plate 4, which has observational conditions somewhat closer to what can be expected from the MaNGA sample than plate 11. Galaxies p4-19B (Appendix Figure \ref{maps_thirteen}) and p4-19C (Figure \ref{maps_fourteen}) would fall in the primary MaNGA sample, with \N{19} IFU coverage of 1.4 and 1.3 \Rekpc~respectively. We take galaxy p4-19C as an example of galaxies in this group. It is an early-type galaxy, that from the stellar age population map has an intermediate-age (6 Gyr in both \logageL and \logageM) core with a very flat gradient out to large radii. The \metalL and \metalM map shows a clear negative gradient; going from approximately half-solar metallicity to [Z/H] = -- 0.5. The stellar mass density map is relatively smooth and radially symmetric.

The poorer exposure times results in somewhat poorer stellar population radial recovery than the group $\alpha$ and $\beta$ galaxies. Errors are on the order of 0.06, 0.09 dex in age, 0.10, 0.21 dex in metallicity, and 0.06, 0.09 in stellar mass density for the profiles of p4-19C and p4-19B respectively. Stellar mass density profile errors are the least affected compared to group $\alpha$ which suggests that the quality of data lost by the lower exposure time affects age and metallicity determinations more significantly than surface mass density determinations. The corresponding respective errors on the gradients are 0.07, 0.11 dex / \Rekpc in age and 0.08 -- 0.21 dex / \Rekpc in metallicity on average. Thus a 2 hour exposure, whilst giving about a factor 1.5 higher in error on the radial information obtained compared to MaNGA-like 3 hour exposures, can still give significant age gradients for many galaxies but would not be able to significantly test many metallicity gradients.

Plate 11 has two galaxies that would be selected in MaNGA, p11-61A and p11-19B, in the primary and secondary samples respectively. We do not show their stellar population maps since they represent much poorer data quality than what we expect from MaNGA, and the high pixel-to-pixel variance of the maps make them difficult to interpret. Instead we describe them briefly here. 

Galaxy p11-61A is visually identifiable as an inclined disk galaxy, with \N{61} IFU coverage of 1.5 \Rekpc, with a red core with bluer surroundings. This is matched to an older (\logageL = 5 Gyr / \logageM = 10 Gyr compared to surrounding \logageL = 0.5 Gyr / \logageM = 3 Gyr, metal-rich (\metalM = 0.2 dex compared to -- 0.1 dex surroundings) bulge with a 1 kpc radial size, giving a strong negative age gradient. However, the stellar population maps and particularly the dust attenuation map shows a high degree of scatter that is not spatially confined to a particular region, suggesting that we are suffering from the effects of poor observational conditions.

Similarly galaxy p11-19B, a face-on blue spiral galaxy, shows a high degree of scatter in its stellar population maps, particularly in its age map. This galaxy has an average age of 500 Myr, and also has a 2 Gyr red component in the South at the same redshift as the central galaxy visible in the SDSS image.

The stellar population radial profile recovery is somewhat better in p11-61A, despite both showing signs in their maps of artificial irregularity in their age and metallicities resulting from uncertainty in the spectral fit. This could be the result of galaxy p11-61A being a highly compact object compared to every other object in the sample and thus its fluxes on each individual fiber being higher. Since this galaxy is therefore a much brighter per unit area than other galaxies in the sample we expect errors on its gradients to be lower than the other galaxy in this subsample, p11-19B.

The radial profile errors are 0.09, 0.27 dex in age, 0.11, 0.36 dex in metallicity and 0.07, 0.16 dex in stellar mass density for p11-61A and p11-19B respectively. We note that galaxy p11-19B has a much larger radial extent than other galaxies in the sample, which could explain the larger errors on the profile, but reduces errors in the gradients. Errors on the gradients are 0.07, 0.16 dex in age and 0.11, 0.36 dex for p11-61A and p11-19B respectively.

\subsection{Non-MaNGA galaxies with low-quality data: Group \large{$\delta$}}

This group has the highest membership of any of the identified groups above, containing seven galaxies, four in plate 11 and three in plate 4. The galaxies in this subsample are: p11-127A, p11-127B, p11-19A, p11-19C, p4-127A, p4-127B, and p4-61A. Rather than describe each galaxy in detail, we discuss some interesting features from this group. We later discuss the errors on the radial profiles for these galaxies as part of the sample as a whole and so refer the reader to section \ref{staterror} for this discussion.

\begin{figure*}
\centering
\begin{subfigure}[t]{0.46\linewidth}
	\includegraphics[width=\linewidth]{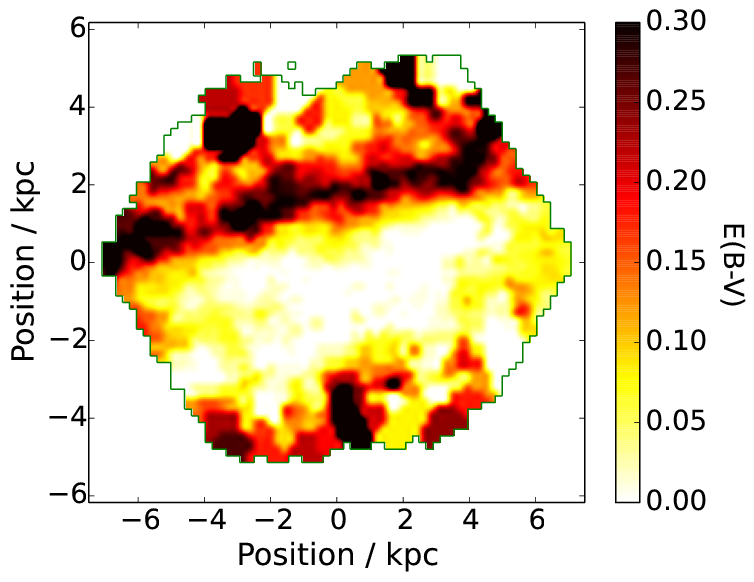}
	\caption{Dust attenuation map of galaxy p4-127A.}
	\label{bad_nonmanga_1}
\end{subfigure}
\begin{subfigure}[t]{0.46\linewidth}
	\includegraphics[width=\linewidth]{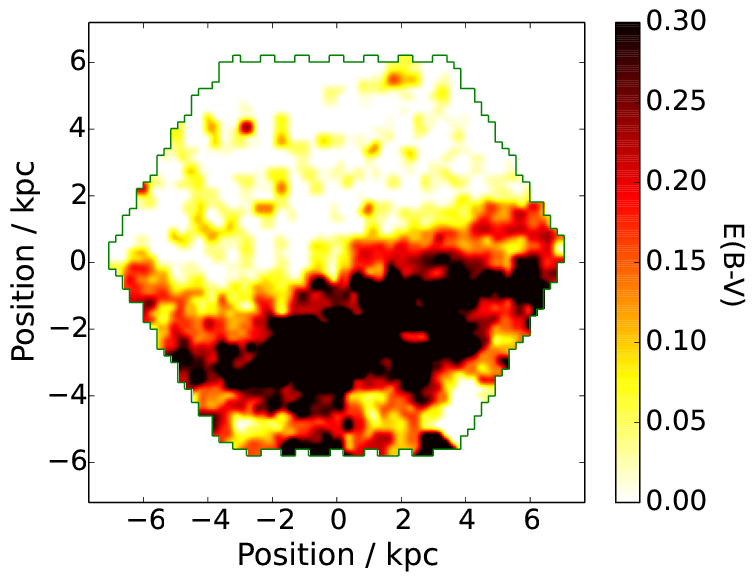}
	\caption{Dust attenuation map of galaxy p11-127A.}
	\label{bad_nonmanga_2}
\end{subfigure}
\caption{{\bf Group $\delta$ galaxies p4-127A and p11-127A} as in table \protect\ref{tab:sample}. These two galaxies have very clear dust lanes that show evidence of more recent star formation than their surroundings. Both of these galaxies have been observed under poorer conditions and lower exposures and dithering setup than expected for MaNGA, although p4-127A has been observed under worse conditions and setup than p11-127A, translating to a more speckled appearance in its maps.}
\end{figure*}

A feature common to p4-127A and p11-127A is the presence of a very strong dust lane across a large portion of the observation. In both cases, this feature is shown clearly in the dust attenuation maps (Figure \ref{bad_nonmanga_1} and \ref{bad_nonmanga_2} for p4-127A and p11-127A respectively) and corresponds to somewhat younger ages and less metal-rich (approximately \logageL = 3 Gyr younger and \metalL = 0.2 dex less compared to surroundings). Both of these cases give a clear interpretation of a dusty disk that has undergone more recent star formation than its surroundings. We also note that p11-127A has more variance in its derived stellar population properties, giving a `speckled' appearance in the maps, compared to p4-127A, which can be attributed to the poorer observational conditions in the former case.

\begin{figure*}
\centering
\begin{subfigure}[t]{0.46\linewidth}
	\includegraphics[width=\linewidth]{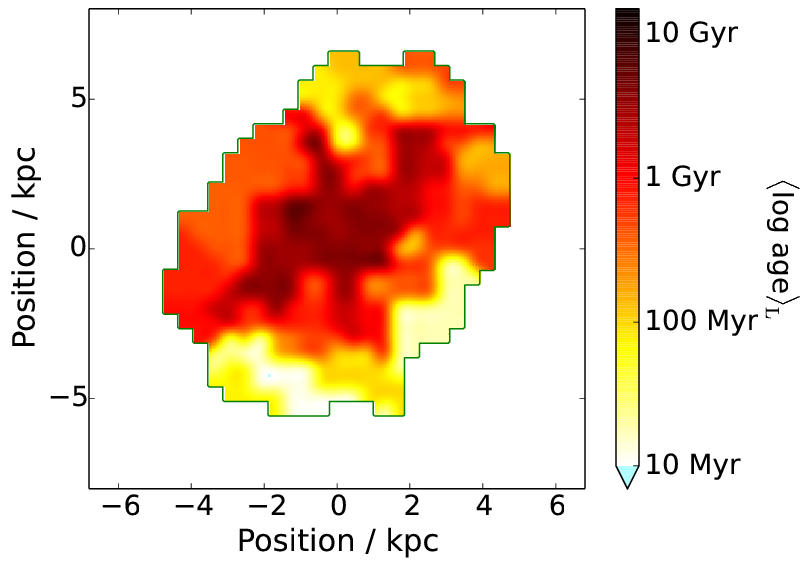}
	\caption{Light-weighted stellar age map of galaxy p11-19A.}
	\label{bad_nonmanga_3}
\end{subfigure}
\begin{subfigure}[t]{0.46\linewidth}
	\includegraphics[width=\linewidth]{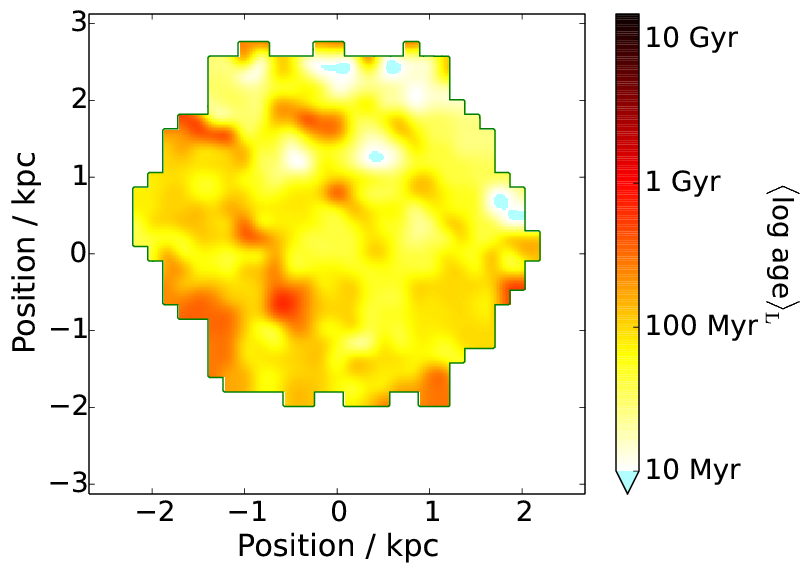}
	\caption{Light-weighted stellar age map of galaxy p11-19C.}
	\label{bad_nonmanga_4}
\end{subfigure}
\caption{{\bf Group $\delta$ galaxies p11-19A and p11-19C} as in table \protect\ref{tab:sample}. These galaxies have been observed under much poorer conditions and much lower exposures and dithering setup than expected for MaNGA. Both of their maps show an unclear distribution of stellar population properties, with in general high amounts of variance between neighbouring pixels. }
\end{figure*}

The observational conditions combined with the \N{19} IFU for galaxies p11-19A and p11-19C render the interpretation of stellar population maps (such as the age maps in Figures \ref{bad_nonmanga_3} and \ref{bad_nonmanga_4}) very difficult, however one can draw comparisons with the qualitatively similar galaxies in the higher-quality subsamples, galaxies p4-127A and p4-19A respectively. This gives an idea of the conclusions one might be able to draw from these galaxies if they were observed under better conditions, and conversely how and how much conclusions get blurred by observing galaxies in poor conditions. The variance of stellar population parameters of p11-19A and p11-19C additionally translates to poor constraints on these galaxies' radial profiles, and so this is a clear limitation of using such low (1 hour) exposure times under poor observational conditions.

The main effect of the poor observational conditions is the added Voronoi cell-to-cell variance which washes out smaller-scale structure. The identification of galaxy features such as disk structure is not possible in these cases.

\section{Results}\label{sampleresults}
\subsection{Radial gradients}\label{radialgradients}

\begin{figure*}
\centering
\begin{subfigure}{\linewidth}
	\includegraphics[width=0.5\linewidth]{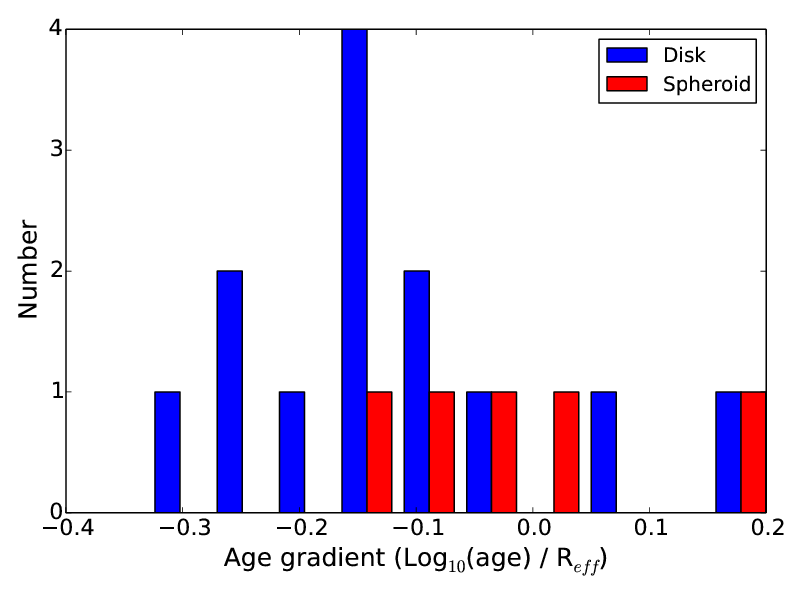}
	\includegraphics[width=0.5\linewidth]{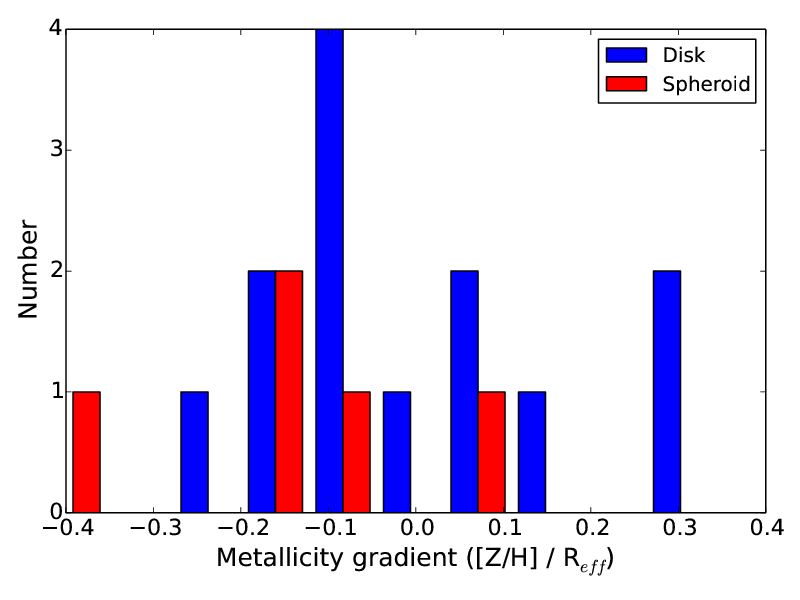}
\end{subfigure}
\caption{Summary of all of the age (left) and metallicity (right) gradients deduced for the P-MaNGA dataset, using visual morphological classification from SDSS imaging data.}
\label{fig:gradsummary}
\end{figure*}

The age and metallicity gradients for the prototype dataset are summarized in the histograms of figure \ref{fig:gradsummary}. We re-iterate from the beginning of this section that all gradients and radial profiles are calculated from mass-weighted stellar population properties rather than light-weighted properties, since these more precisely trace the evolutionary history of the galaxies observed, and thus are more appropriate for assessing formation and evolutionary mechanisms.

Both age and metallicity gradients seem to draw from different distributions for spheroid / early-type galaxies and disk / late-type galaxies. Age gradients for late-types have an overall strongly negative with a mean of -- 0.39 dex) and a wide standard deviation of  0.36 dex, compared to early-type galaxies which have a flat (mean -- 0.05) age gradient and 0.24 dex standard deviation. Conversely, metallicity gradients for late type galaxies are flat (mean 0.00 dex), with 0.17 dex standard deviation, and early-type galaxies have a negative gradient of mean -- 0.15 dex with a 0.14 dex standard deviation.

Our findings on metallicity and age gradients agree very well with previous analyses of early-type galaxies, such as in \cite{1999ApJ...527..573K}, \cite{2000A&A...360..911S}, \cite{2000AJ....119.2134T}, \cite{2003A&A...407..423M}, \cite{2005ApJ...622..244W}, \cite{2007A&A...463..455A}, \cite{2007MNRAS.377..759S}, \cite{2008MNRAS.389.1891R}, and \cite{2010MNRAS.408...97K}, in particular the lack of a significant age gradient in early-type galaxies.

For our late-type galaxy sample we do not attempt to separate the light from the bulge and from the disk as it is beyond the scope of this work. Thus our overall gradients, particularly in age, cannot be compared directly to previous work that has done this, for example \cite{2014A&A...570A...6S} and \cite{2009MNRAS.395...28M}, and work based on local galaxies such as \cite{2009ApJ...695L..15W} and \cite{2007AJ....133.1138B}. However we note that our metallicity gradient estimates agree well in these cases. In this paper we conclude that bulge-disk separation is the main driver for the negative age gradients seen in P-MaNGA galaxies and so our conclusions agree well with the literature. For previous work assessing the age and metallicity gradients of late-type galaxies including the disk and bulge, such as \cite{2011MNRAS.415..709S} and \cite{2008ApJ...683..707Y}, we find good agreement with our gradients and overall conclusions.

\begin{figure}
%\begin{center}
\includegraphics[width=8.5cm]{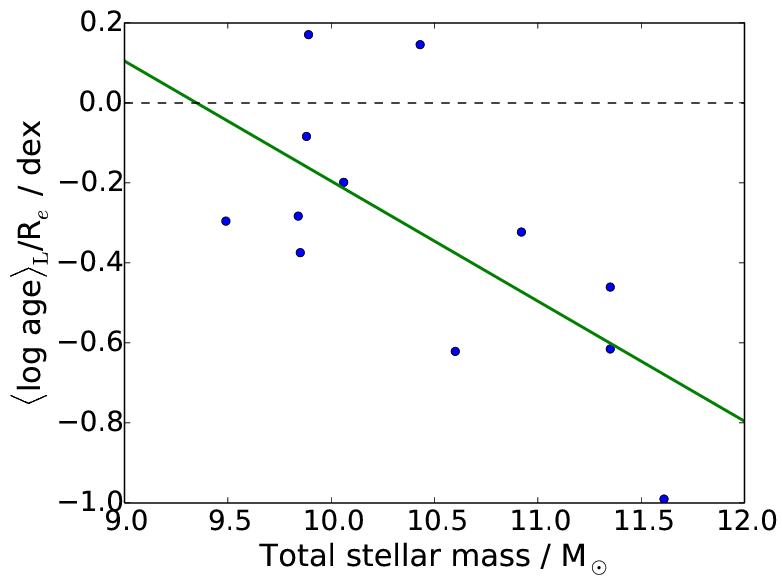}
\caption{A scatterplot showing the relation between photometric stellar masses and \logageL as derived in this paper, for the late-type galaxies in the P-MaNGA sample as listed in Table \protect\ref{tab:sample}. This relation is plotted in order to compare to the conclusions of \protect\cite{2014A&A...562A..47G} (see text). In dashed black we show zero gradient line, and in green we show a linear regression line.}
\label{age_mass_plot}
%\end{center}
\end{figure}

With our limited sample size we cannot draw significant conclusions on the relationships of stellar population gradients as a function of other physical properties. However in Figure \ref{age_mass_plot}, to compare with recent work by \cite{2014A&A...562A..47G} we plot total stellar mass as measured in P-MaNGA against \logageL for the late-type galaxies to see if our results are consistent. Total stellar mass estimates are calculated from $u,g,r,i,z$~photometry following the method in \cite{2006ApJ...652...85M} and \cite{2013MNRAS.435.2764M} for SDSS-III/BOSS galaxies (Pforr et al., in preparation). `Photometric' stellar masses do not suffer from aperture effects as will be described in Section \ref{masscomparison}. In section \ref{masscomparison} we also test the relationships between gradients and total stellar mass.

We find that at low masses, below $\sim 10^{10}$ M$_\odot$, \logageL~gradients are generally close to flat, but that above $\sim 10^{10.5}$ M$_\odot$ the \logageL~gradients are very strongly negative. This matches well with \cite{2014A&A...562A..47G}, who find that low-mass disk galaxies show flat age gradients but at higher masses the gradients become strongly negative. They also test early-type galaxies but with our very small sample, we are unable to see a clear correlation with total stellar mass. In our sample we find no clear evidence for relationships between \logageM, \metalL~or \metalM~with total stellar mass, suggesting that the main driver for the \logageL~---~M$_\odot$ relation is the radial transition between bulge and recent star formation in the disk.

\subsection{Beam smearing}\label{beam_smearing}

Results from IFU experiments can be threatened by the risk of `beam smearing', an effect that spreads out the signal for a given pixel position into nearby pixels. Beam smearing is the combination of three observational effects: atmospheric point-spread function (PSF), sampling size of the IFU fibers, and dithering; the process of offsetting the telescope from the target by small amounts such that the fibers collectively observe whole footprint, including gaps between fibers. The effect of beam smearing should be greatest for the smaller IFUs. 

Given that our largest sample of galaxies are late-type galaxies and these have generally a large range of negative gradients, we can test to see if decreasing the IFU size results in poorer age gradient recovery, for these galaxies. Metallicity gradients are not tested since there are both positive and negative gradients in the late-type sample which may be hard to separate between, and our early-type sample is too small to derive meaningful conclusions in this area.

 \begin{figure*}
\centering
\begin{subfigure}{\linewidth}
	\includegraphics[width=0.5\linewidth]{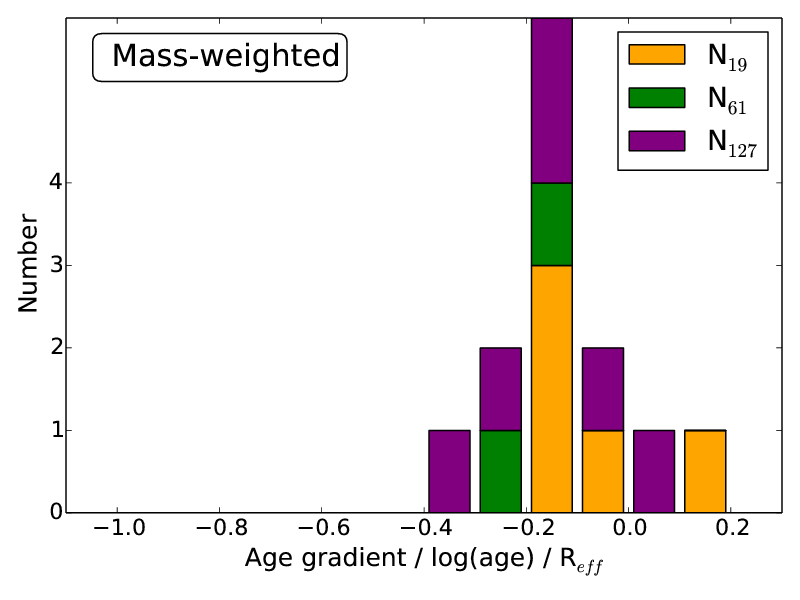}
	\includegraphics[width=0.5\linewidth]{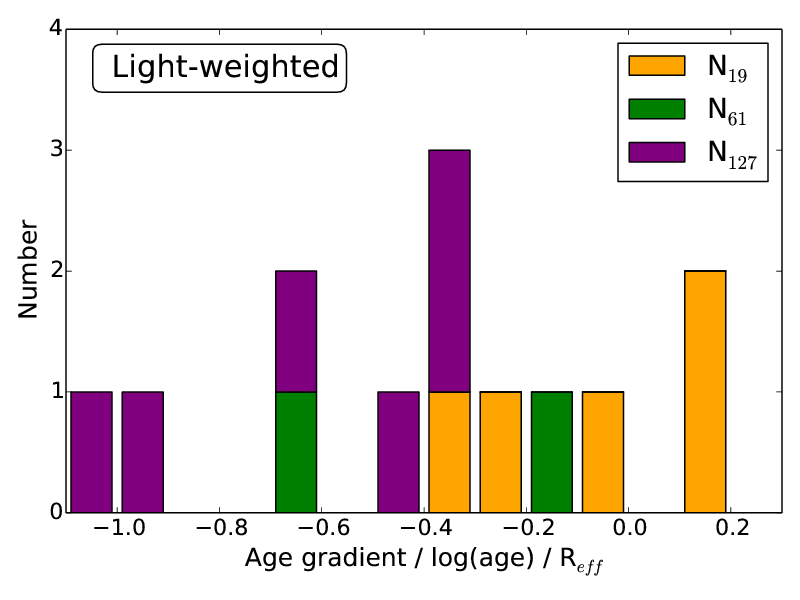}
\end{subfigure}
\caption{Histograms of age gradients as in Figure \protect\ref{fig:gradsummary}, selecting only the late-type galaxies in the sample, but as a function of IFU size. In the left panel we show mass-weighted age gradients as in the other profile and gradient Figures in this paper. In the right panel we show light-weighted age gradients since these will trace bright stellar populations, and thus are important for testing the effect of beam smearing.}
\label{beam_age}
\end{figure*}

In Figure \ref{beam_age} we plot the age gradient histograms for the late type galaxies shown in Figure \ref{fig:gradsummary}, but as a function of IFU size: \N{19}, \N{61} and \N{127} fiber IFUs. In the left-panel we plot the mass-weighted radial age gradients as we have plotted in Figure \ref{fig:gradsummary}, where we find no correlation between mean gradient and IFU size. The mean age gradients are -- 0.12, -- 0.19, and -- 0.13 dex for \N{19}, \N{61} and \N{127} observations respectively. 

We also plot light-weighted age gradients for comparison in the right panel, since these will be more sensitive to bright stellar populations and so should be more directly related to beam smearing effects. We therefore expect larger (in magnitude) gradients to be present since small amounts of star formation in an otherwise more passive galaxy can cause large changes in the light-weighted profiles. In this case the separation is much easier to see, with \N{19} observations giving much flatter gradients (mean -- 0.08 dex) than \N{127} observations (mean -- 0.55 dex), and \N{61} lying in between but closer to the \N{127} observations (mean -- 0.43 dex). We note that the 2 galaxies with the largest negative age gradients, around --1.0 \logageL/\Rekpc, correspond to two late-type galaxies with both spiral or disk features and very prominent bulges, p9-127A and p11-127A, clearly identifiable visually and covered within P-MaNGA \N{127} footprint. Hence it is expected that these galaxies should have very large negative age gradients in their light-weighted profiles, compared to other types of galaxies or smaller IFU sizes where the morphological features may be more difficult to separate.

We therefore conclude that beam smearing does appear to affect the lowest IFU sizes much more in the light-weighted properties, and thus is more sensitive to bright stellar populations such as from very young stars, but this effect is mitigated by calculating mass-weighted properties. 

We do not go further to quantify this effect due to small sample size, and thus overlap with other effects such as radial extent and other physical galaxy properties. A detailed analysis of \N{127} MaNGA data, downgraded to the other IFU sizes, will be necessary for robust quantification, which will be subject of future work once a larger sample is available. Additionally with the much larger MaNGA sample we will be able to carry out this analysis with comparable inclinations, morphologies and colour gradients.

\subsection{Statistical error as a function of observational conditions}\label{staterror}

In each of the P-MaNGA galaxies analysed we have derived stellar population property statistical errors on each of their pixels, as described in section \ref{analysis tools}, which we then propagate to standard errors on each of the elliptical bins in the radial profiles. We explore how these errors vary as a function of the observation conditions and setup used to observe each galaxy.

To measure the error on the gradient we need to find the standard error at the median radial bin. However, this decreases as a function of the number of pixels used increases, and does not include the covariance between pixels, thereby underestimating the uncertainty on the fit. In this work we do not estimate covariance between pixels due to the prototype nature of the observations. To compare these errors across each of the P-MaNGA galaxy, we therefore take the standard deviation averaged across all the pixels at each radial bin, assume the number of independent observations is equal to the number of grid points analysed divided by 16, as explained in Section \ref{fittingmethod}, and find the median of this value. To be concise, we shall refer to this as the {\it median standard error}. This value remains constant as the number of pixels in each bin is changed, allowing us to compare this value across each galaxy directly. 

Corresponding errors on the gradients are calculated from the errors on the profile points by using bootstrap resampling of 10,000 points for each profile point recovered and recalculating the linear gradient on each resample. We note that the error calculated assumes a linear fit, hence does not describe {\it how linear} the radial profiles are. We assume that the errors on each point are Gaussian, which typically is an accurate approximation of the error obtained from the spectral fit with \firefly. We note that errors on the gradient will not directly correspond to the errors on the profile since each galaxy is observed with a different radial extent.

\begin{figure}
\centering
\includegraphics[width=1.0\linewidth]{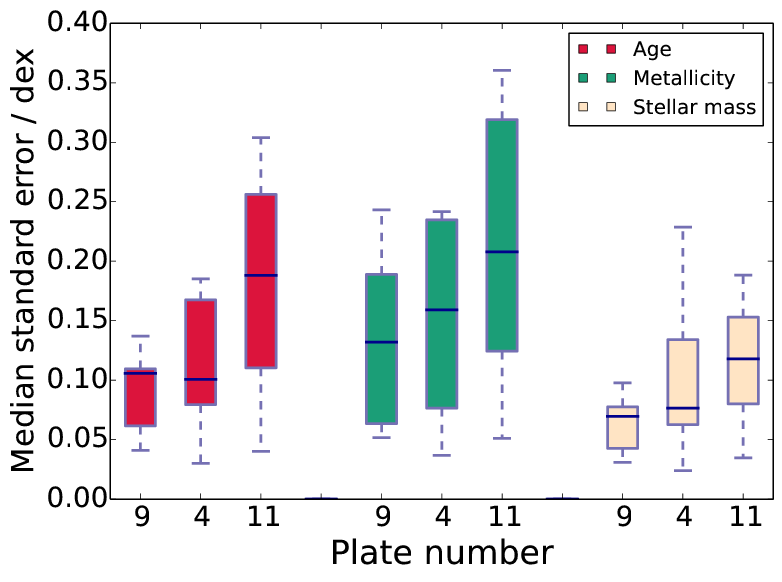}
\includegraphics[width=1.0\linewidth]{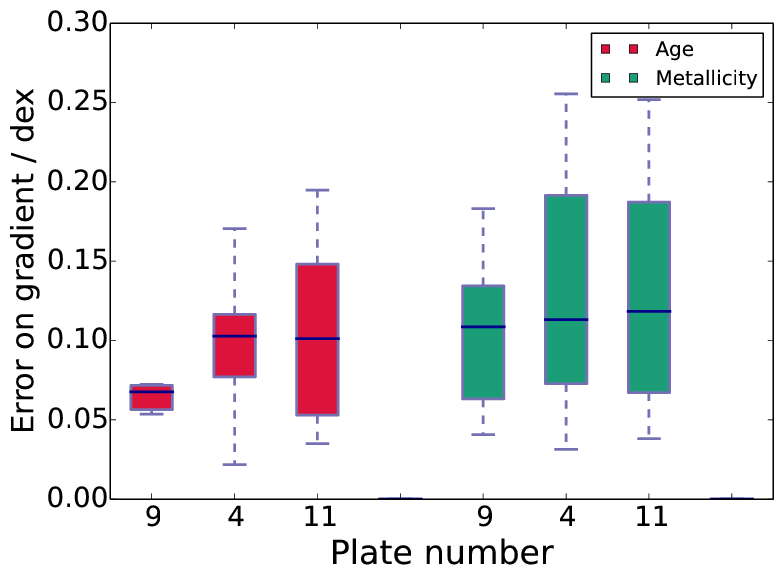}
\caption{Median standard errors (top panel), and errors in the linear gradient (bottom panel), of stellar population profiles as a function of property and observational conditions grouped by plate number, see Table \protect\ref{gradient_summary}. Errors in the gradient are calculated through bootstrap resampling of the median standard errors. Each box plot shows the median and interquartile range of the corresponding subsample, each of which contains the median standard error and the errors on the gradients of six galaxies. Stellar mass density gradients not shown as they are not appropriate to fit with a linear profile.}
\label{fig:boxplots}
\end{figure}

The upper panel of Figure \ref{fig:boxplots} presents the median standard error for mass-weighted stellar age, metallicity and surface mass density as a function of plate, each of which corresponds to a set of observational conditions summarised in Table \ref{plate_table}. We can clearly see that for all three properties, plate 9 provides the smallest errors, and plate 11 provides the highest errors, on the radial profiles. Each box plot has a high interquartile range due to the small sample size, as there are only 6 galaxies in each bin. We remind the reader that only plate 9 has been observed with MaNGA-quality conditions and setup (3 hour exposure), whereas plate 4 and plate 11 have less than MaNGA-quality observations (2 hour exposure, and 1 hour exposure with 1.5 airmass, respectively).

The corresponding errors on the age and metallicity gradients are shown in the lower panel, also summarised in Table \ref{plate_table}. We note that this panel does not linearly relate to the upper panel due to each of the galaxies having different radial extents, as well as some parts of the profile being more important than others to constrain the errors on the gradient. We do not calculate stellar mass density gradients as they are not appropriate to fit with a linear profile. Again plate 9 provides the best gradient determination, with the smallest interquartile ranges in the gradient errors. We thus deduce that plate 9 galaxies provide a more stable and reliable gradient determination for both age and metallicity. There is less evidence to separate plate 4 and plate 11 galaxies in this case, other than plate 11's age gradients seeming to have a larger interquartile range and thus is provides the least stable errors estimations as a function of galaxy observed. Metallicity gradients show little separation in their median value for all three plates, despite plate 9 having a lower interquartile range, suggesting that {\it average} metallicity gradients are on average less sensitive to observational quality than age gradients but do become less stable with each galaxy, with worsening observations. Due to the small size of P-MaNGA's sample, we do not further separate on galaxy type, radial extent, or suggest stronger conclusions on the errors on the profiles and gradients. We summarise the errors in Table \ref{gradient_summary}.

\begin{table*}
\centering
{\footnotesize
\begin{tabular}{c | c | c | c c c | c c}
Field & Groups & Conditions and setup & \multicolumn{3}{c}{Median error in profile / dex} &  \multicolumn{2}{c}{Error on gradient / dex/\Rekpc} \\ 
& & & Age & Metallicity & Mass density & Age & Metallicity \\ \hline
9 & $\alpha$/$\beta$ & Exposure 3.0 hr, seeing 1\farcs7 & 0.10 & 0.13 & 0.07 & 0.07 & 0.11 \\
4 & $\gamma$/$\delta$ & Exposure 2.0 hr, seeing 1\farcs3 & 0.10 & 0.16 & 0.08 & 0.10  & 0.12  \\
11 & $\gamma$/$\delta$ &  Exposure 1.0 hr, seeing 2\farcs0, (airmass $\sim$1.5)& 0.19 & 0.22 & 0.12 & 0.10 & 0.12 \\
\end{tabular}
\caption{Summary of errors on radial profiles and gradients as a function of observational conditions and setup.\label{gradient_summary}}}
\end{table*}

\subsection{Stellar mass estimates}\label{masscomparison}

Given that our full spectral fitting also allows the calculation of stellar masses, we make a qualitative comparison with previous estimates for the P-MaNGA galaxies. 

\begin{figure}
%\begin{center}
\includegraphics[width=8.5cm]{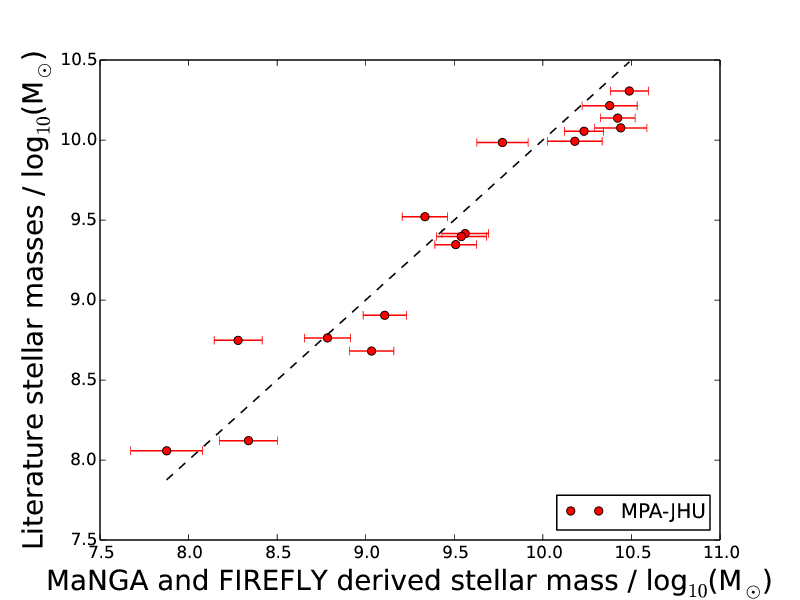}
\includegraphics[width=8.5cm]{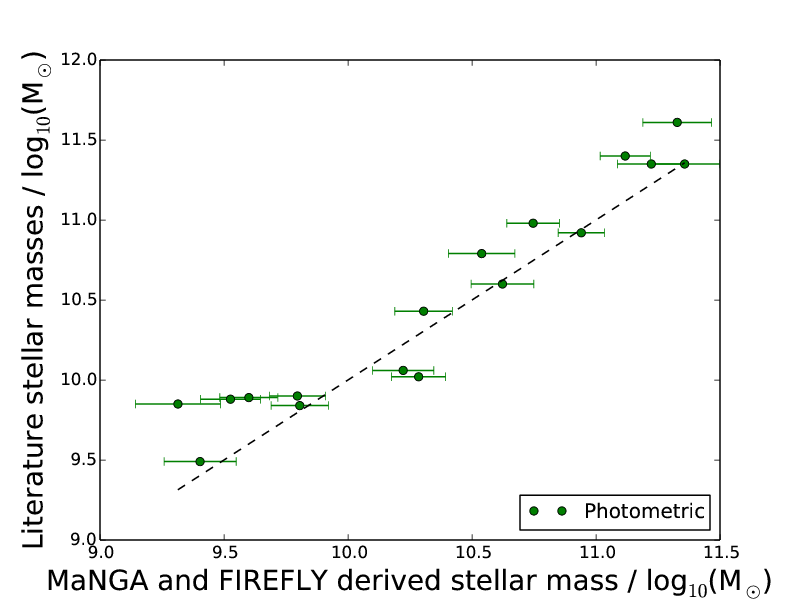}
\caption{A comparison of the total stellar mass derived for each of the P-MaNGA galaxies between this work, the MPA-JHU catalogue in the SDSS Data Release 8 pipeline, and the `Photometric'-based stellar masses based on \protect\cite{2012MNRAS.422.3285P}. In black we show a one-to-one dashed line. We note that the two panels have different scales due to using a 3'' aperture in the MPA-JHU comparison, but the full P-MaNGA aperture in the Photometric comparison.}
\label{fig:totalmass}
%\end{center}
\end{figure}

First we compare to the widely used MPA-JHU catalogue of the SDSS Data Release 8 pipeline (see Table \ref{tab:sample}) in Figure \ref{fig:totalmass}. These masses are based on broadband \textit{ugriz}-filter SED fitting scaled to 3 arcsecond fiber sizes. 
This spatial coverage is much smaller than our P-MaNGA, which extends up to 1.5 effective radii. Hence we integrate the surface mass density only up to this radius for each galaxy. We find good overall agreement with the MPA-JHU masses with a systematic offset of about 0.1 dex and a scatter of 0.2 dex.

We note that the presence of galaxy gradients may complicate this simple scaling of the $M/L$~ratio as this quantity depends on both age and metallicity. In addition, the stellar population modelling and fitting technique is different. The MPA-JHU masses are based on \cite{2003MNRAS.344.1000B} stellar population models and a coarser grid of ages than this work, using (unpublished) models based on the MILES stellar library and a Kroupa IMF. Hence, the stellar library and most importantly the IMF are the same, but the population models and age grids used are different. Other effects could also introduce systematics. The subtraction of stellar mass loss which is performed in our modelling, may corresponds to 30-40\% by mass depending on the IMF (see \cite{1998MNRAS.300..872M}, \cite{2005MNRAS.362..799M}) and it is not clear to us whether and how this is accounted for in the MPA-JHU calculations. 

Furthermore, masses obtained via spectral fitting may be different from those from broad-band SED fit particularly when the former refer to low S/N spectra. \cite{2013MNRAS.435.2764M} (see Appendix) find that the stellar mass~of SDSS-III/BOSS galaxies at redshift $0.4-0.6$~were smaller when computed via broad band \textit{ugriz} SED fitting  by ~$0.25$~dex~compared to those obtained via PCA spectral fitting by \cite{2012MNRAS.421..314C}. The hypothesis is that the spectral fitting of low S/N spectra may lead to high-ages, which have a higher M/L ratios. \cite{2012MNRAS.421..314C} quantify this effect by applying full spectral fitting to high S/N empirical spectra and conclude in this direction. In Wilkinson \& Maraston (in preparation) we try to quantify this effect via mock galaxies with known input $M^{*}$.

This source of systematics is in common with the second stellar mass comparison plot in Figure \ref{fig:totalmass}, namely `photometric'-based stellar masses calculated with the same model and fitting set up as those output of the SDSS-III/BOSS data release and pipeline \citep{2013AJ....145...10D}.
In brief, these are also broad band \textit{ugriz} SED fitting performed with the publicly available code Hyper-Z customed with 
\cite{2005MNRAS.362..799M} models, Kroupa IMF and our same account of mass loss. The template star formation histories encompass a range of exponentially declining $\tau$-models, ranging from 0.1 to 3 Gyrs. No dust reddening is applied to the fitting in order to minimise the age-dust degeneracy which can lead to too young solutions which underestimate the true stellar mass (see \cite{2012MNRAS.422.3285P}). We note that a clump of galaxies in this lower-left of the plot which seem even quite far from the line are all identified as star-forming late-type galaxies in this sample, and so we expect a $\tau$ to not be able to capture these more complex star formation histories, explaining this discrepancy. Our stellar masses are on average 0.15~dex smaller than these photometric total masses. This offset will mainly come from the difference in aperture. Since P-MaNGA covers on average 1.5 \Rekpc~rather than the full extent of the galaxy as in the photometric masses, we expect our observations to cover $\sim 75$ per cent of the light. This would give a 0.12 dex underestimate of the stellar masses, consistent with the offset we find in our comparison.

\subsection{Effect of input stellar library}\label{librarycomparison}

\begin{figure*}
\centering
\begin{subfigure}{\linewidth}
	\includegraphics[width=0.33\linewidth]{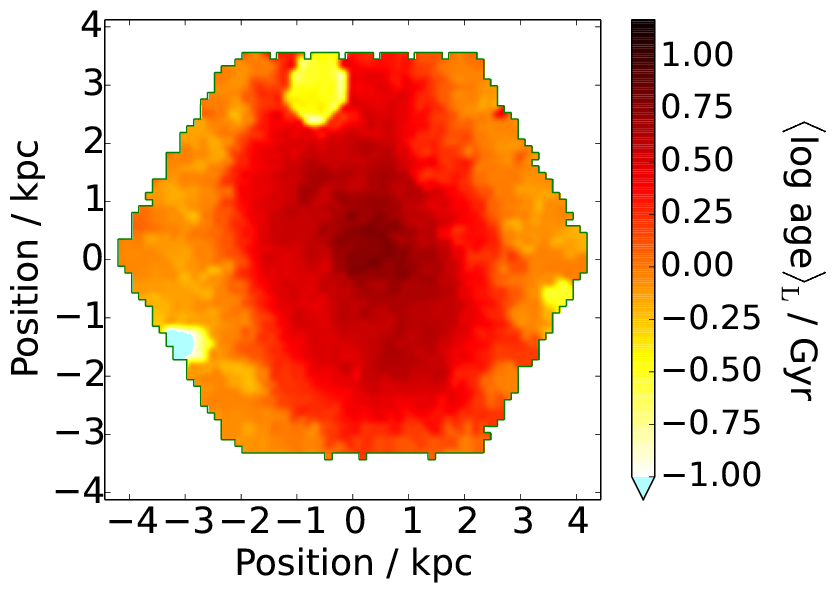}
	\includegraphics[width=0.33\linewidth]{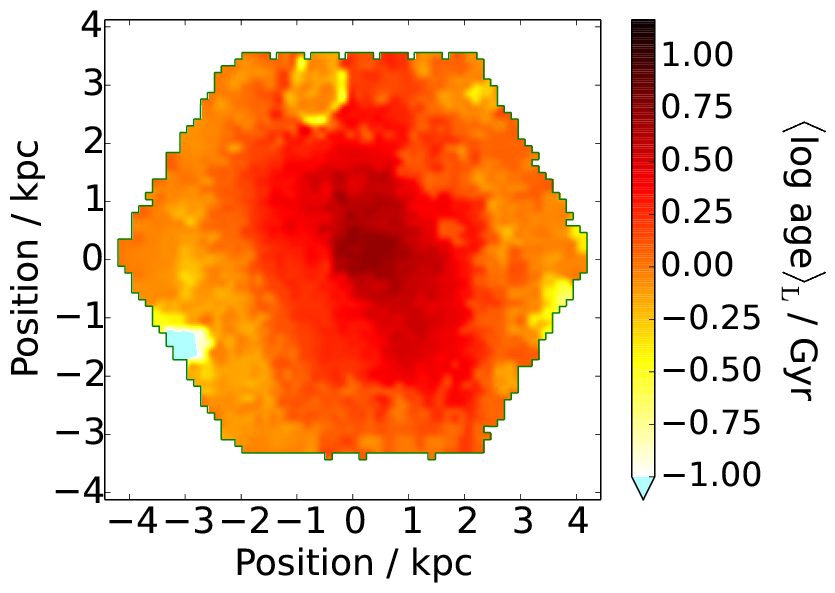}
	\includegraphics[width=0.33\linewidth]{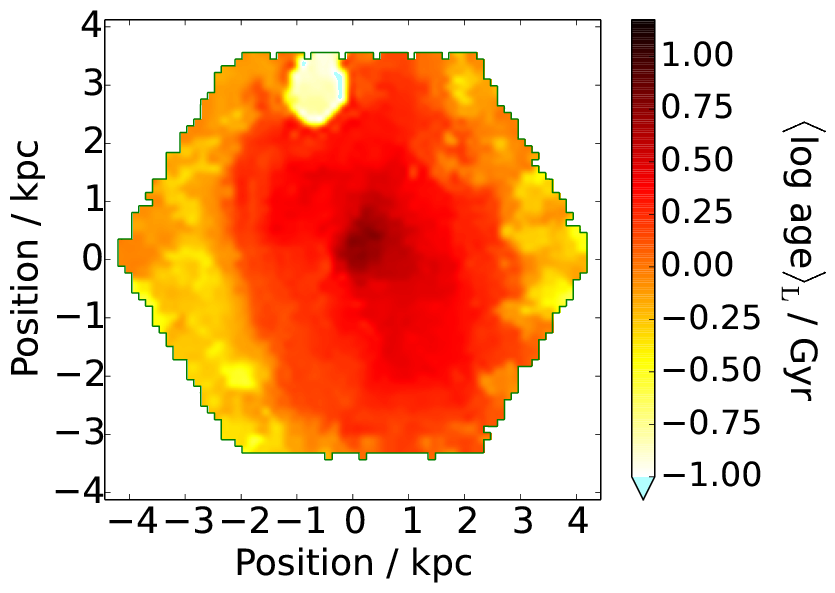}
	\caption{Light-weighted age maps of p9-127A as a function of stellar library; MILES, STELIB, and ELODIE are shown in the left, middle, and right panels respectively.}
\end{subfigure}
\begin{subfigure}{\linewidth}
	\includegraphics[width=0.33\linewidth]{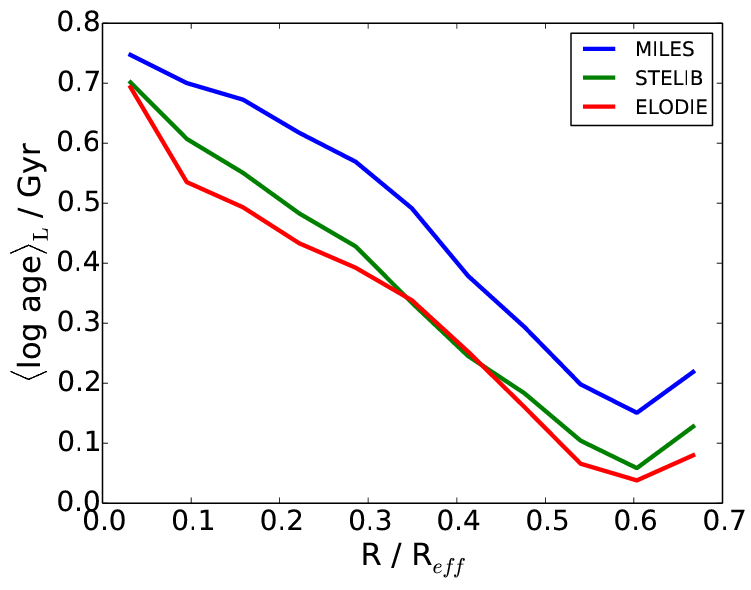}
	\includegraphics[width=0.33\linewidth]{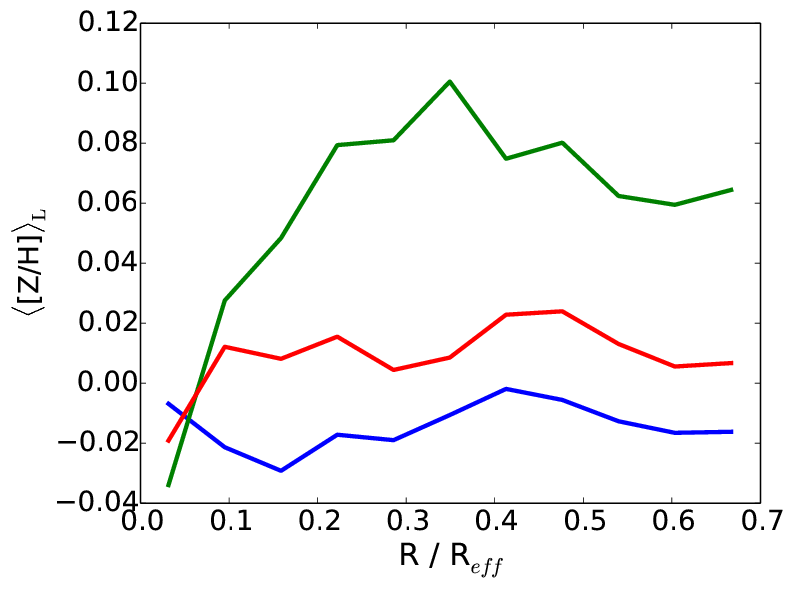}
	\includegraphics[width=0.33\linewidth]{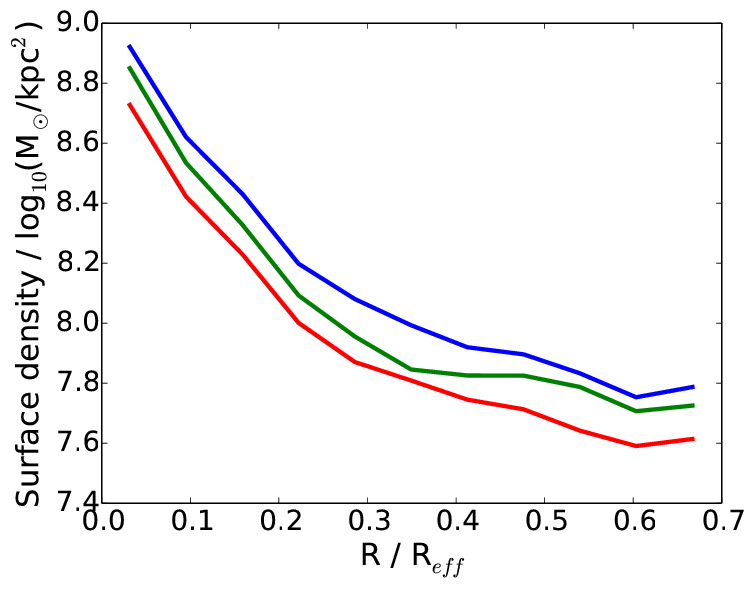}
	\caption{Radial profiles of stellar age, metallicity and surface mass density respectively as a function of stellar library.}
\end{subfigure}
\caption{The effect on galaxy properties of the stellar library input in the stellar population model. In all cases, models were fit with identical stellar population parameter grids (see section \protect\ref{sec:models}). Maps are composed of Voronoi binned cells of properties, corresponding to binned spectra that cover enough sky area such that the median signal-to-noise at any given cell is as close as possible to 5. Typically this means that the outer Voronoi cells of the galaxy cover a more extended area than the inner cells. The radial properties are computed from spectra summed in annuli that have radial extent that approximately scales logarithmically, in order to give the surface mass density (right-hand side) profiles.}
\label{comparemodels}
\end{figure*}

As anticipated in Section~\ref{sec:models}, we perform spectral fitting with models based on three different empirical stellar libraries. We use the same example galaxy \galaxyfour for which we show in Figure \ref{comparemodels} \logageL and \logageM maps, light-weighted age and metallicity profiles and stellar mass density profiles. In both this section and Section \ref{califacomp}, we use light-weighted properties for the radial profiles, as these are more sensitive tracers of processes that we wish to test recovery of, such as recent star formation or very low metallicity stellar populations. However we also discuss the differences between the light- and mass-weighted age maps.

Note that in this comparison we have restricted the model parameters to be identical, hence results obtained with the MILES-based models here will be different from those obtained with the full parameter range. This corresponds to a wavelength range of 3900 -- 6800 \AA, ages varying between 30 Myr and 15 Gyr, and metallicities of 0.5, 1.0, and 2.0 times solar metallicity. Hence for example, the metallicity gradient for MILES-based models is somewhat positive in this case compared to the negative gradient for the same galaxy fit with the full parameter range, for which we are able to use lower metallicity stellar population components.

These plots illustrate behaviours of the model comparisons that are found among all test galaxies, namely:

\noindent -- Overall distributions of stellar ages are similar in structure, however the average age values can vary by an order of 3 Gyr in the worst cases (see for example, the radial extent of high ages in the MILES stellar age map in Figure \ref{comparemodels}, compared to the other two models). Age gradients are the same for the three models.

\noindent -- Metallicity maps and profiles also show very similar structures among the three models, though STELIB-based models correspond to a slight increase of 0.1 dex in [Z/H] in the outer parts of its maps compared to the other models. The metallicity gradients are flat for MILES and ELODIE-based models but the STELIB-based models have a clear increase in the inner 0.2 \Rekpc, giving a positive metallicity gradient across the whole galaxy of 0.1 dex / \Rekpc. 

\noindent -- We note that ELODIE-based models have slightly redder spectra than STELIB-based models at these approximate parameters, explaining why STELIB-based models fits for higher metallicity and slightly higher age. Age-metallicity degeneracy causes this colour offset to mix between age and metallicity offsets. This is visible in Figure \ref{comparemodels} where the larger offsets in either the age or metallicity profiles at a given radius correspond to a smaller offset in the other profile.

\noindent -- Stellar mass maps obtained also show very similar structure, with 0.1 dex variation between models visible in the gradients. This is interpreted as different models fitting differently according to age-metallicity degeneracy, with the older ages (as in the MILES-based models) corresponding to higher stellar mass.

These systematic effects between libraries may be important and so we provide the fitting parameters obtained for all 3 libraries used. These conclusions allow us to have estimates for systematic errors in the modelling to bear in mind when considering comparisons with other fitting codes, methods, models and surveys. A more quantitative and comprehensive analysis of this will be subject of future work based on a larger set of MaNGA galaxies.

\subsection{Comparison with the CALIFA galaxy survey and with PPXF}\label{califacomp}

In the P-MaNGA sample there are three galaxies that overlap with the CALIFA survey \citep{2012A&A...538A...8S}. Two of these lie in group $\delta$, whose poor data quality in the P-MaNGA survey make a direct comparison between the results of the survey difficult to interpret. However, one galaxy, p9-127A, lies in group $\beta$, enabling a good opportunity to explore how MaNGA-like observations compare to CALIFA observations, even though this galaxy will not be in MaNGA due to its redshift being too low.

The fitting code used in CALIFA is STARLIGHT \citep{2005MNRAS.358..363C},  and the stellar population models are a blend of \cite{2010MNRAS.404.1639V} and \cite{2005MNRAS.357..945G} for STARLIGHT/CALIFA. 

\begin{figure*}
\centering
\begin{subfigure}{\linewidth}
	\includegraphics[width=0.33\linewidth]{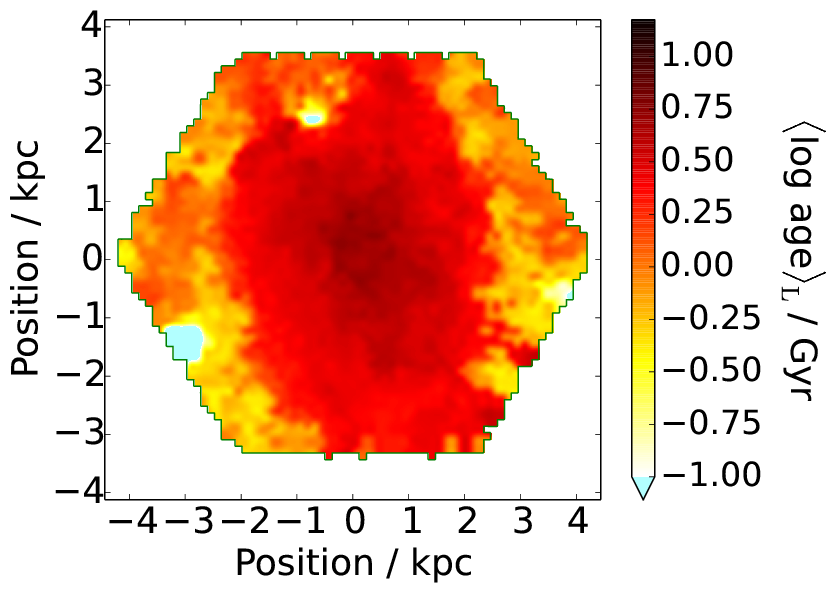}
	\includegraphics[width=0.33\linewidth]{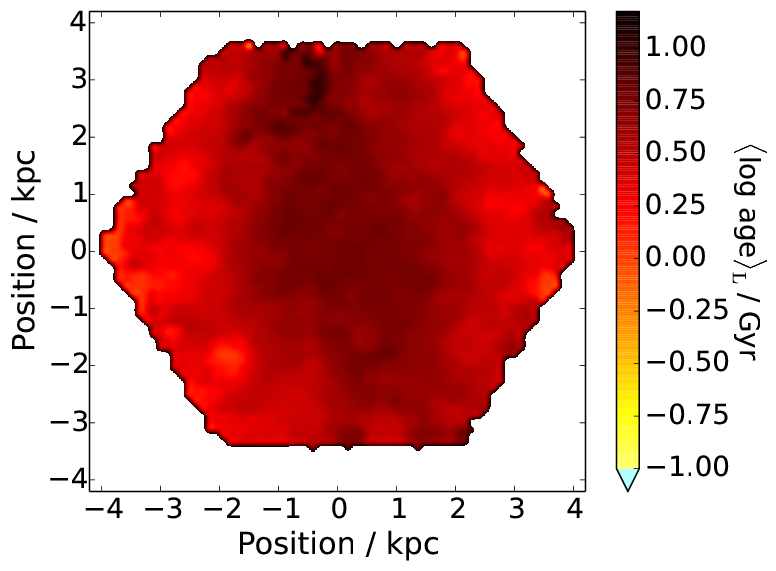}
	\includegraphics[width=0.33\linewidth]{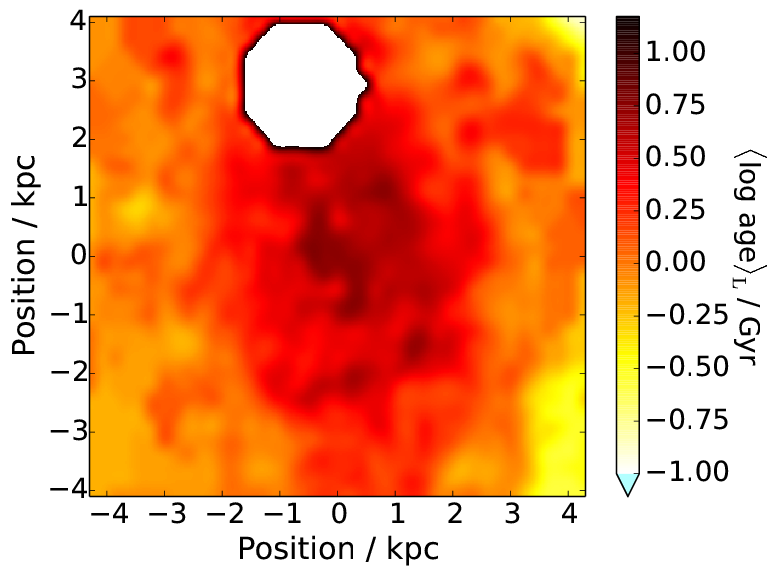}
	\caption{Stellar age maps for CALIFA 277 / p9-127A. In the left panel are the results from this paper. In the middle panel are the results from fitting to P-MaNGA with PPXF. In the right panel, results from fitting CALIFA data with STARLIGHT are shown as derived in \protect\cite{2013A&A...557A..86C}.}
\end{subfigure}
\begin{subfigure}{\linewidth}
	\includegraphics[width=0.33\linewidth]{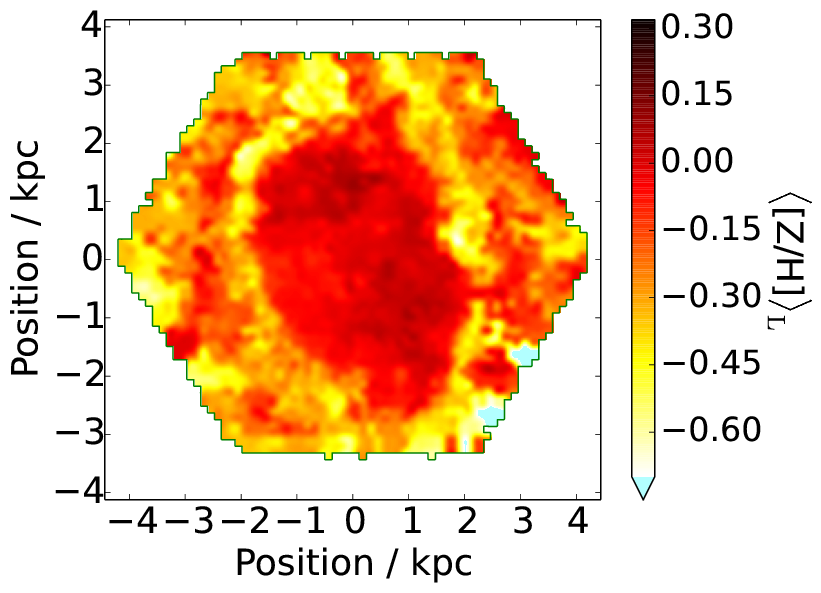}
	\includegraphics[width=0.33\linewidth]{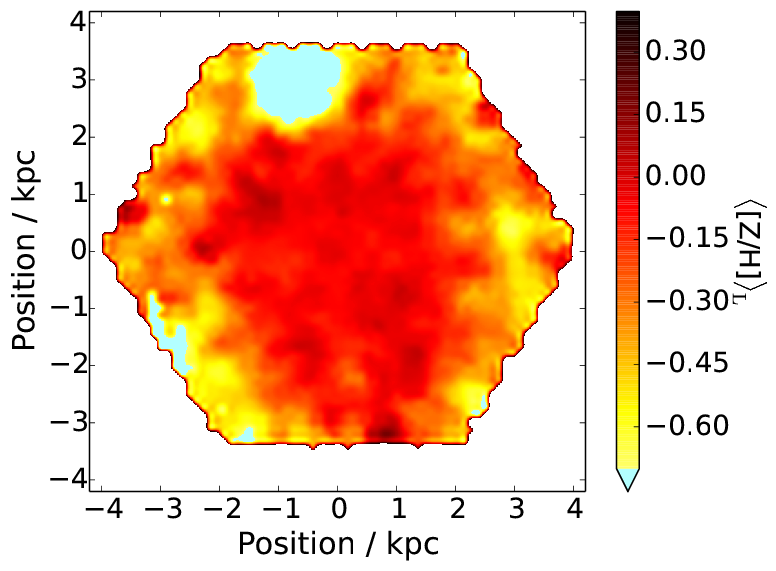}
	\includegraphics[width=0.33\linewidth]{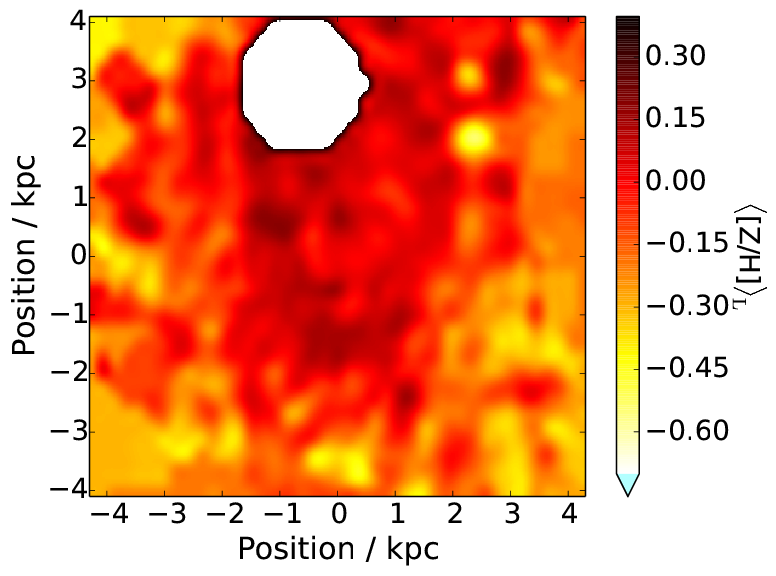}
	\caption{As above but for metallicity.}
\end{subfigure}
\begin{subfigure}{\linewidth}
\centering
\includegraphics[width=0.40\linewidth]{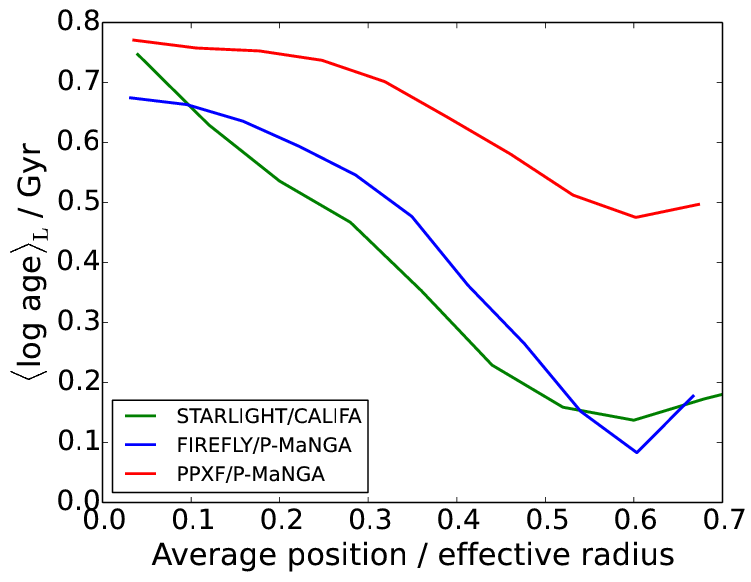}
\includegraphics[width=0.40\linewidth]{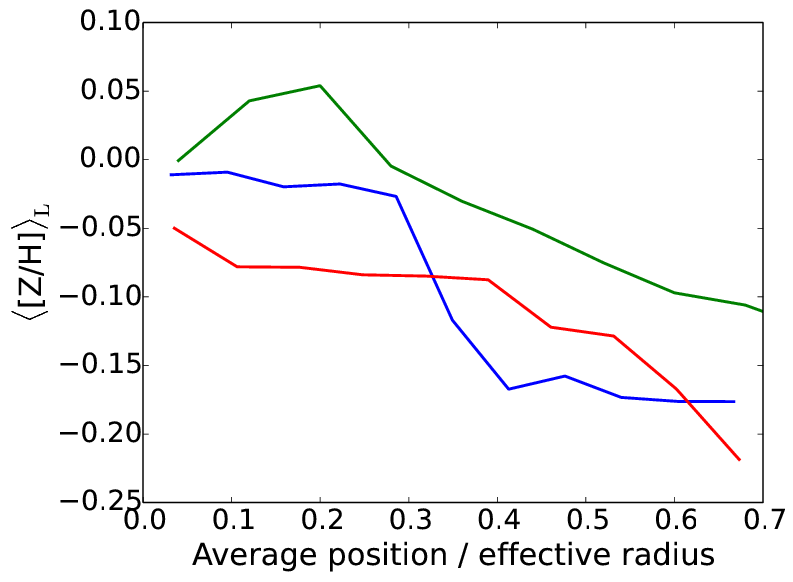}
\caption{Radial profiles of age (left-hand panel) and metallicity (right-hand panel), colour-coded by the survey and fitting code used.}
\end{subfigure}
\caption{Derived stellar population maps and profiles for CALIFA-277, known in this paper as p9-127A, a target shared by both the P-MaNGA and CALIFA surveys at observational conditions similar to those expected for MaNGA. The galaxy as observed with CALIFA is analysed with the spectral fitting code STARLIGHT, and is compared to the results from P-MaNGA observations with both \protect\firefly~analysis and analysis with PPXF as described in section \protect\ref{califacomp}.}
\label{fig:califa}
\end{figure*}

In the left hand panels of Figure \ref{fig:califa} we plot the light-weighted stellar age and metallicity of p9-127A as measured by applying CALIFA data cubes (CALIFA 277 in \cite{2013A&A...557A..86C}) but with this paper's plotting scheme. These can be compared directly with the corresponding plots applying \firefly~to P-MaNGA data cubes. We also compare to PPXF fits of the same P-MaNGA spectra as used in this paper's analysis in the middle panels of Figure \ref{fig:califa}.

We find a remarkable agreement in the structures of maps between STARLIGHT/CALIFA and \firefly/P-MaNGA, particularly between the stellar age and stellar mass map. The CALIFA metallicity map appears to show much finer structure in its core than the P-MaNGA map, although the general high metallicity core with lobes of lower metallicity is matched in both codes and surveys. This finer structure could be artificial, arising from uncertainties in deconvolving the CALIFA fiber spectra into the datacube, since CALIFA has a lower fiber density across the P-MaNGA radial footprint analysed in this galaxy. 

We look quantitively at these differences by binning the properties in elliptical annuli in the right panels of Figure \ref{fig:califa}, where we have plotted the corresponding properties from the left hand panel as radial profiles. Again we note remarkable agreement in stellar age ($<$ 0.1 dex at all points), bearing in mind that the statistical errors plotted will be much lower than the systematic errors from flux calibration, covariance between pixels, and stellar population model uncertainties.
We find a more complex structure, and generally lower values, in metallicity in the CALIFA profile than the P-MaNGA profile, differing by up to 0.1 dex in [Z/H].  We note that the stronger transition between high and low metallicity identified in \firefly~and, as discussed below, in PPXF, is consistent with the bulge/disk separation visible in the SDSS image of the galaxy. The agreement here is very encouraging since we use a different set of models, base stellar population components, wavelength range, fitting code and survey compared to \cite{2013A&A...557A..86C}.

In the middle panels of Figure~\ref{fig:califa} we show the \logageL and \metalL maps extracted with the full-spectrum fitting PPXF code on the same spectra fitted with \firefly. For this comparison PPXF used as input a grid of SSP models from \cite{2010MNRAS.404.1639V}, based on the MILES stellar library. Ages were sampled with 24 models logarithmically  spaced between 0.08 and 14 Gyr, while for each age metallicity was samples by 6 models with [M/H]~=~[-1.71, -1.31, -0.71, -0.40, 0.00, 0.22], for a total of 144 model templates. The fit was restricted to the wavelength  range from 3600 to 7400 \AA. To suppress the noise in the inferred population parameter we used linear regularization (keyword REGUL in PPXF) in the two-dimensional age and metallicity distribution. The level of regularization was chosen in such a way that the $\chi^2$ of the regularized fit increases by $\Delta\chi^2\approx\sqrt{2 N_{\rm pix}}$ from the unregularized best-fitting $\chi^2$. This provides the smoothest age-metallicity distribution consistent with the observed spectrum. The results are very weakly sensitive to the choice of the regularization parameter. The stellar kinematics was fitted simultaneously to the stellar population. We adopted a 6 degree multiplicative polynomial to account for possible inaccuracies in the relative flux calibration and remove the effect of dust attenuation on the continuum shape. No additive polynomials were used to avoid affecting the line strength of the SSP models.

The profiles and maps of PPXF fits find good spatial agreement with \firefly. The systematically higher ages, by about 0.2 going out to 0.4 dex, and the greater smoothness of the age maps compared to both STARLIGHT/CALIFA and \firefly/P-MaNGA could arise from differences in the ages of the SSP templates used. For example, the other two codes include younger ages (in our case down to 6.5 Myr, see Table~\ref{tab:modelparams}). The metallicity of the PPXF fits agrees with the other codes to within 0.1 dex. 
This suggests that the greatest sensitivity on the results obtained is from the range of SSPs used, based on this example. We will investigate these differences in detail for a larger sample in future work.

In conclusion we find an overall good, $\sim$ 0.1 dex, agreement between the CALIFA/STARLIGHT and P-MaNGA/\firefly, and across all three cases consistent recovery of the spatial features of the galaxy observed, and higher differences in age are present when comparing with P-MaNGA/PPXF which may just come by construction. Matching general profiles are recovered in all three cases, but differences caused by the different analysis codes, stellar population templates used and models are significant and larger than the random errors in the analysis of detailed features.

\section{Comparisons with other P-MaNGA analyses}\label{analcomp}

In this section we compare the analyses in this paper with other current analyses of the P-MaNGA data.

\subsection{Emission line ratios}

Emission line ratios are a powerful tool to study the state of the ionised gas in galaxies. The Baldwin-Phillips-Terlevich (BPT, \cite{Baldwin1981, Veilleux1987}) diagram, making use of the line ratios [NII] $\rm \lambda 6584 / H \alpha$ and $\rm [OIII] \lambda 5007 / H \beta$, is the most popular diagnostic used to distinguish ionisation due to star-formation from ionisation due to other sources, including Active Galactic Nuclei (AGN), Low Ionisation Nuclear Emission-line Regions (LINERs) and shocks. A detailed analysis of the ionised gas content of the P-MaNGA galaxies using the BPT diagram, complemented with other diagnostics, is presented in \cite{2014arXiv1410.7781B}.

\begin{figure*}
\centering
\begin{subfigure}{\linewidth}
	\includegraphics[width=0.5\linewidth]{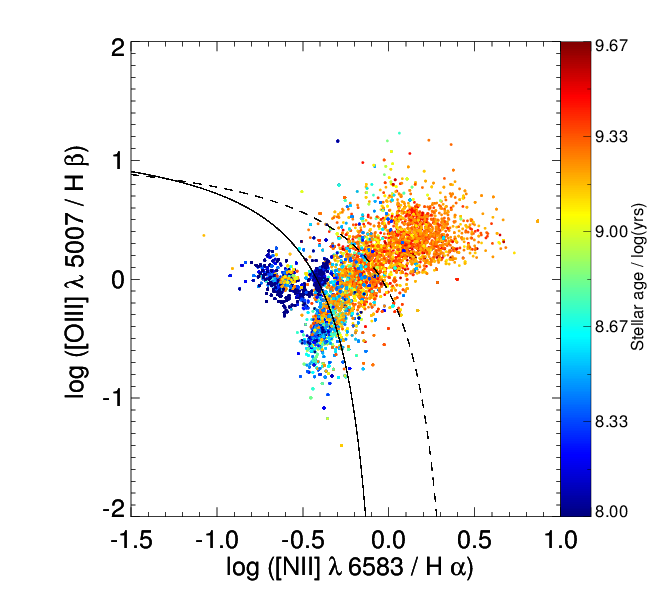}
	\includegraphics[width=0.5\linewidth]{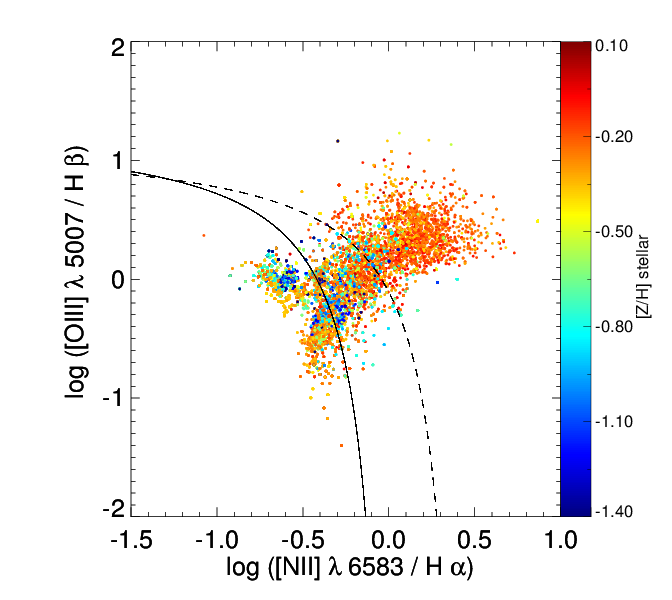}
\end{subfigure}
\caption{Emission line ratios as derived in \protect\cite{2014arXiv1410.7781B} as a function of light-weighted stellar ages (left) and metallicities (right) as derived in this paper. Overplotted are demarcation lines from \protect\cite{Kewley2001} and \protect\cite{Kauffmann2003a}. Each pixel from all of their subsample of P-MaNGA galaxies is plotted following the binning scheme used in \protect\cite{2014arXiv1410.7781B}.}
\label{fig:fran_plots}
\end{figure*}

In this section we compare the results of the BPT classification with stellar ages and metallicities derived in this paper from full spectral fitting. We make use of the emission lines maps created by \cite{2014arXiv1410.7781B} and plot in Fig. \ref{fig:fran_plots} the position of all the regions in the 14 galaxies considered by \cite{2014arXiv1410.7781B} in the BPT diagram. \cite{2014arXiv1410.7781B} discuss a comparison with D4000 and $H \delta$ absorption. Here we extend this analysis by directly comparing with stellar population ages and metallicities. We make use of the demarcation lines of \cite{Kewley2001} and \cite{Kauffmann2003a}. We refer the reader to \cite{2014arXiv1410.7781B} for a careful discussion of the assumptions and caveats implicit in the use of these demarcation lines. We expect galactic regions (Voronoi bins) which lie below the demarcation Kauffmann 2003 line to harbour on-going star-formation in H\textsc{ii} regions. This is confirmed by the study of the stellar population ages, as evident in the left-hand-side of Figure \ref{fig:fran_plots}, where the regions lying below the Kauffmann 2003 line present younger stellar populations, consistently with a significant population of O and B stars, capable of ionising classical H\textsc{ii} regions. 

We wish to remark on the cluster of young points ($t\sim100$~Myr) which lie just across the Kauffmann 2003 line: these all belong to a single galaxy (p11-19C) which is experiencing a star-burst. This galaxy is found by \cite{2014arXiv1410.7781B} to be Nitrogen enriched and hence appearing to cross over into the `Composite' region of the BPT diagram, between the two demarcation lines, although its ionisation is fully compatible with star-formation.

It is very interesting to note that, as we move away from the star-forming sequence in the BPT diagram towards the `right-wing', inhabited by AGN and LINERs, we see progressively older stellar populations. There is a tantalising hint that the regions lying at higher $\rm [NII] / H \alpha$, generally associated with LINERs, are older than the regions at lower $\rm [NII] / H \alpha$, generally associated with AGN. Larger statistics and a bona-fide AGN sample is needed to make further progress in the study of stellar populations of AGN and LINER hosts, but the results from these preliminary observations seem to confirm trends already observed in SDSS for the global galaxy population \citep{Kauffmann2003a, Kewley2006, CidFernandes2010}. This analysis also confirms the more simplistic conclusions reached using stellar populations indices ($\rm D_N(4000)$ and $\rm EW(H \delta_A)$) in \cite{2014arXiv1410.7781B} and \cite{2015arXiv150207040L}.

The right-hand-side of Figure \ref{fig:fran_plots} shows BPT diagram colour-coded by stellar metallicity. We see that AGN hosts have more metal-rich stellar populations, most likely because they are more massive, which would then give a mass-metallicity relationship as described in e.g. \cite{2010MNRAS.404.1775T}. In the star forming regions, on the other hand, we see a variety of stellar metallicities from relatively metal-poor to metal-rich. The pixels with metal-poor populations are mostly found at high $[OIII] / H \beta$ ratio, coinciding with regions of low metallicity in the gas \citep{1978ApJ...222..821S}.

\subsection{Absorption and emission line diagnostics}
\cite{2015arXiv150207040L} perform emission- and absorption-line fitting to obtain 2D maps and radial gradients of the $D_{4000}$~Angstrom break and the equivalent widths of $H\delta_A$ absorption and $H\alpha$ emission for the P-MaNGA dataset. They categorise the dataset into two groups; `centrally quiescent/passive', where the $D_{4000}$~break in the central pixel of the P-MaNGA data cube of the galaxy observed is above 1.6, and `centrally starforming', where the $D_{4000}$~ break is below 1.6. They find that their centrally star-forming galaxies generally have very weak radial variation in these diagnostics, whereas the gradients of centrally quiescent galaxies are significant. 

\begin{figure*}
\centering
\begin{subfigure}{\linewidth}
	\includegraphics[width=0.5\linewidth]{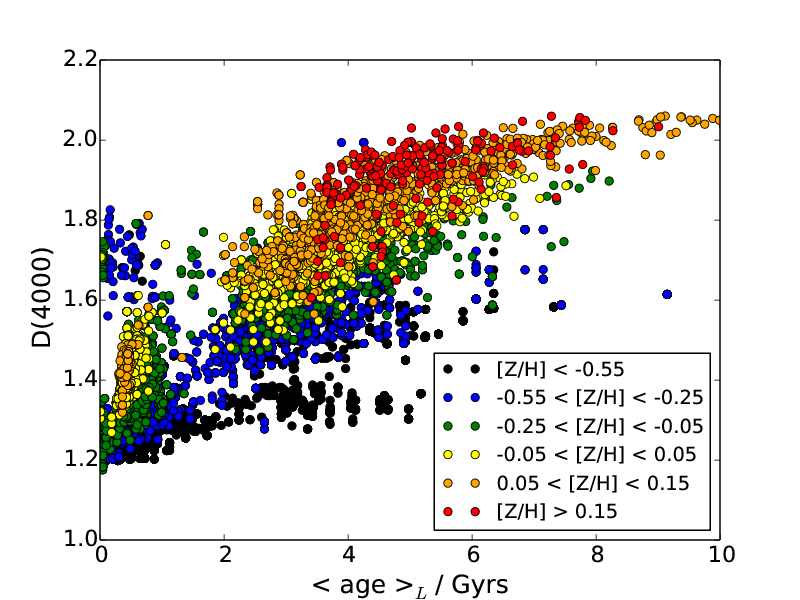}
	\includegraphics[width=0.5\linewidth]{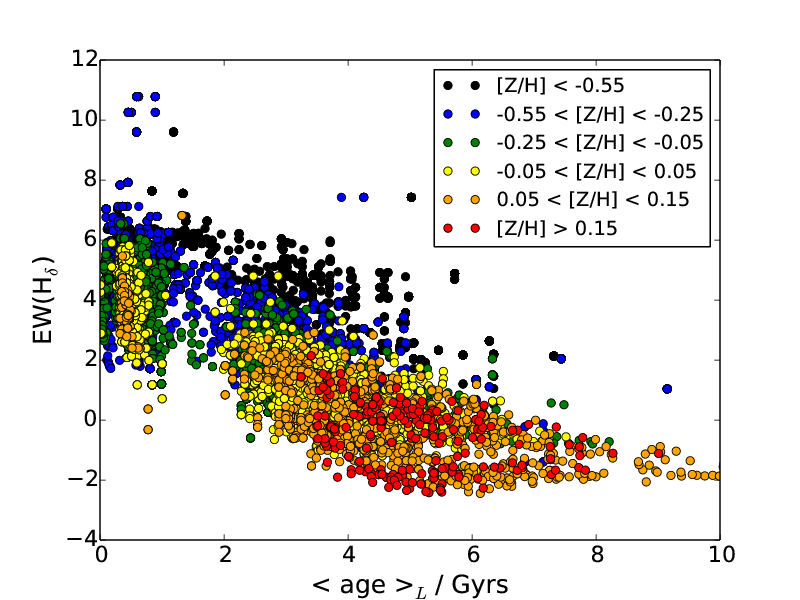}
\end{subfigure}
\caption{$D_{4000}$~and EW($H\alpha$) measurements from \protect\cite{2015arXiv150207040L} as a function of the light-weighted stellar ages and metallicities (colour-coded) derived in this paper. Each pixel from all of their subsample of P-MaNGA galaxies is plotted following the binning scheme used in \protect\cite{2014arXiv1410.7781B}.}
\label{fig:cheng_age}
\end{figure*}

We compare their results to the ages and metallicities obtained from spectral fitting in two sets of figures, first to assess how well their indices trace our results, and secondly to see whether we recover similar gradients as a function of galaxy group. We note that plate 11 galaxies, identified in this paper as being the most difficult to recover smooth gradients from, are excluded from their analysis due to being unable to obtain reliable diagnostic measurements. Figure \ref{fig:cheng_age} shows the pixel-by-pixel $H\delta_A$ and $D_{4000}$~ as a function of this paper's recovered ages and metallicities.

As expected, we see a clear dependence of both $D_{4000}$~and EW($H\alpha$) with both age (\cite{Kauffmann2003a}, \cite{2014MNRAS.441.2717K}) and metallicity (\cite{1995AJ....109.1433D}, \cite{1997ApJS..111..377W}, \cite{2005MNRAS.362..799M}), as well as some scatter which could be due to other dependencies such as dust attenuation and element abundance ratios (\cite{2004MNRAS.351L..19T}, \cite{2012A&A...538A...8S}) as well as observational errors and statistical errors in the spectral fitting. We confirm a clear dependence on age, which becomes flatter as metallicity decreases, for both indices (see \cite{2005MNRAS.362..799M}). We see that there is an approximately 0.1 in $D_{4000}$~ and 1.0 in EW($H\delta_A$) scatter. 

We note that there are two branches of high metallicity, high age points in the $H\delta_A$ plot that are not visible in the $D_{4000}$ plot. The bifurcation is due to most of the high metallicity points coming from a single galaxy in the sample, galaxy p9-127B, which is the dust lane clumpy galaxy shown in Figure \ref{fig:baddust}. The galaxy has high metallicity (above \metalL $>$ 0.05), high age (\logageL $>$ 5 Gyr) and high $D_{4000}$ ($>$ 1.8) across all of the bulge and most of the disc, except the dust lane region where these values are much lower. However, the EW($H\delta_A$) has a large asymmetry between the two sides of the core along the major axis that transitions quickly between very low ($\sim -2$) to more moderate ($\sim 0$) values, giving rise to the appearance of two fairly distinct branches in the $H\delta_A$ plot. The physical explanation for why this asymmetry exists in this galaxy in  $H\delta_A$ and not the other properties is not currently known.

\begin{figure*}
\centering
\begin{subfigure}{\linewidth}
	\includegraphics[width=0.5\linewidth]{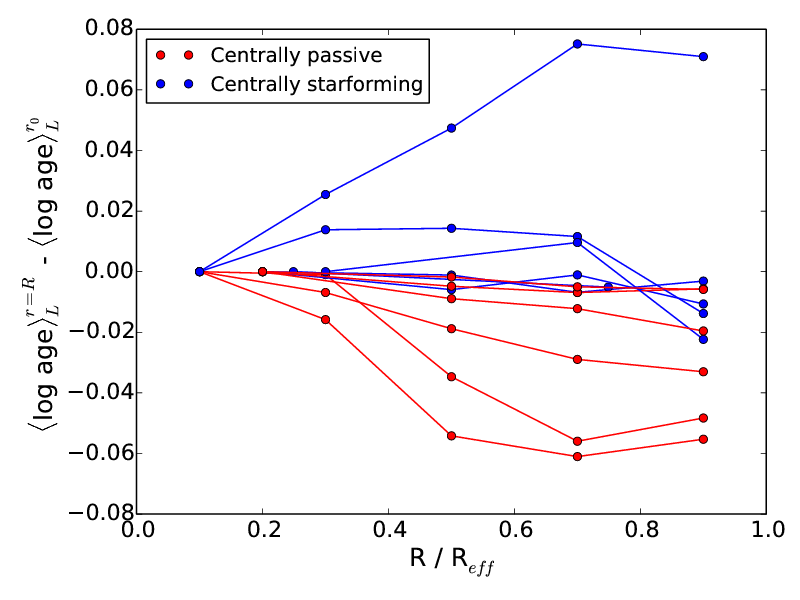}
	\includegraphics[width=0.5\linewidth]{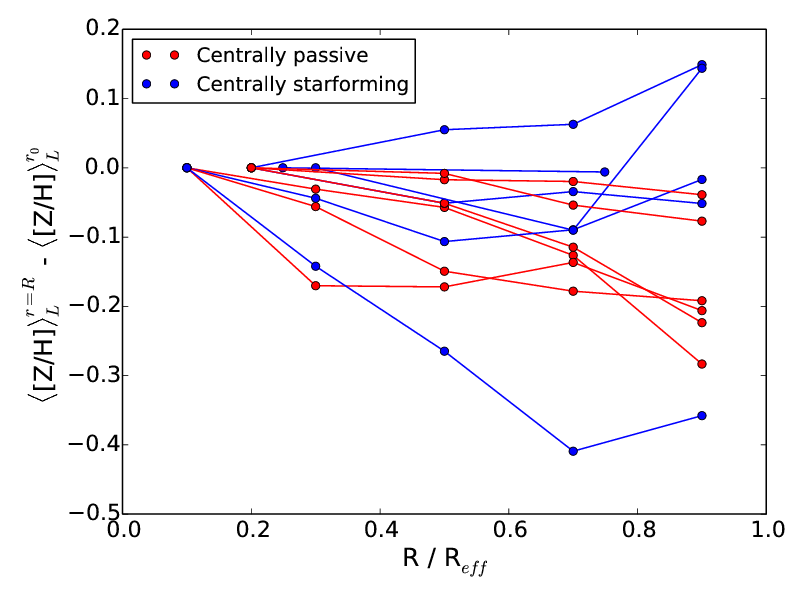}
	\includegraphics[width=0.5\linewidth]{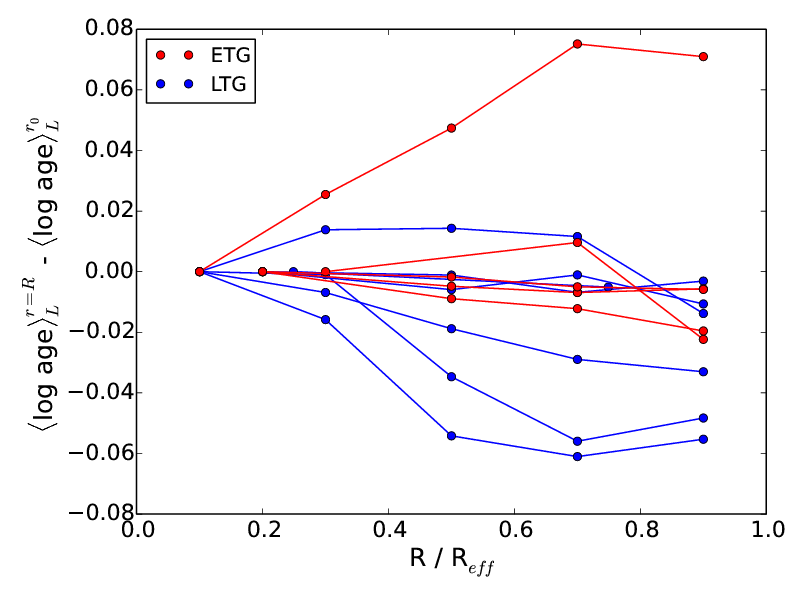}
	\includegraphics[width=0.5\linewidth]{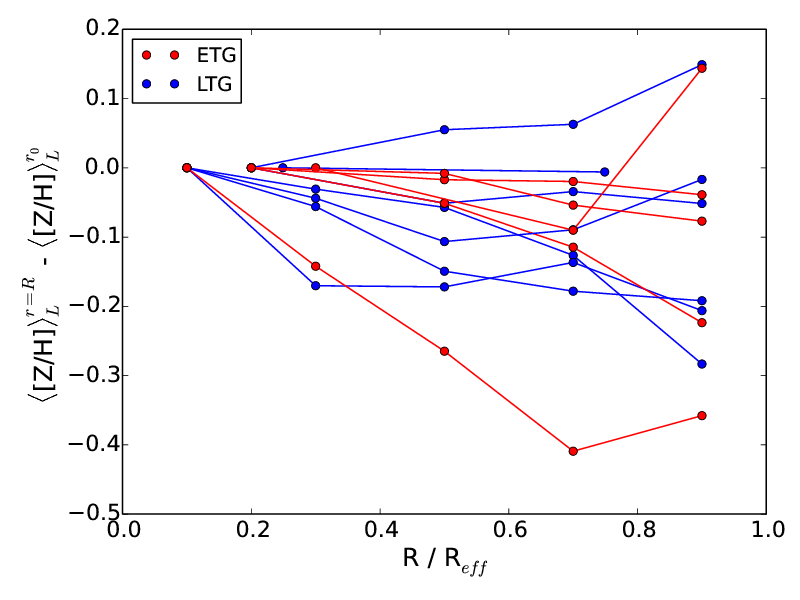}
\end{subfigure}
\caption{Light-weighted age and metallicity gradients up to 1 effective radius, plotted as the change in age / metallicity from the first elliptical annular bin, where each line represents the profile from a P-MaNGA galaxy. In the upper panels we use the classifications of centrally quiescent/passive or centrally starforming as described in \protect\cite{2015arXiv150207040L}. In the lower panels we use the morphological classifications in this paper (as in Table \protect\ref{tab:sample}). Bins are 0.2 effective radius in size to match \protect\cite{2015arXiv150207040L} as much as possible, however we impose a condition that each bin must contain at least 10 pixels from the corresponding data cube. If this condition is not met we sample at wider bins until the criterion is met, in order to ensure a reliable mean age or metallicity at each point.}
\label{fig:cheng_grad}
\end{figure*}

In figure \ref{fig:cheng_grad} we use the same plotting and binning conventions used in \cite{2015arXiv150207040L}, binning our properties in 0.2 effective radius annular bins and measuring the difference in age and metallicity determined from the first bin (r$_0$). We use classifications based on measured $D_{4000}$ in the central pixel of each galaxy in the upper panels, and use the same morphological classification as used in Table \ref{tab:sample} in the lower panels.

Encouragingly we see that age gradients are clearly the main driver of the split between centrally quiescent and centrally star forming galaxies, with metallicity gradients showing no clear divide between the two. We emphasise that this result is not in conflict with the gradient results presented earlier. The centrally quenched galaxies in \cite{2015arXiv150207040L} contain a mix of different morphologies, and in particular include objects in which the combination of a passive central bulge with a star forming disc result in the age gradients measured. If we separate again by morphology as in Section \ref{radialgradients}, we recover the conclusions presented in that section as shown by Figure \ref{fig:gradsummary}. We note that compared to the gradients derived in Figure \ref{fig:gradsummary}, we only analyse up to 1 \Rekpc~in this figure to match \cite{2015arXiv150207040L}, hence we do not probe the outer regions of the early type galaxies which will typically have a lower metallicity stellar component and therefore a more negative gradient.  

We find that age gradients in late type galaxies are typically negative whereas the early type galaxies are typically flat. The metallicity gradients of late-type galaxies show no clear separation, while the early type galaxies have on average more negative gradients. This shows that the clear distinction in age and metallicity gradients between centrally quiescent and centrally star forming galaxies as discussed in \cite{2015arXiv150207040L} is mostly driven by bulge-disc transition. Such a transition will be more clearly identifyiable in the full MaNGA sample.

\section{Conclusions}\label{conclusions}

We have shown and discussed the first application of full spectral fitting of stellar population models to P-MaNGA galaxy data, a prototypical data release preceeding the start of the actual MaNGA IFU galaxy survey (\cite{2015ApJ...798....7B}). We embarked in this work for several reasons, namely exploring the capabilities of our fitting and modelling method to unveil the physical parameters of galaxies and study our success as a function of observational conditions, exposure times, and galaxy type. As a hot spot of our approach, we introduce a method to deal with dust attenuation that by removing the large scale features of spectra such as the continuum before fitting, allow us to fit unattenuated spectra and to then derive very accurate dust values. We calibrate the method by using galaxies which display dust lanes in their optical images as well as mock galaxies with known dust input. We show that we can accurately recover the dust displacement in the stellar population maps. This now allows us to calculate other stellar population properties that are more accurate and physically plausible, such as radially symmetric stellar mass maps.

We present our calculations as 2D maps of age, metallicity, dust and stellar mass and their corresponding radial profiles and gradients. Since the P-MaNGA data contain a varied selection of galaxy types, we show that we are able to recover many different features and structures in this data, such as spiral arms, additional galaxy components, clear radial profiles for spheroidal galaxies, and artefacts in the data. This will be a superlative tracer of galaxy evolution.

We derive radial profiles of age, metallicity and stellar mass for all of the P-MaNGA data. We show that the mass profiles are in general very smooth as a function of radius. Age and metallicity profiles can show unevenness, which depends on the quality of observations. We notice that the maps of poorly observed or noisy galaxies display an increased variance on the Voronoi bins which create fake structures, but most importantly erase the signature of real structures as spiral arms. 

Importantly for galaxy evolution studies, we derive the radial gradients of stellar population properties. We find that metallicity gradients are negative for spheroids (-- 0.15 dex / \Rekpc), compared to the flat gradients of late-type galaxies, which is consistent with previous knowledge. Age gradients are also negative for late type galaxies (-- 0.39 dex / \Rekpc), though almost negligible for spheroids, which is again consistent with the literature. We calculate statistical errors on the radial profiles and gradients obtained as a function of observational quality, comprised of S/N, airmass and exposure times, which we will use to calibrate analysis of MaNGA data.

Additionally we demonstrate that the stellar mass density profiles remain relatively unchanged as a function of three stellar population model input stellar spectral libraries, even if we find $\sim$ 0.1 dex differences in the age and metallicity profiles.  Nonetheless the overall set of gradients for the P-MaNGA data sample is consistent across the 3 model libraries. Additionally we show that the recovery of spatial features in age and metallicity is consistent across model, fitting code, fitting parameters and the CALIFA survey.

Lastly we show how comparisons with analyses of the absorption and emission features MaNGA data can shed light on galaxy evolution questions such as inside-out growth and/or quenching, and studies of AGN. 

This work will proceed with the analysis of the full MaNGA dataset in the future, and constitutes the basis of our understanding of the performances of full spectral fitting on MaNGA-type IFU data. 

\section{Acknowledgements}

Funding for SDSS-III and SDSS-IV has been provided by the Alfred P.~Sloan Foundation and Participating Institutions.  Additional funding for SDSS-III comes from the National Science Foundation and the U.S.~Department of Energy Office of Science.  Further information about both projects is available at {\tt www.sdss3.org}.

SDSS is managed by the Astrophysical Research Consortium for the Participating Institutions in both collaborations.  In SDSS-III these include the University of Arizona, the Brazilian Participation Group, Brookhaven National Laboratory, Carnegie Mellon University, University of Florida, the French Participation Group, the German Participation Group, Harvard University, the Instituto de Astrofisica de Canarias, the Michigan State/Notre Dame/JINA Participation Group, Johns Hopkins University, Lawrence Berkeley National Laboratory, Max Planck Institute for Astrophysics, Max Planck Institute for Extraterrestrial Physics, New Mexico State University, New York University, Ohio State University, Pennsylvania State University, University of Portsmouth, Princeton University, the Spanish Participation Group, University of Tokyo, University of Utah, Vanderbilt University, University of Virginia, University of Washington, and Yale University.

The Participating Institutions in SDSS-IV are Carnegie Mellon University, Colorado University, Boulder, Harvard-Smithsonian Center for Astrophysics Participation Group, Johns Hopkins University, Kavli Institute for the Physics and Mathematics of the Universe Max-Planck-Institut fuer Astrophysik (MPA Garching), Max-Planck-Institut fuer Extraterrestrische Physik (MPE), Max-Planck-Institut fuer Astronomie (MPIA Heidelberg), National Astronomical Observatory of China, New Mexico State University, New York University, The Ohio State University, Penn State University, Shanghai Astronomical Observatory, United Kingdom Participation Group, University of Portsmouth, University of Utah, University of Wisconsin, and Yale University.

The authors would like to thank Sebastian F. Sanchez for his very useful and detailed comments and to Roberto Cid Fernandes for providing access to the CALIFA data used in Figure \ref{fig:califa}.

Numerical computations were done on the Sciama High Performance Compute (HPC) cluster which is supported by 
the ICG, SEPNet and the University of Portsmouth.

MC acknowledges support from a Royal Society University Research Fellowship.

{\footnotesize{
  \bibliographystyle{mn2e}
  \bibliography{bibdesk}
 }}

\vspace{1cm}
\noindent$^{1}$~Institute of Cosmology and Gravitation, University of Portsmouth, Portsmouth, PO1 3FX, UK\\
	$^{2}$~European Southern Observatory, Karl-Schwarzschild-Stra\ss e 2, D-85748 Garching bei Muenchen, Germany\\
	$^{3}$~School of Physics and Astronomy, University of St Andrews, North Haugh, St Andrews KY16 9SS, UK\\
	$^{4}$~Sub-department of Astrophysics, Department of Physics, University of Oxford, Denys Wilkinson Building, Keble Road, Oxford OX1 3RH, UK\\
	$^{5}$~Kavli Institute for Cosmology, University of Cambridge, Madingley Road, Cambridge CB3 0HA, UK\\
	$^{6}$~Cavendish Laboratory, University of Cambridge, 19 J. J. Thomson Avenue, Cambridge CB3 0HE, UK\\
	$^{7}$~Department of Astronomy, University of Wisconsin-Madison, 475 N. Charter Street, Madison, WI, 53706, USA\\
	$^{8}$~Center for Cosmology and Particle Physics, Department of Physics, New York University, 4 Washington Place, New York, NY 10003\\
	$^{9}$~Kavli Institute for the Physics and Mathematics of the Universe, Todai Institutes for Advanced Study, the University of Tokyo, Kashiwa, Japan 277-8583 (Kavli IPMU, WPI)\\
	$^{10}$~Yale Center for Astronomy and Astrophysics, Yale University, New Haven, CT, 06520, USA\\
	$^{11}$~Dunlap Institute for Astronomy and Astrophysics, University of Toronto, 50 St. George Street, Toronto, Ontario M5S 3H4, Canada\\
	$^{12}$~McDonald Observatory, Department of Astronomy, The University of Texas at Austin, 2515 Speedway Stop C1402, Austin, Texas 78712-1206, USA\\
	$^{13}$Universit\'e Lyon 1, Observatoire de Lyon, Centre de Recherche Astrophysique de Lyon and Ecole Normale Sup\'erieure de Lyon, 9 avenue Charles Andr\'e, F-69230 Saint-Genis Laval, France\\
	$^{14}$~Department of Physics \& Astronomy, University of Iowa, 751 Van Allen Hall, Iowa City, IA 52242, USA\\
	$^{15}$~Space Telescope Science Institute, 3700 San Martin Drive, Baltimore, MD 21218, USA\\
	$^{16}$~Partner Group of Max-Planck Institute for Astrophysics, Shanghai Astronomical Observatory, Nandan Road 80, Shanghai 200030, China\\
	$^{17}$~Department of Physics and Astronomy, University of Kentucky, 505 Rose Street, Lexington, KY 40506-0055, USA\\
	$^{18}$~Apache Point Observatory and New Mexico State University, P.O. Box 59, Sunspot, NM, 88349-0059, USA

\section{Appendix}

In this Appendix we provide the stellar population maps and radial profiles of the P-MaNGA galaxies described briefly in the Results section, but not shown there to avoid disrupting the flow of the paper with detail about individual galaxies. As in the Results section, we split our analyses by groups as described in Table \ref{tab:sample}.

\subsection{Additional maps: Group \large{$\alpha$}}

\begin{figure*}
\centering
\vspace{0.5cm}
\hspace{0.3cm}
\begin{subfigure}{0.30\linewidth}
	\hspace{0.4cm}
	\includegraphics[width=0.72\linewidth]{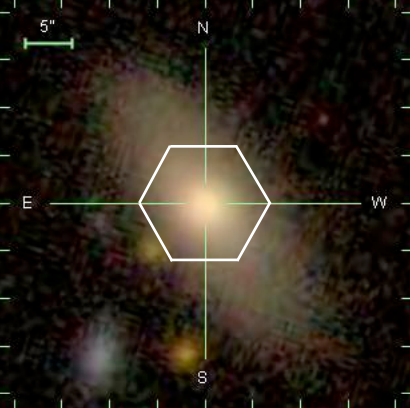}
	\caption{SDSS image with the P-MaNGA footprint.}
\end{subfigure} 
\begin{subfigure}{0.32\linewidth}
	\includegraphics[width=\linewidth]{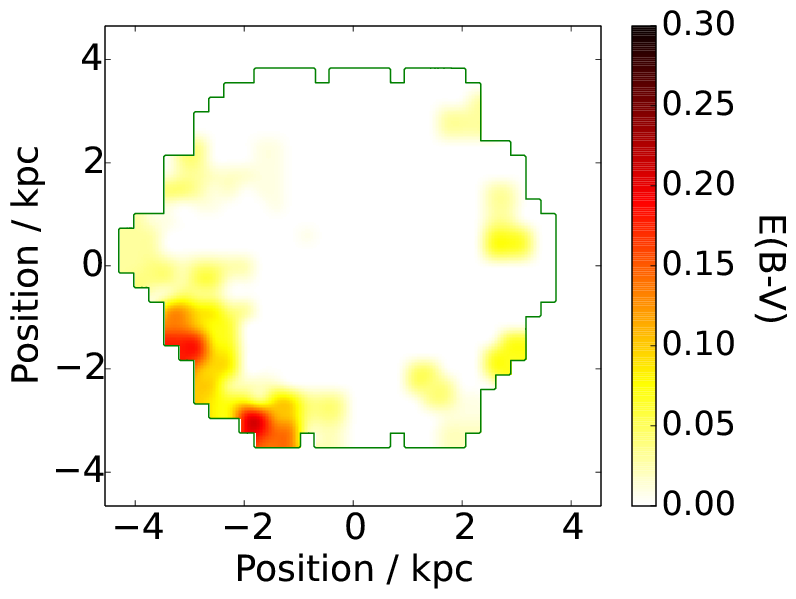}
	\caption{Dust extinction, E(B-V).}
\end{subfigure}
\begin{subfigure}{0.32\linewidth}
	\includegraphics[width=\linewidth]{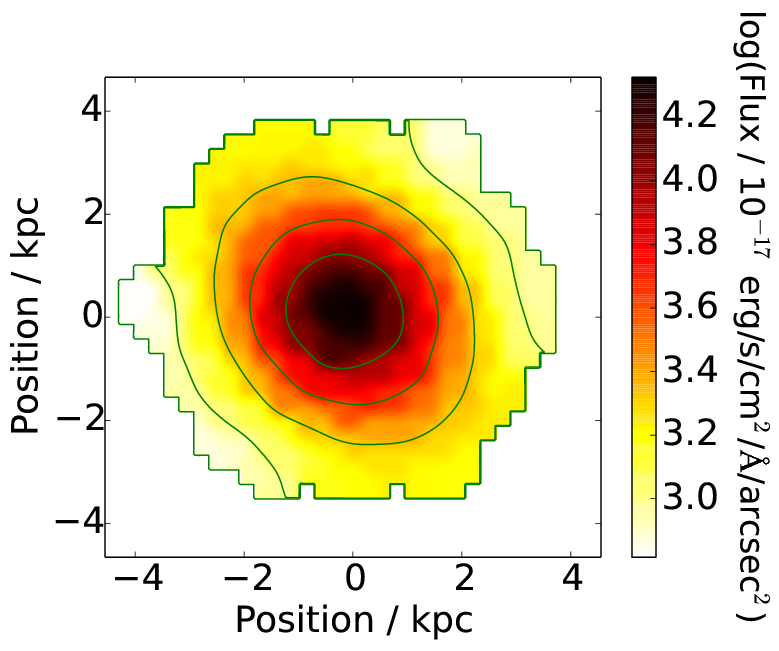}
	\caption{Flux map with isoflux contours (green).}
\end{subfigure}\hspace{0.1cm}
\begin{subfigure}{0.32\linewidth}
	\includegraphics[width=\linewidth]{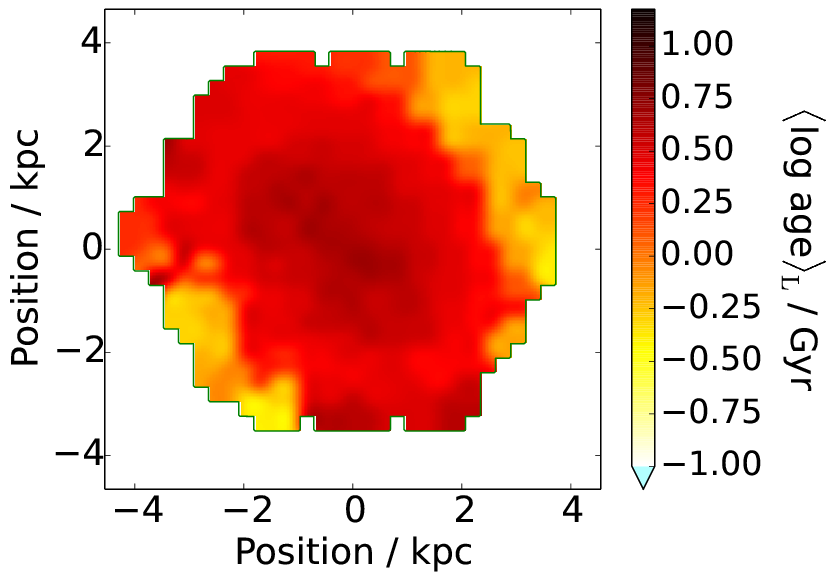}
	\caption{Luminosity-weighted stellar age.}
\end{subfigure}
\begin{subfigure}{0.32\linewidth}
	\includegraphics[width=\linewidth]{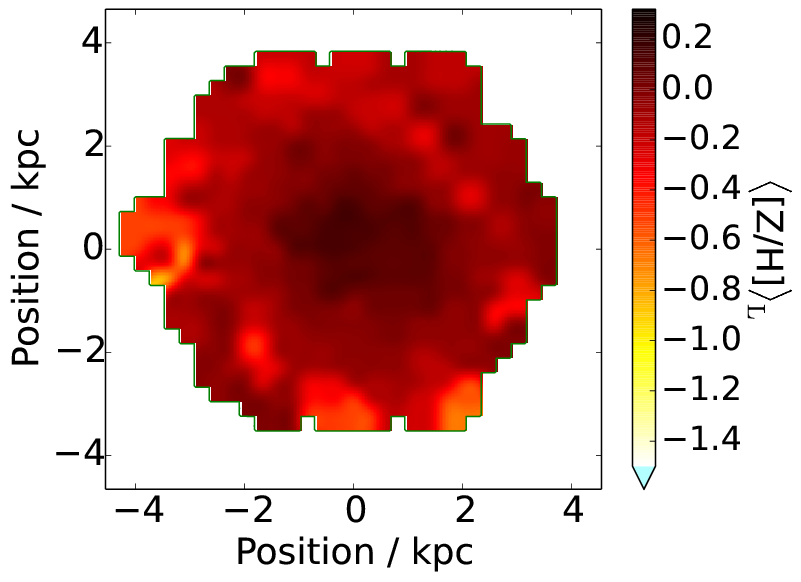}
	\caption{Luminosity-weighted metallicity.}
\end{subfigure}
\begin{subfigure}{0.32\linewidth}
	\includegraphics[width=\linewidth]{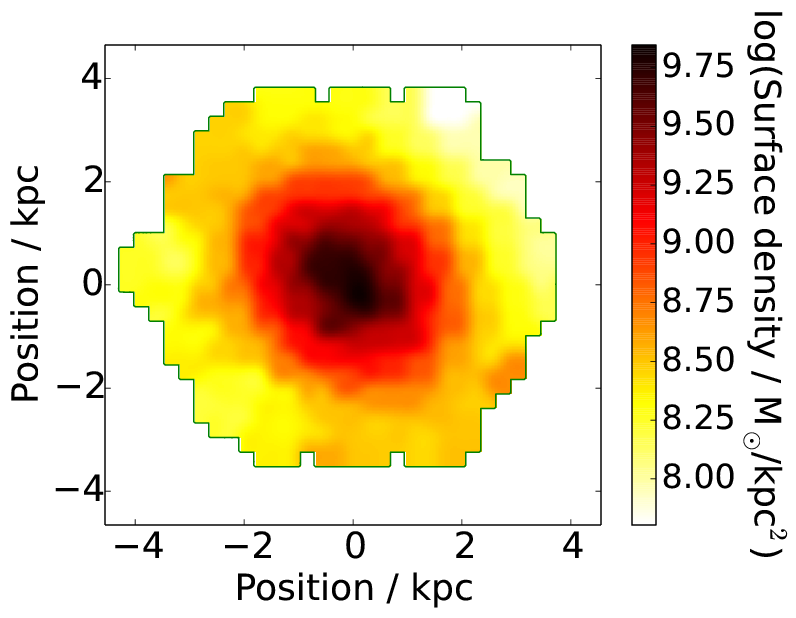}
	\caption{Stellar mass.}
\end{subfigure}
\begin{subfigure}{0.32\linewidth}
	\includegraphics[width=\linewidth]{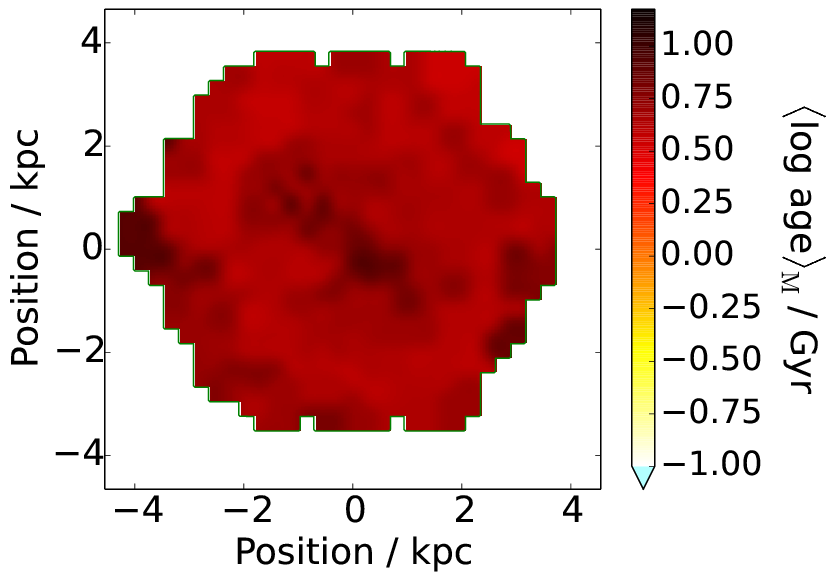}
	\caption{Mass-weighted stellar age.}
\end{subfigure}
\begin{subfigure}{0.32\linewidth}
	\includegraphics[width=\linewidth]{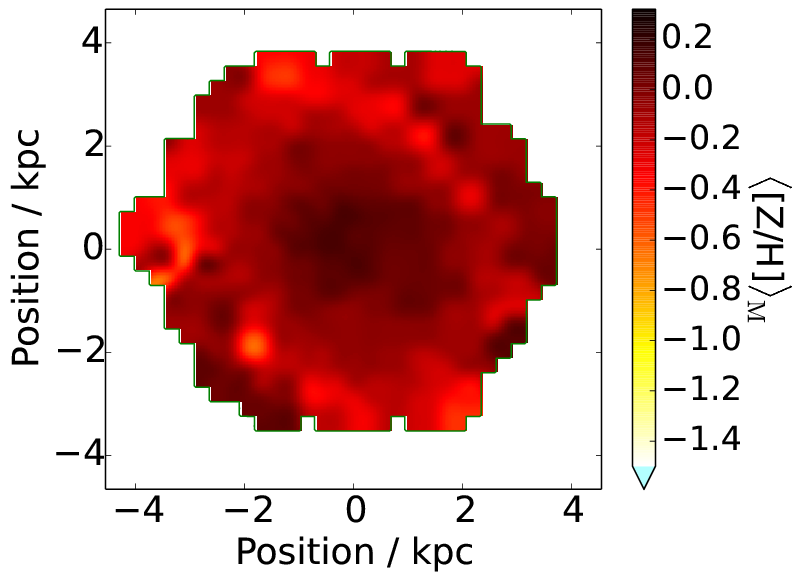}
	\caption{Mass-weighted metallicity.}
\end{subfigure}
\begin{subfigure}{0.32\linewidth}
	\includegraphics[width=\linewidth]{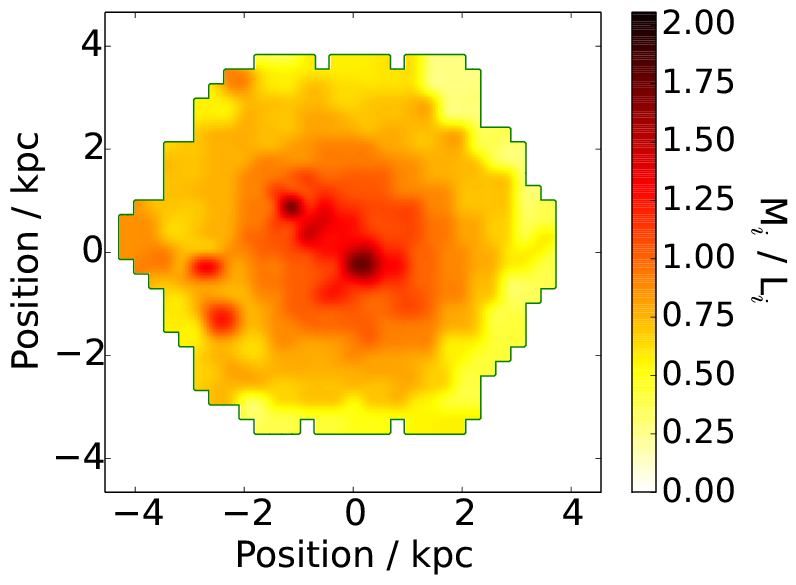}
	\caption{Stellar mass-to-light ratio in the SDSS $i$-band.}
\end{subfigure}
\begin{subfigure}{0.3\linewidth}
	\includegraphics[width=\linewidth]{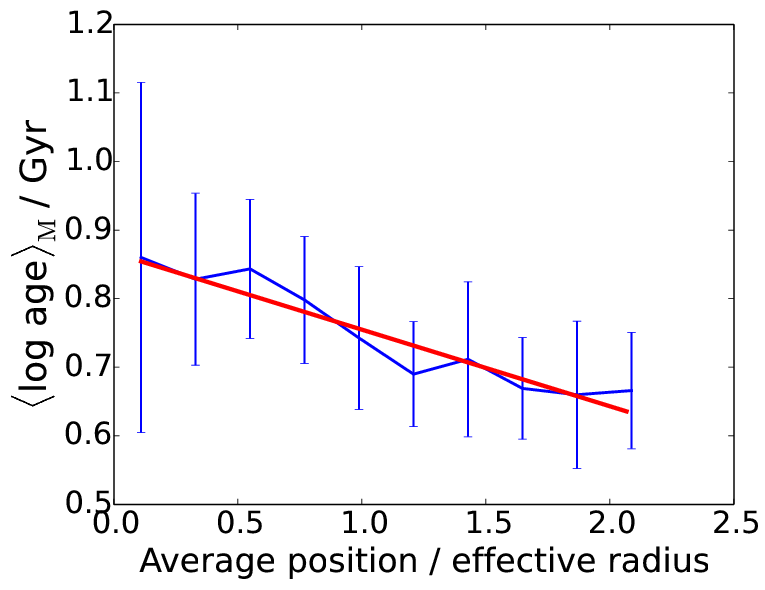}
	\caption{Radial age profile.}
\end{subfigure}\hspace{0.2cm}
\begin{subfigure}{0.3\linewidth}
	\includegraphics[width=\linewidth]{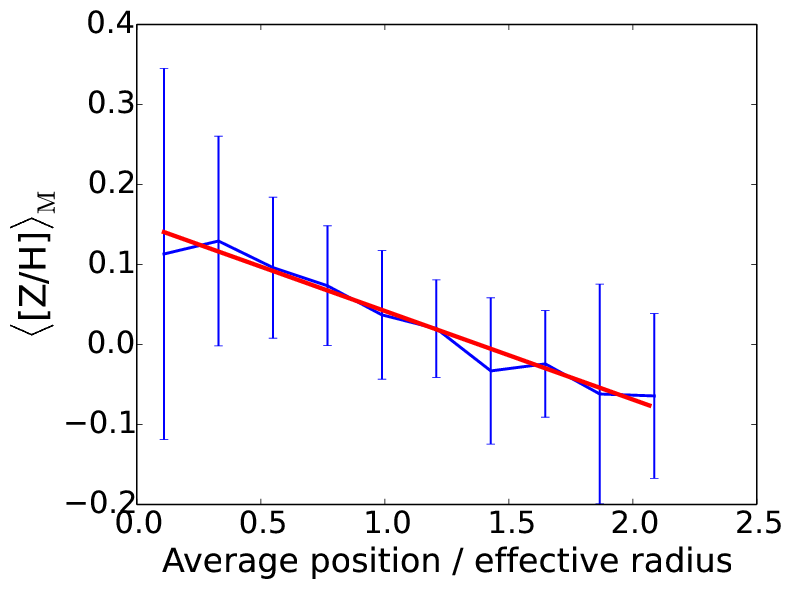}
	\caption{Radial metallicity profile.}
\end{subfigure}\hspace{0.4cm}
\begin{subfigure}{0.3\linewidth}
	\includegraphics[width=\linewidth]{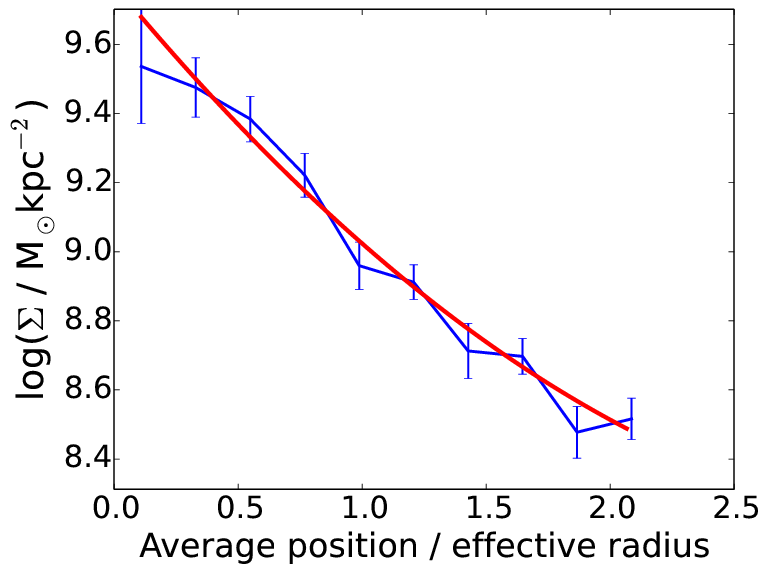}
	\caption{Stellar mass surface density gradient profile.}
\end{subfigure}\hspace{0.9cm}
\caption{{\bf Group $\alpha$, galaxy \galaxyseventeen} as in table \protect\ref{tab:sample}. Stellar population maps and profiles analyzed using MILES-based models with their full parameter range, as described in detail in Figure \protect\ref{maps_eighteen}. This galaxy, like \galaxyeighteen in Figure \protect\ref{maps_eighteen}, has been observed with good observational conditions with a MaNGA-like exposure time and dithering, and is selected in the MaNGA primary sample. Therefore out of the P-MaNGA dataset, these observations are the most similar to the expected output of the MaNGA primary sample.}
\label{maps_seventeen}
\end{figure*}

Galaxy p9-19D (Appendix Figure \ref{maps_seventeen}) has been observed with MaNGA-like data quality with 3 hour exposures, decent (1``.7) seeing conditions, and is in the MaNGA primary sample selection. It is an early-type galaxy that has been observed with the \N{19} IFU setup. We see a high level of radial symmetry in the age, metallicity and stellar mass density maps. It has a \logageL = 6 Gyr / \logageM = 8 Gyr, somewhat super-solar metallicity core going down to \logageL = \logageM =  5 Gyr with approximately solar metallicity at above 1 effective radius, and shows little signs of dust. It therefore has quite clear, albeit weak, negative age and metallicity gradients.

\begin{figure*}
\centering
\vspace{0.5cm}
\hspace{0.3cm}
\begin{subfigure}{0.30\linewidth}
	\hspace{0.4cm}
	\includegraphics[width=0.72\linewidth]{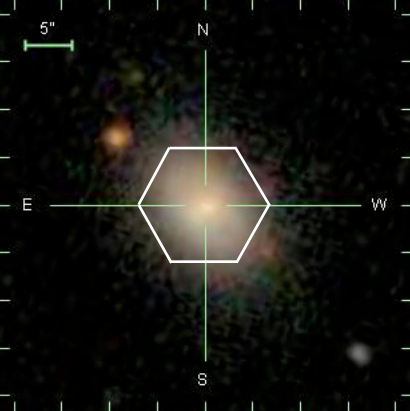}
	\caption{SDSS image with the P-MaNGA footprint.}
\end{subfigure} 
\begin{subfigure}{0.32\linewidth}
	\includegraphics[width=\linewidth]{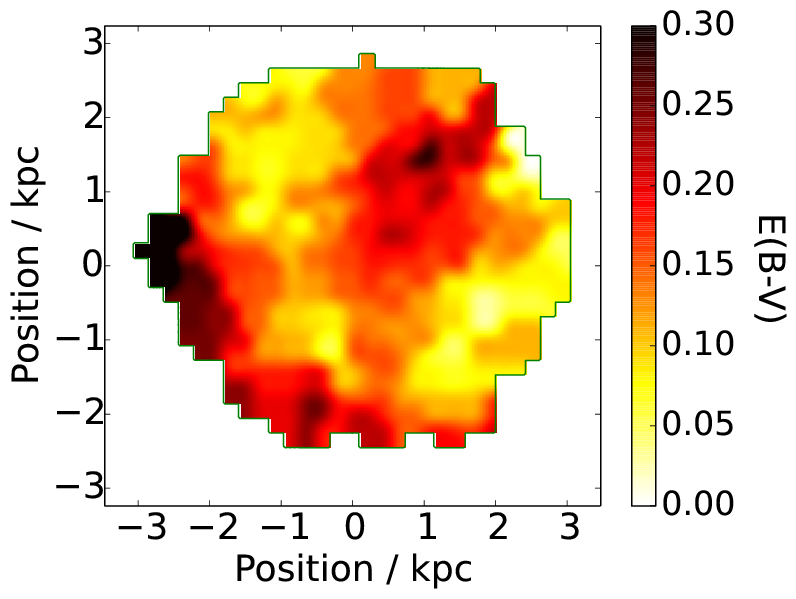}
	\caption{Dust extinction, E(B-V).}
\end{subfigure}
\begin{subfigure}{0.32\linewidth}
	\includegraphics[width=\linewidth]{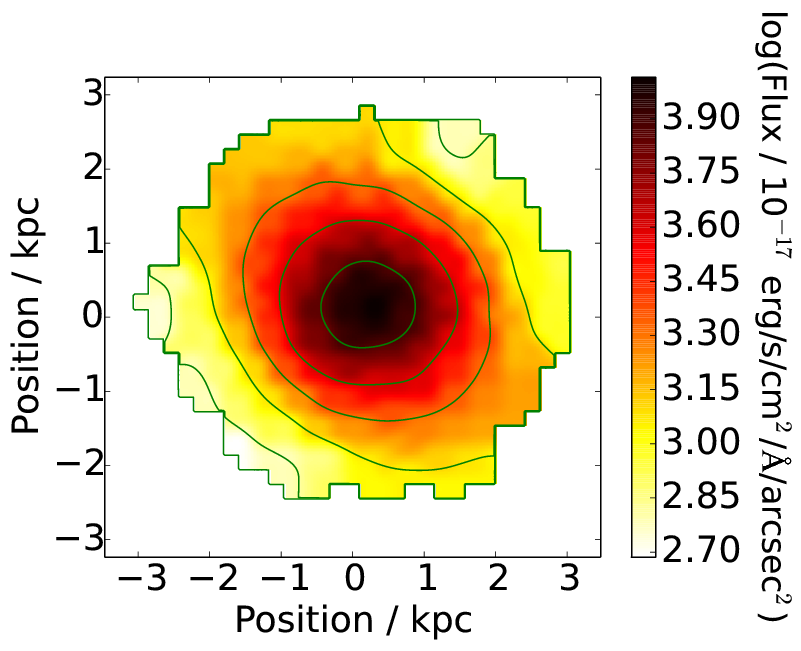}
	\caption{Flux map with isoflux contours (green).}
\end{subfigure}\hspace{0.1cm}
\begin{subfigure}{0.32\linewidth}
	\includegraphics[width=\linewidth]{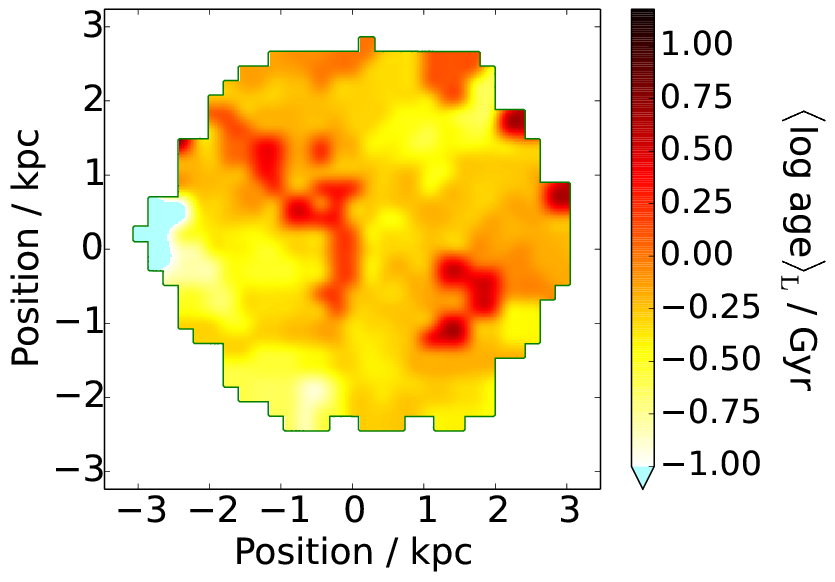}
	\caption{Luminosity-weighted stellar age.}
\end{subfigure}
\begin{subfigure}{0.32\linewidth}
	\includegraphics[width=\linewidth]{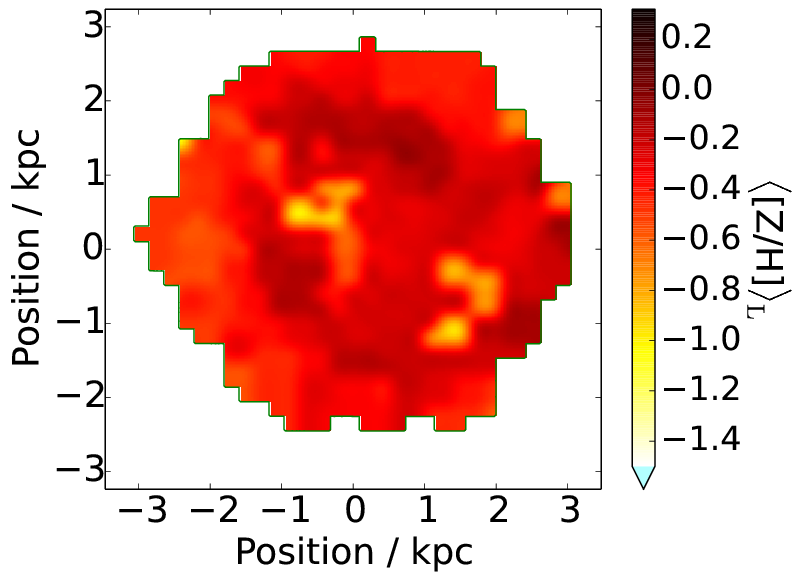}
	\caption{Luminosity-weighted metallicity.}
\end{subfigure}
\begin{subfigure}{0.32\linewidth}
	\includegraphics[width=\linewidth]{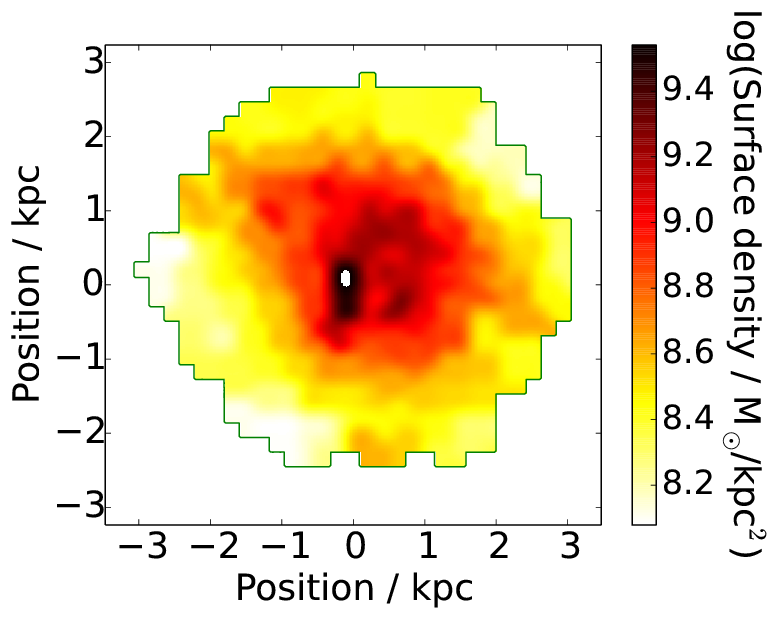}
	\caption{Stellar mass.}
\end{subfigure}
\begin{subfigure}{0.32\linewidth}
	\includegraphics[width=\linewidth]{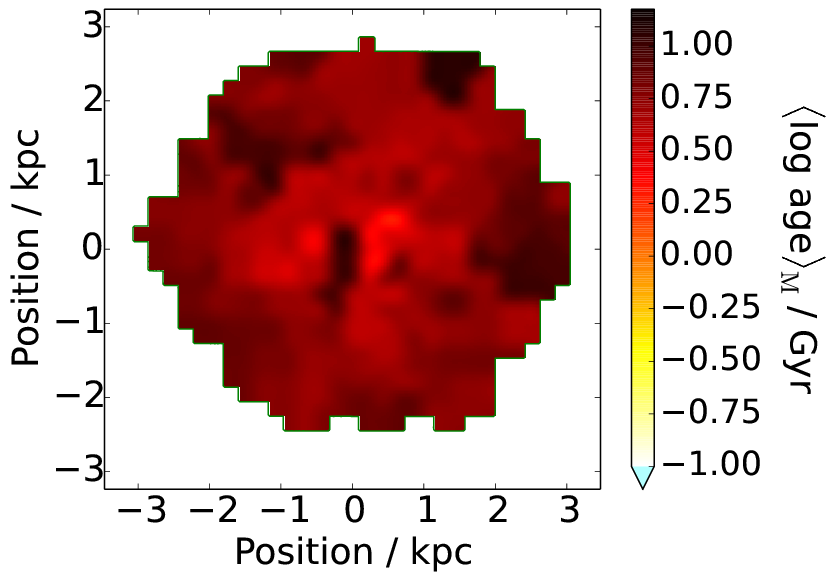}
	\caption{Mass-weighted stellar age.}
\end{subfigure}
\begin{subfigure}{0.32\linewidth}
	\includegraphics[width=\linewidth]{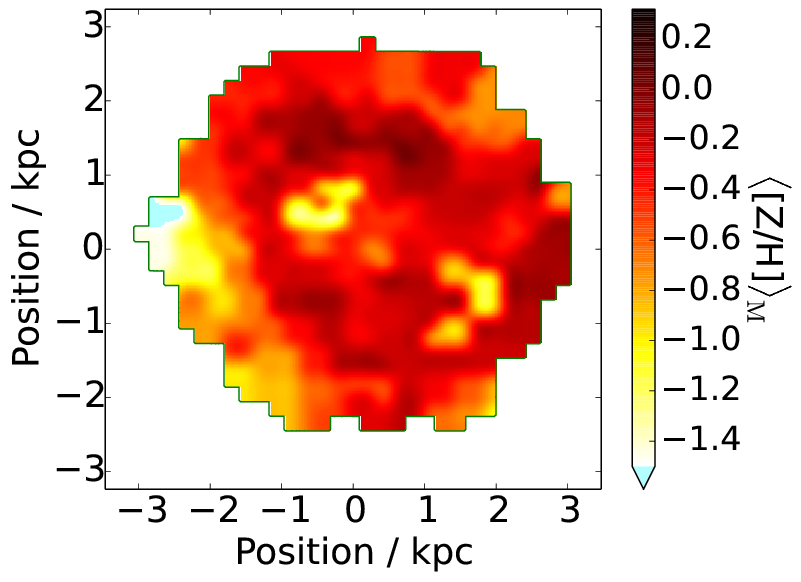}
	\caption{Mass-weighted metallicity.}
\end{subfigure}
\begin{subfigure}{0.32\linewidth}
	\includegraphics[width=\linewidth]{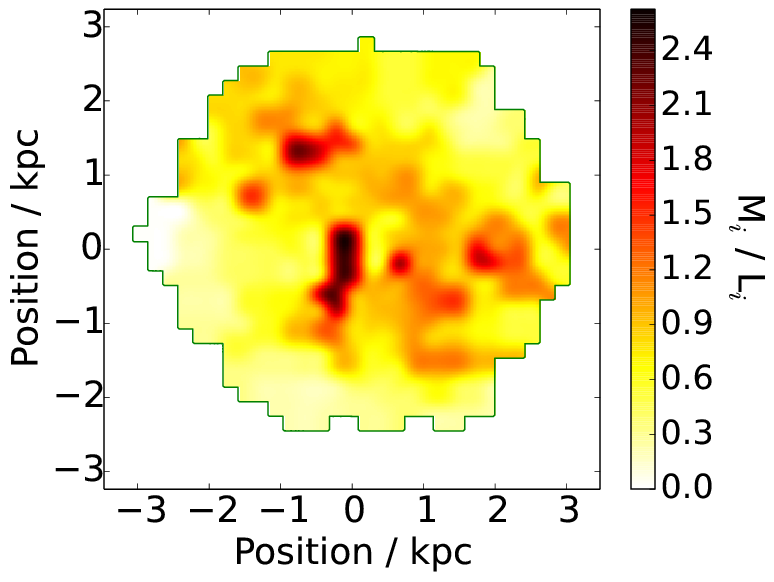}
	\caption{Stellar mass-to-light ratio in the SDSS $i$-band.}
\end{subfigure}
\begin{subfigure}{0.3\linewidth}
	\includegraphics[width=\linewidth]{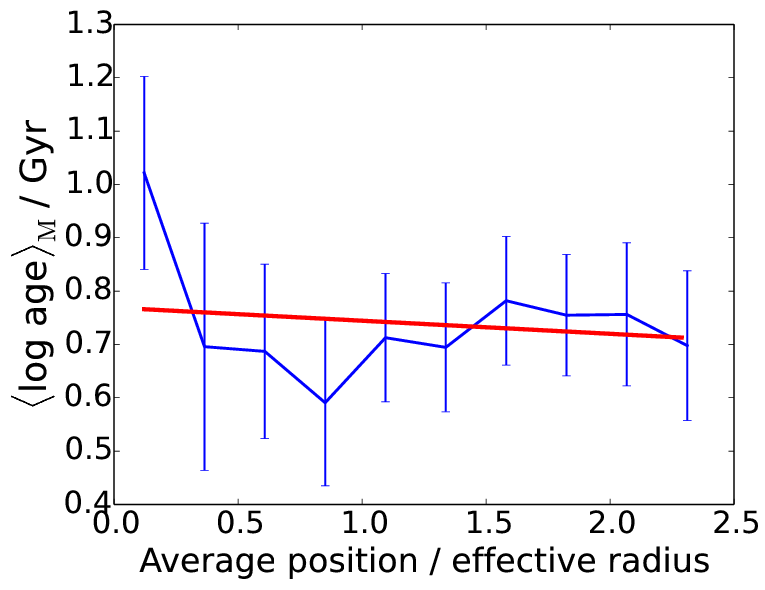}
	\caption{Radial age profile.}
\end{subfigure}\hspace{0.2cm}
\begin{subfigure}{0.3\linewidth}
	\includegraphics[width=\linewidth]{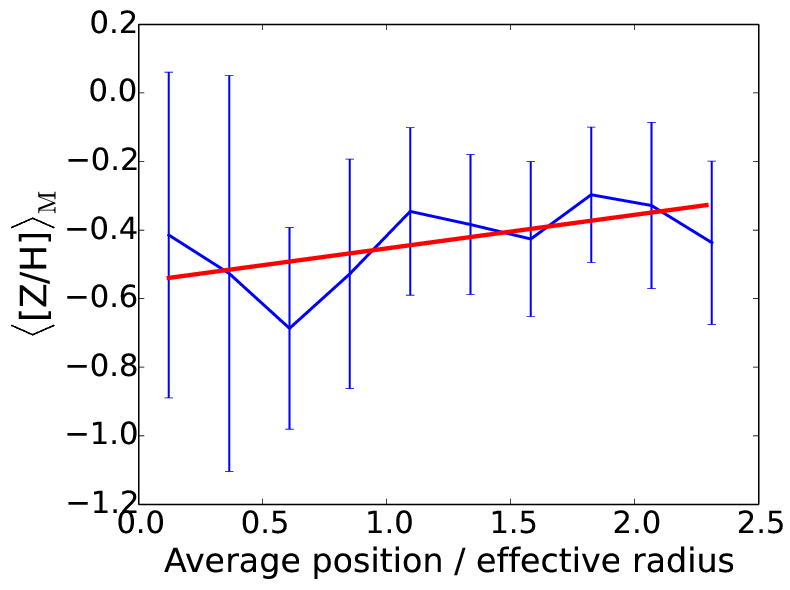}
	\caption{Radial metallicity profile.}
\end{subfigure}\hspace{0.4cm}
\begin{subfigure}{0.3\linewidth}
	\includegraphics[width=\linewidth]{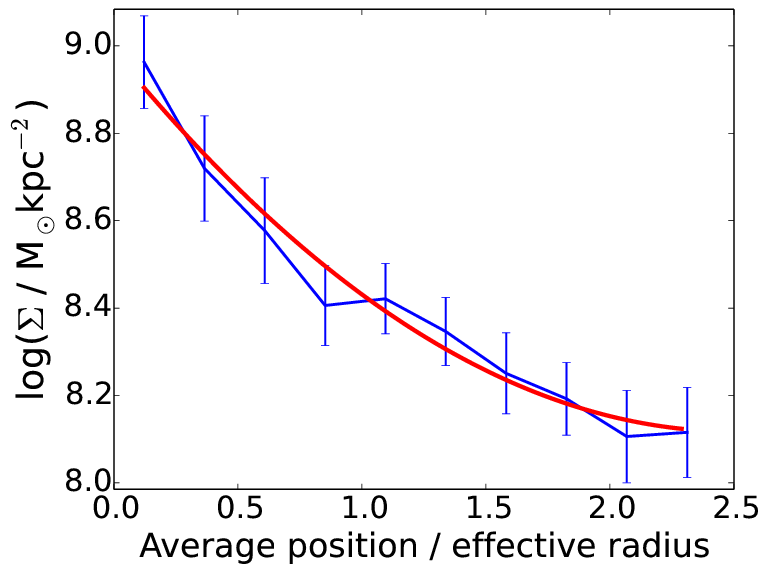}
	\caption{Stellar mass surface density gradient profile.}
\end{subfigure}\hspace{0.9cm}
\caption{{\bf Group $\alpha$, galaxy \galaxyfifteen} as in table \protect\ref{tab:sample}. Stellar population maps and profiles analyzed using MILES-based models with their full parameter range, as described in detail in Figure \protect\ref{maps_eighteen}. This galaxy has been observed under good observational conditions with MaNGA-like exposure time and dithering, but covers a larger radial extent than would be targeted in the MaNGA primary sample. However it is selected by the MaNGA secondary sample.}
\label{maps_fifteen}
\end{figure*}

\begin{figure*}
\centering
\vspace{0.5cm}
\hspace{0.3cm}
\begin{subfigure}{0.30\linewidth}
	\hspace{0.4cm}
	\includegraphics[width=0.72\linewidth]{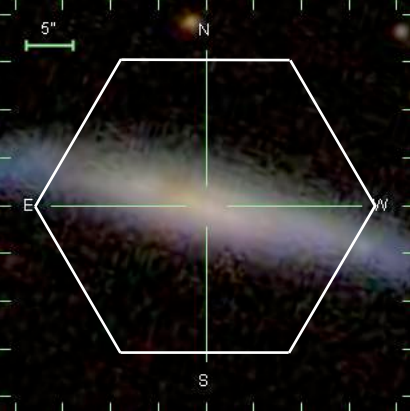}
	\caption{SDSS image with the P-MaNGA footprint.}
\end{subfigure} 
\begin{subfigure}{0.32\linewidth}
	\includegraphics[width=\linewidth]{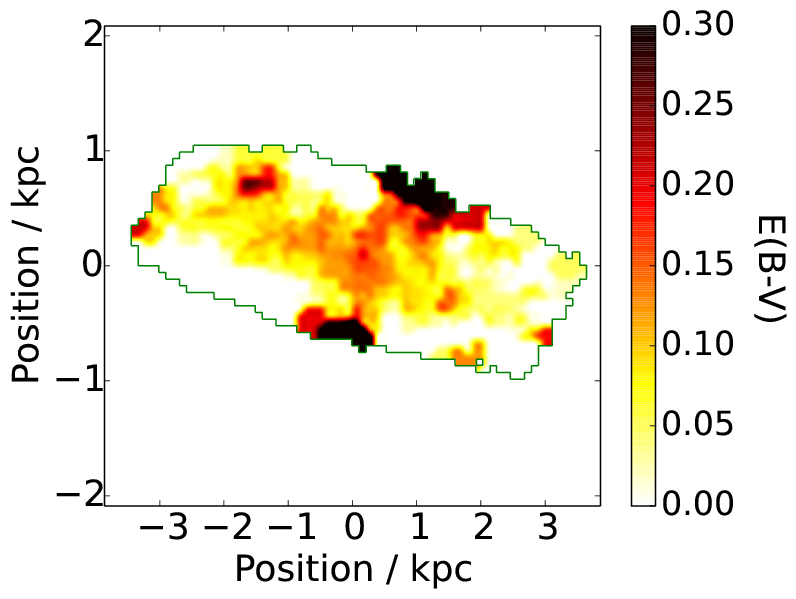}
	\caption{Dust extinction, E(B-V).}
\end{subfigure}
\begin{subfigure}{0.32\linewidth}
	\includegraphics[width=\linewidth]{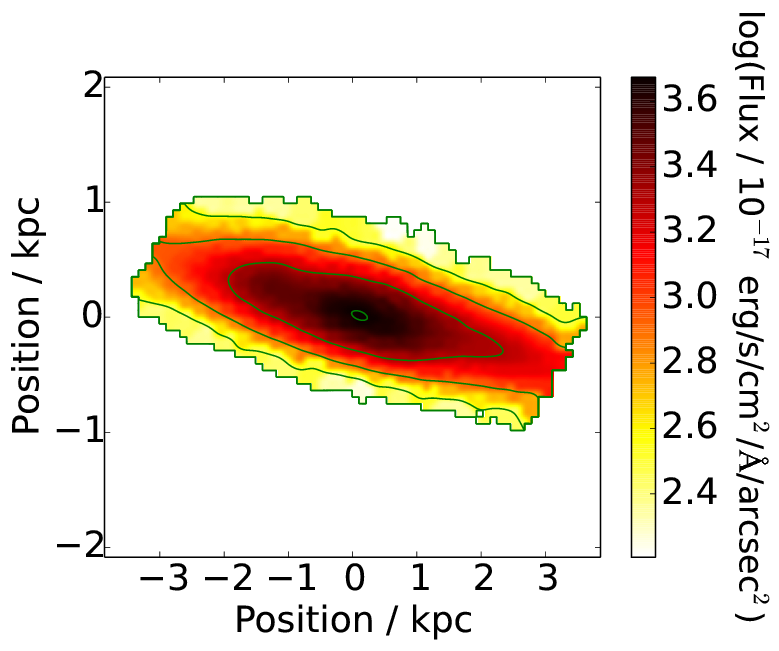}
	\caption{Flux map with isoflux contours (green).}
\end{subfigure}\hspace{0.1cm}
\begin{subfigure}{0.32\linewidth}
	\includegraphics[width=\linewidth]{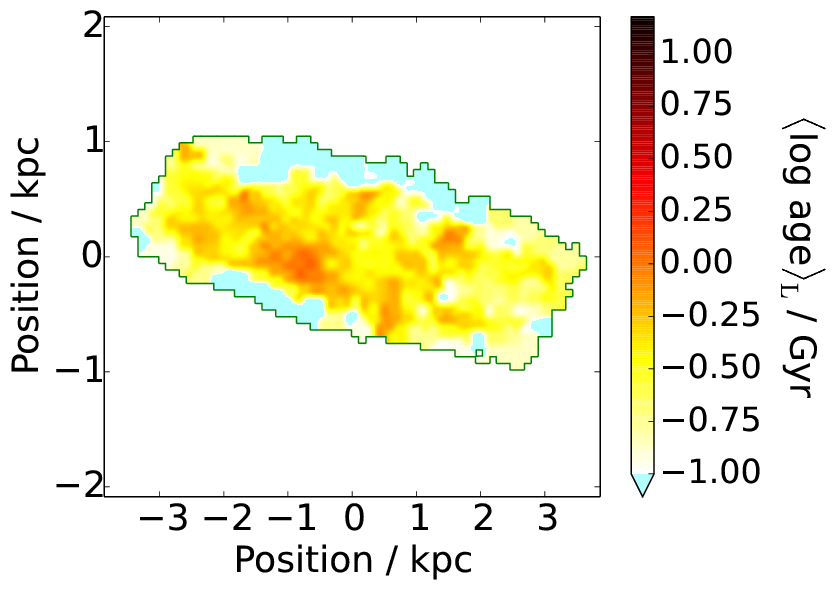}
	\caption{Luminosity-weighted stellar age.}
\end{subfigure}
\begin{subfigure}{0.32\linewidth}
	\includegraphics[width=\linewidth]{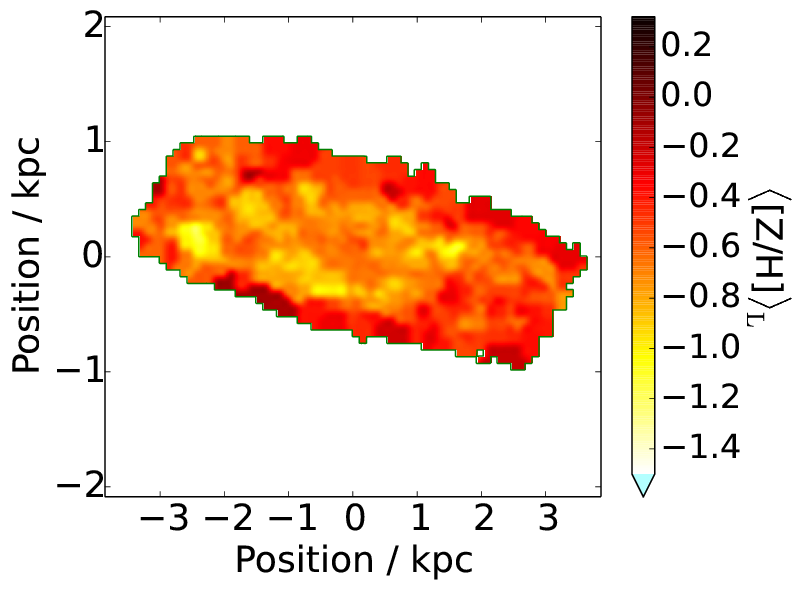}
	\caption{Luminosity-weighted metallicity.}
\end{subfigure}
\begin{subfigure}{0.32\linewidth}
	\includegraphics[width=\linewidth]{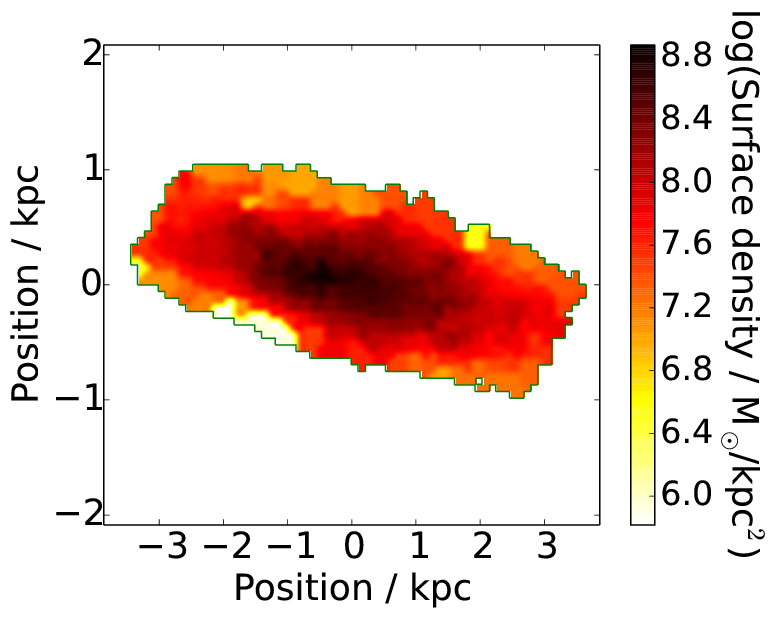}
	\caption{Stellar mass.}
\end{subfigure}
\begin{subfigure}{0.32\linewidth}
	\includegraphics[width=\linewidth]{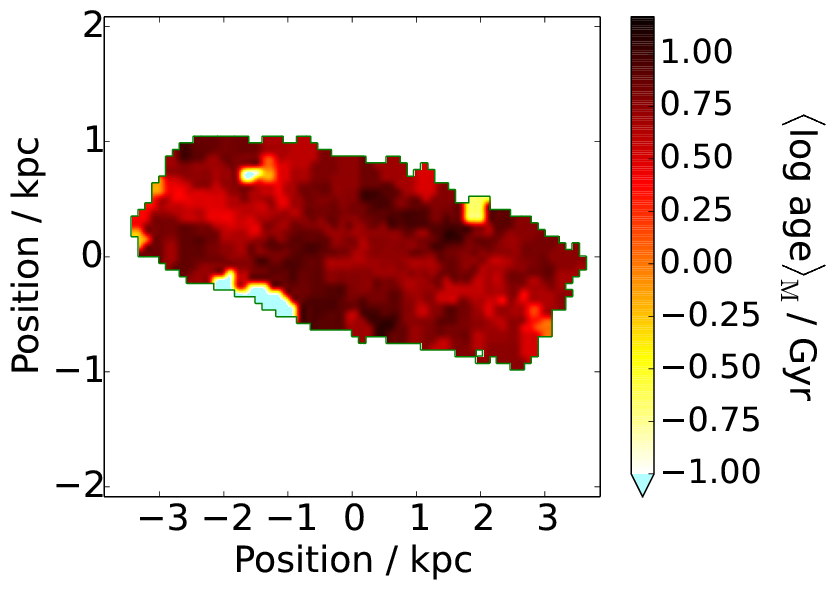}
	\caption{Mass-weighted stellar age.}
\end{subfigure}
\begin{subfigure}{0.32\linewidth}
	\includegraphics[width=\linewidth]{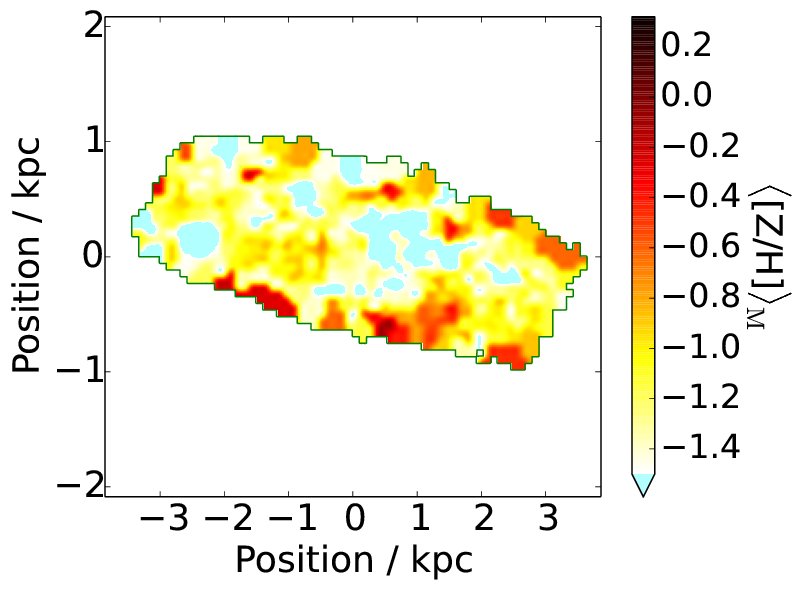}
	\caption{Mass-weighted metallicity.}
\end{subfigure}
\begin{subfigure}{0.32\linewidth}
	\includegraphics[width=\linewidth]{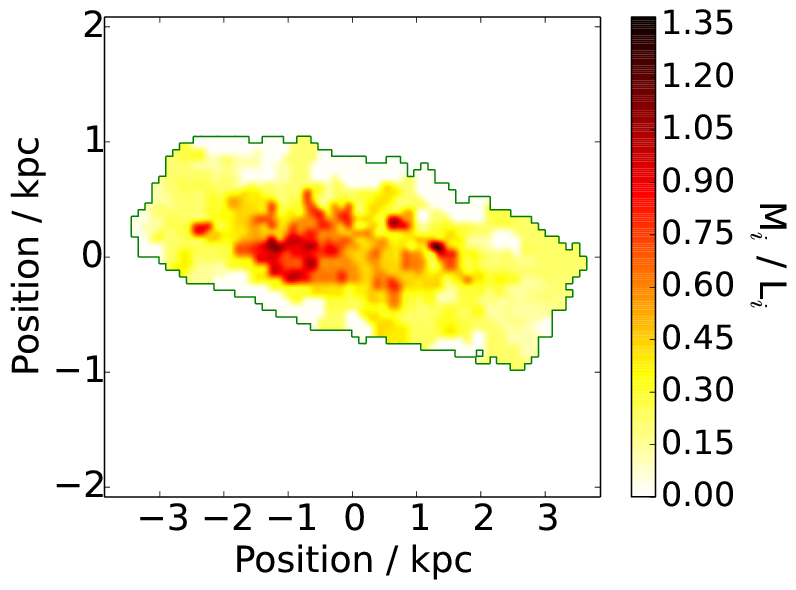}
	\caption{Stellar mass-to-light ratio in the SDSS $i$-band.}
\end{subfigure}
\begin{subfigure}{0.3\linewidth}
	\includegraphics[width=\linewidth]{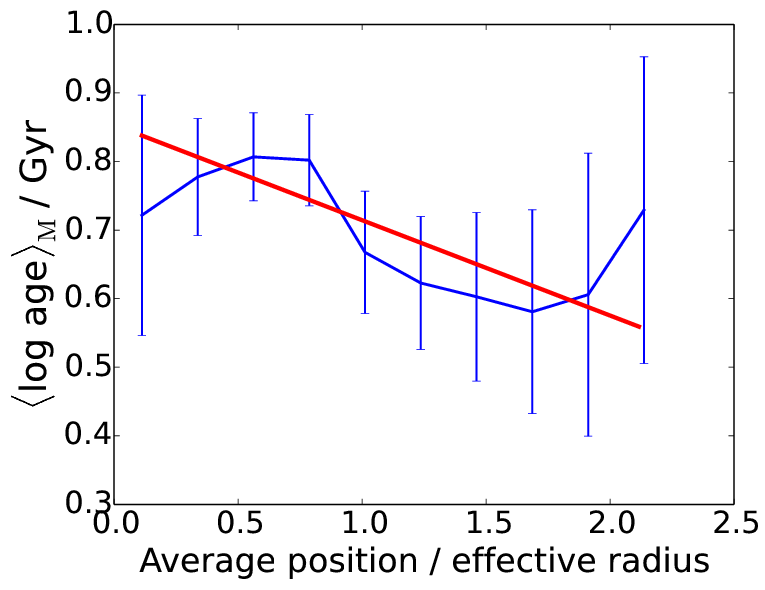}
	\caption{Radial age profile.}
\end{subfigure}\hspace{0.2cm}
\begin{subfigure}{0.3\linewidth}
	\includegraphics[width=\linewidth]{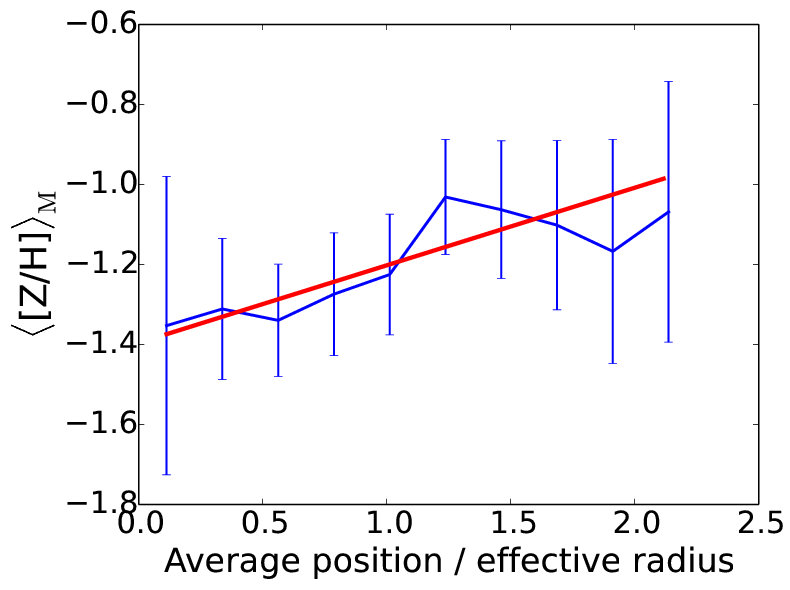}
	\caption{Radial metallicity profile.}
\end{subfigure}\hspace{0.4cm}
\begin{subfigure}{0.3\linewidth}
	\includegraphics[width=\linewidth]{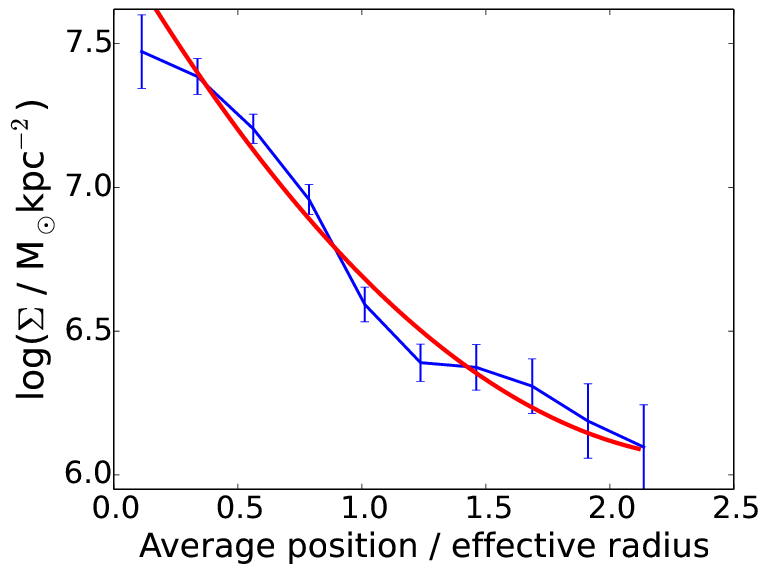}
	\caption{Stellar mass surface density gradient profile.}
\end{subfigure}\hspace{0.9cm}
\caption{{\bf Group $\alpha$, galaxy \galaxyone} as in table \protect\ref{tab:sample}. Stellar population maps and profiles analyzed using MILES-based models with their full parameter range, as described in detail in Figure \protect\ref{maps_eighteen}. This galaxy has been observed under good observational conditions with MaNGA-like exposure time and dithering, but covers a larger radial extent than would be targeted in the MaNGA primary sample, like galaxy \galaxyfifteen (Figure \ref{maps_fifteen}), and is selected by the MaNGA secondary sample.}
\label{maps_one}
\end{figure*}

Galaxies p9-19E (Appendix Figure \ref{maps_fifteen}) and p9-127B (Appendix Figure \ref{maps_one}) have \N{19} and \N{127} IFU observations covering 2.2 and 2.4 effective radii respectively. The early-type galaxy p9-19E shows a much flatter age and metallicity profile and is considerably younger in \logageL than the primary-selected early-type galaxies, though similar \logageM to other galaxies in the sample, with evidence of a possibly younger core in the population maps that is washed out with the intermixed older population in the radial profile. This suggest recent star formation is well-mixed within most of the galaxy, giving the maps a clumpy structure.

Conversely, galaxy p9-127B (Appendix Figure \ref{maps_one}) is an edge-on late-type galaxy with a very clumpy structure, as evidenced clearly in the dust attenuation map. The galaxy as a whole is relatively young (\logageL = 0.8 Gyr / \logageM = 2 Gyr on average), but also has some regions of older populations, on average 1 Gyr older and more metal-poor (--0.50 dex or less) in a ring around the central kpc core. This is visible in the radial profiles in both age and metallicity as a peak and a dip respectively in these properties at low radius. In age this trend turns over to give a negative gradient but in metallicity it rises again at 1.0 \Rekpc~to give a positive gradient on average.

\subsection{Additional maps: Group \large{$\beta$}}

\begin{figure*}
\centering
\vspace{0.5cm}
\hspace{0.3cm}
\begin{subfigure}{0.30\linewidth}
	\hspace{0.4cm}
	\includegraphics[width=0.72\linewidth]{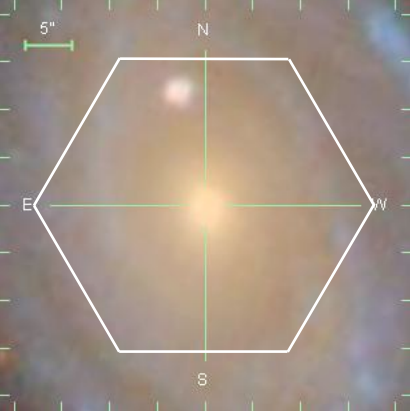}
	\caption{SDSS image with the P-MaNGA footprint.}
\end{subfigure} 
\begin{subfigure}{0.32\linewidth}
	\includegraphics[width=\linewidth]{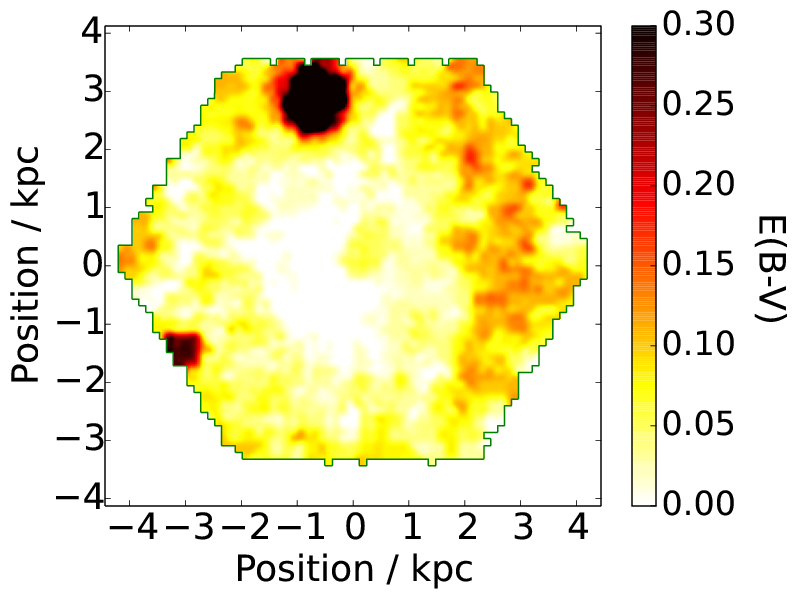}
	\caption{Dust extinction, E(B-V).}
\end{subfigure}
\begin{subfigure}{0.32\linewidth}
	\includegraphics[width=\linewidth]{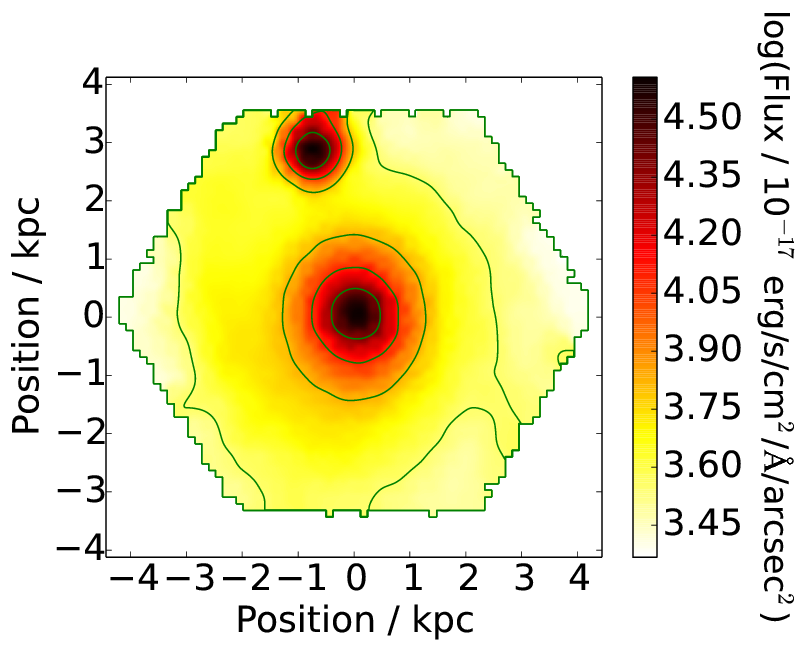}
	\caption{Flux map with isoflux contours (green).}
\end{subfigure}\hspace{0.1cm}
\begin{subfigure}{0.32\linewidth}
	\includegraphics[width=\linewidth]{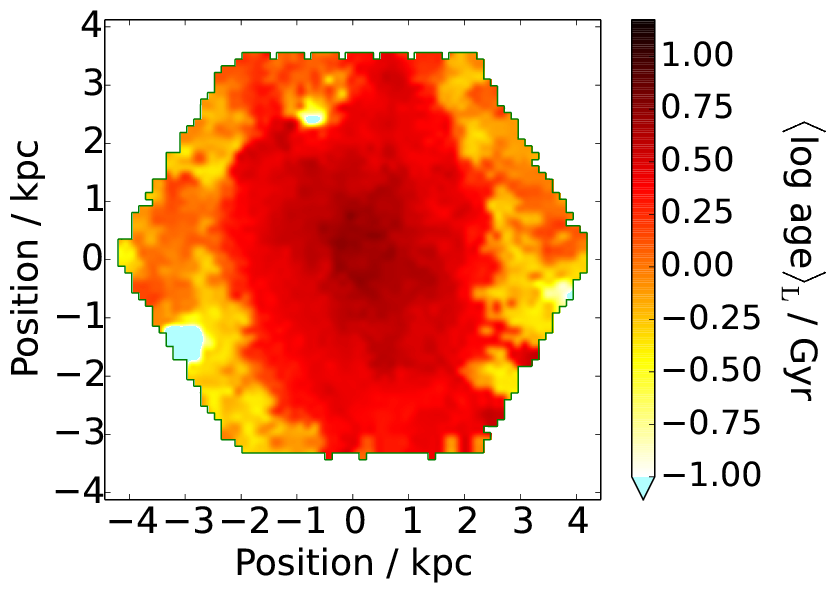}
	\caption{Luminosity-weighted stellar age.}
\end{subfigure}
\begin{subfigure}{0.32\linewidth}
	\includegraphics[width=\linewidth]{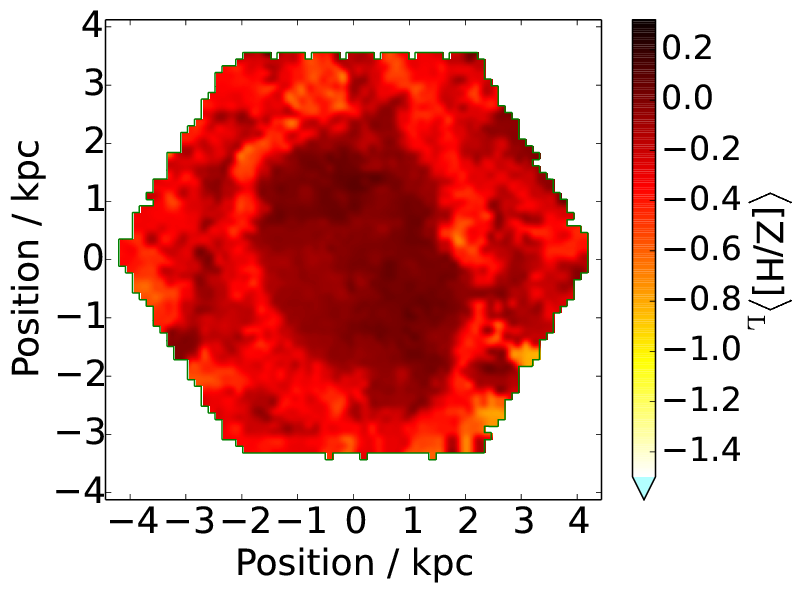}
	\caption{Luminosity-weighted metallicity.}
\end{subfigure}
\begin{subfigure}{0.32\linewidth}
	\includegraphics[width=\linewidth]{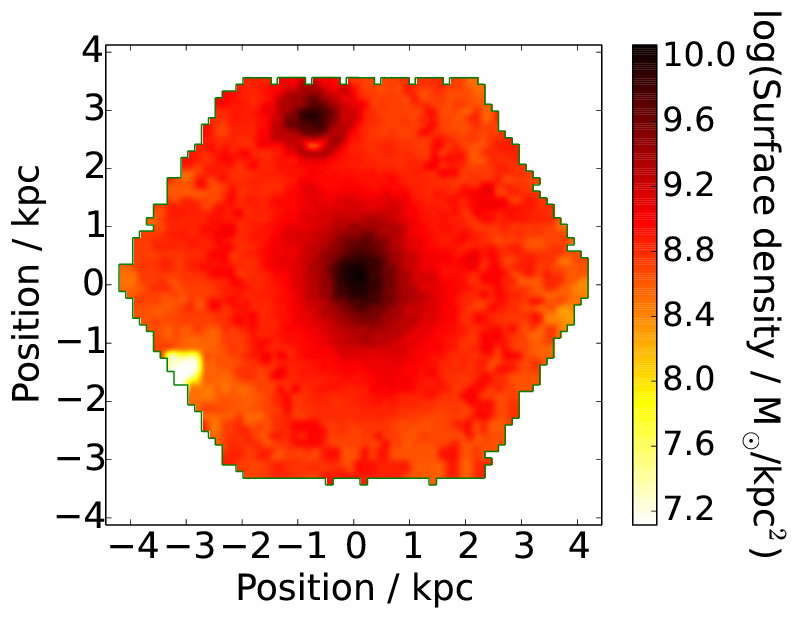}
	\caption{Stellar mass.}
\end{subfigure}
\begin{subfigure}{0.32\linewidth}
	\includegraphics[width=\linewidth]{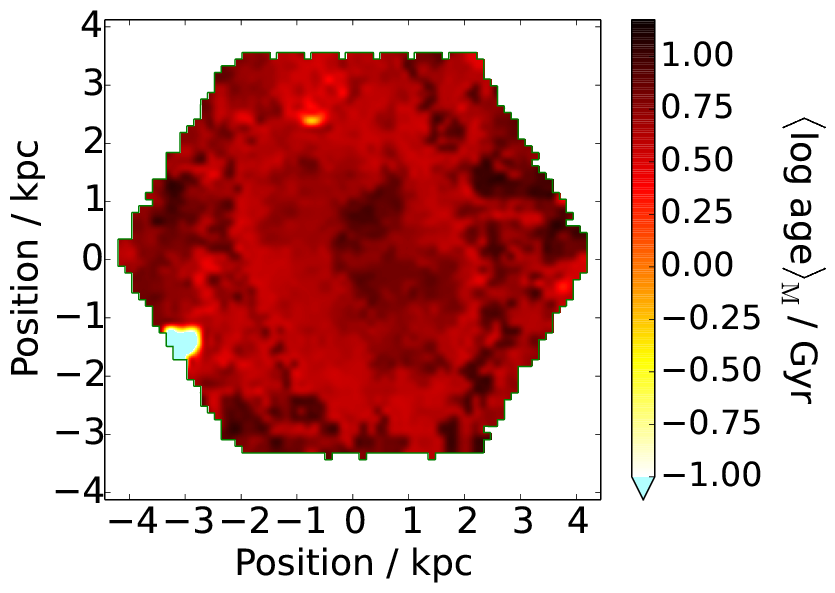}
	\caption{Mass-weighted stellar age.}
\end{subfigure}
\begin{subfigure}{0.32\linewidth}
	\includegraphics[width=\linewidth]{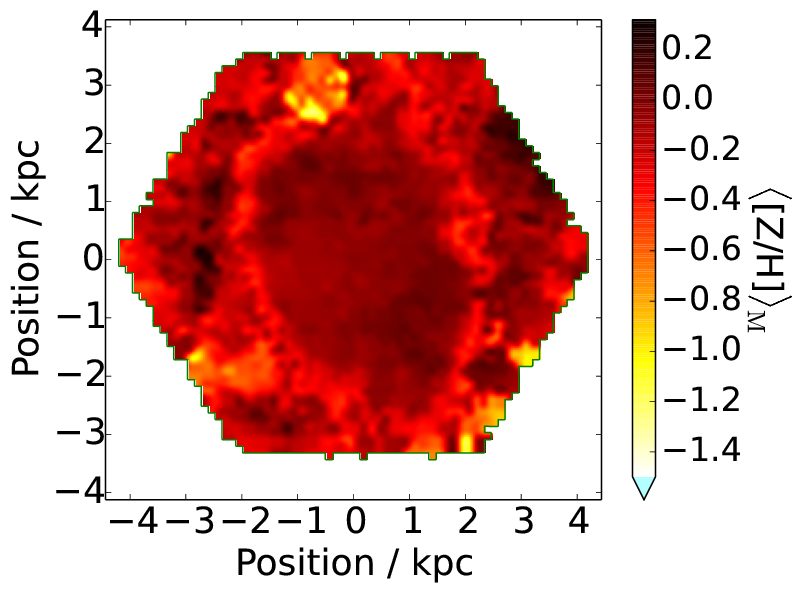}
	\caption{Mass-weighted metallicity.}
\end{subfigure}
\begin{subfigure}{0.32\linewidth}
	\includegraphics[width=\linewidth]{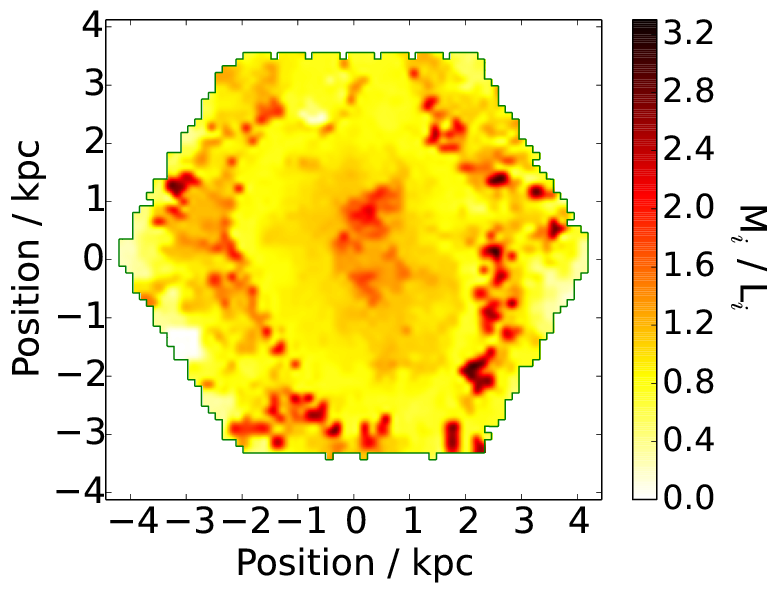}
	\caption{Stellar mass-to-light ratio in the SDSS $i$-band.}
\end{subfigure}
\begin{subfigure}{0.3\linewidth}
	\includegraphics[width=\linewidth]{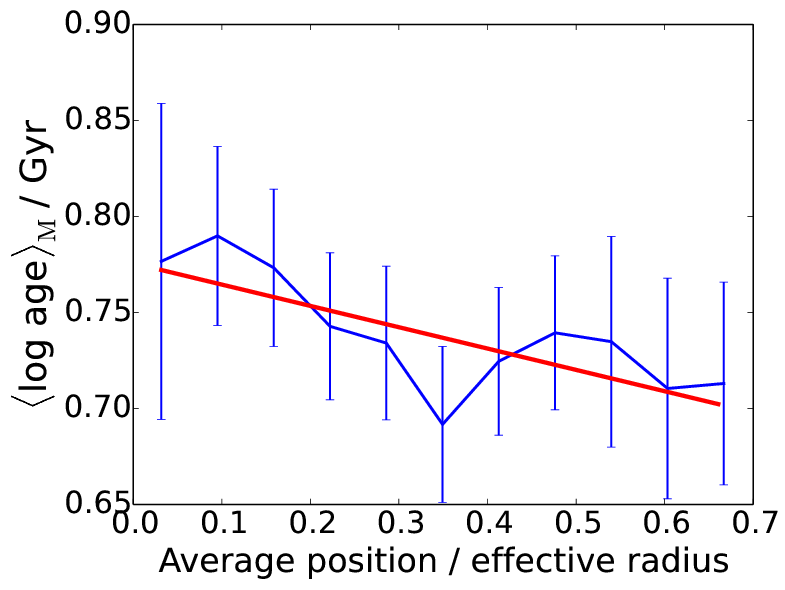}
	\caption{Radial age profile.}
\end{subfigure}\hspace{0.2cm}
\begin{subfigure}{0.3\linewidth}
	\includegraphics[width=\linewidth]{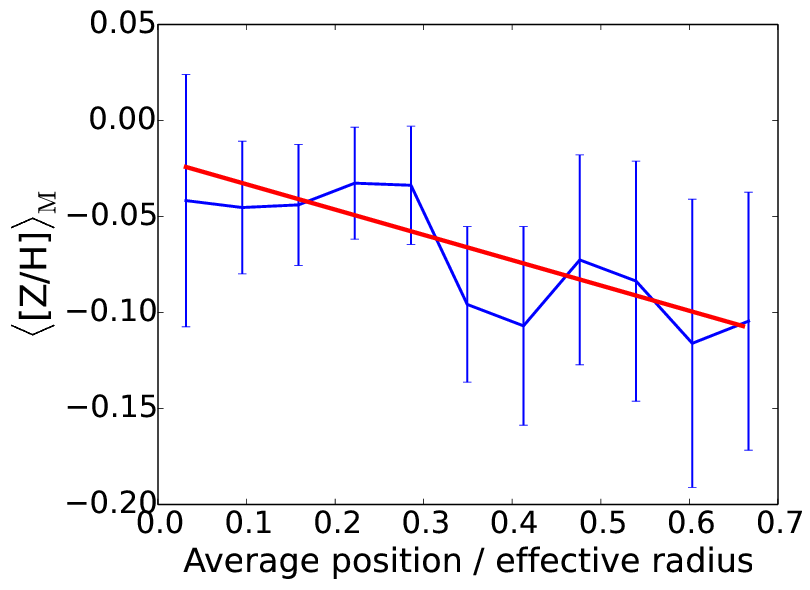}
	\caption{Radial metallicity profile.}
\end{subfigure}\hspace{0.4cm}
\begin{subfigure}{0.3\linewidth}
	\includegraphics[width=\linewidth]{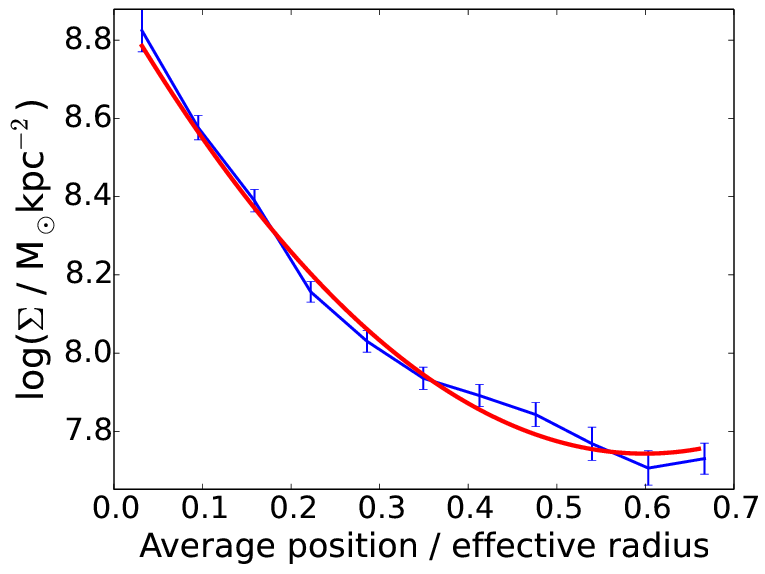}
	\caption{Stellar mass surface density gradient profile.}
\end{subfigure}\hspace{0.9cm}
\caption{{\bf Group $\beta$, galaxy \galaxyfour} as in table \protect\ref{tab:sample}. Stellar population maps and profiles analyzed using MILES-based models with their full parameter range, as described in detail in Figure \protect\ref{maps_eighteen}. This galaxy is taken under MaNGA-like observational conditions and setup, but is not targeted in the main MaNGA samples due to its low redshift, and therefore radial coverage by MaNGA IFUs. Hence this galaxy is observed with a higher amount of structural detail then we should expect to observe with MaNGA.}
\label{maps_four}
\end{figure*}

Galaxy p9-127A (Appendix Figure \ref{maps_four}) is identified very clearly in the light-weighted population maps as a galaxy with an old (8 Gyr) metal-rich ([Z/H] = 0.2) core with a  younger (1-2 Gyr) and less metal-rich ([Z/H]= -- 0.2) outer population, which matches in position with the imaging data as corresponding to the inner part of the spiral arm structure. The mass-weighted properties shows a flatter structure across the galaxy, suggesting that the star formation in the spiral arms are the main driver of the structure in the light-weighted properties. As mentioned earlier, the large difference between mass-weighted and light-weighted age indicates a more continuous star formation history, consistent with spiral arm star formation histories. This spiral structure is also traced in the dust attenuation maps. We note the presence of a foreground star on top of the image, which we mask over in the radial profiles by applying a recessional velocity cut in the data before fitting.

\subsection{Additional maps: Group \large{$\gamma$}}

\begin{figure*}
\centering
\vspace{0.5cm}
\hspace{0.3cm}
\begin{subfigure}{0.30\linewidth}
	\hspace{0.4cm}
	\includegraphics[width=0.72\linewidth]{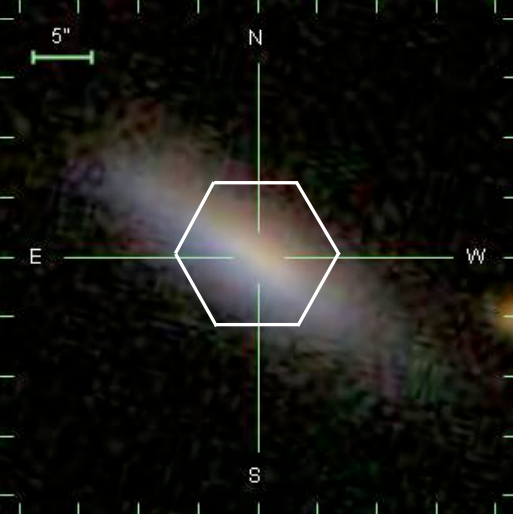}
	\caption{SDSS image with the P-MaNGA footprint.}
\end{subfigure} 
\begin{subfigure}{0.32\linewidth}
	\includegraphics[width=\linewidth]{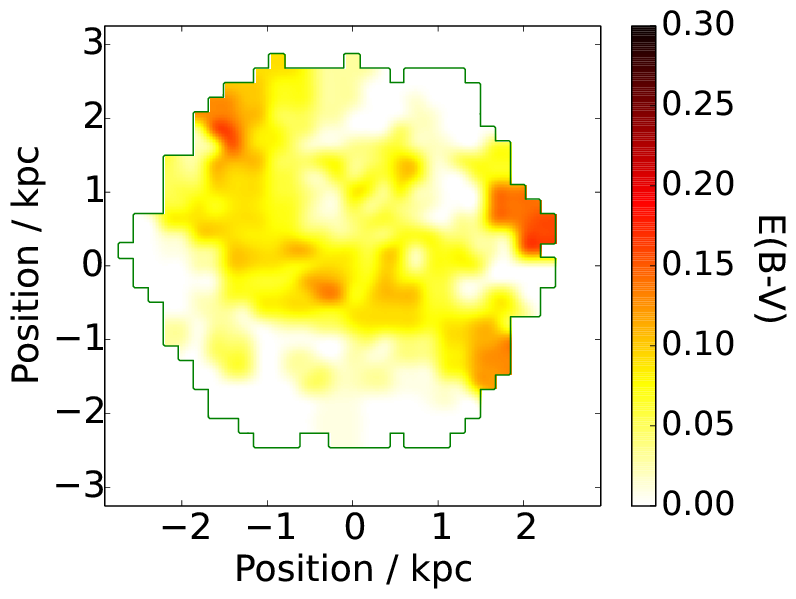}
	\caption{Dust extinction, E(B-V).}
\end{subfigure}
\begin{subfigure}{0.32\linewidth}
	\includegraphics[width=\linewidth]{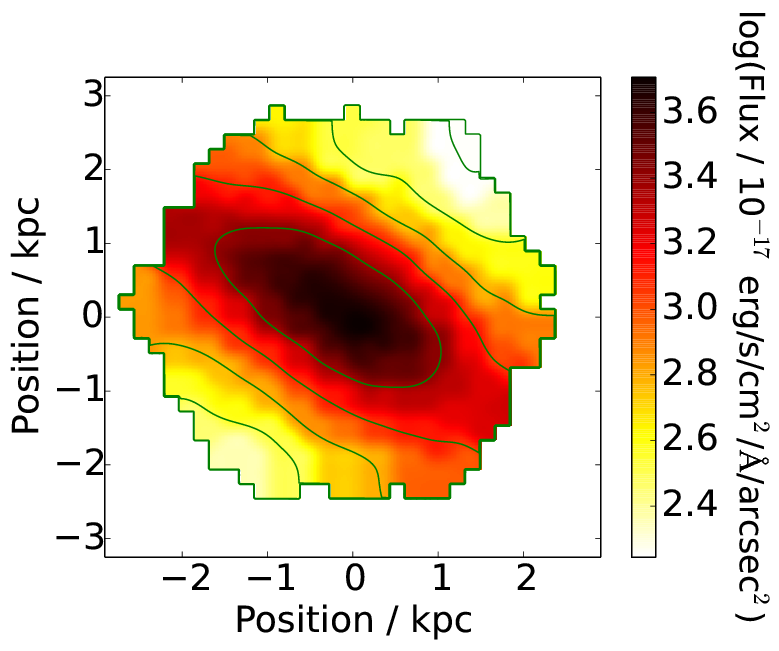}
	\caption{Flux map with isoflux contours (green).}
\end{subfigure}\hspace{0.1cm}
\begin{subfigure}{0.32\linewidth}
	\includegraphics[width=\linewidth]{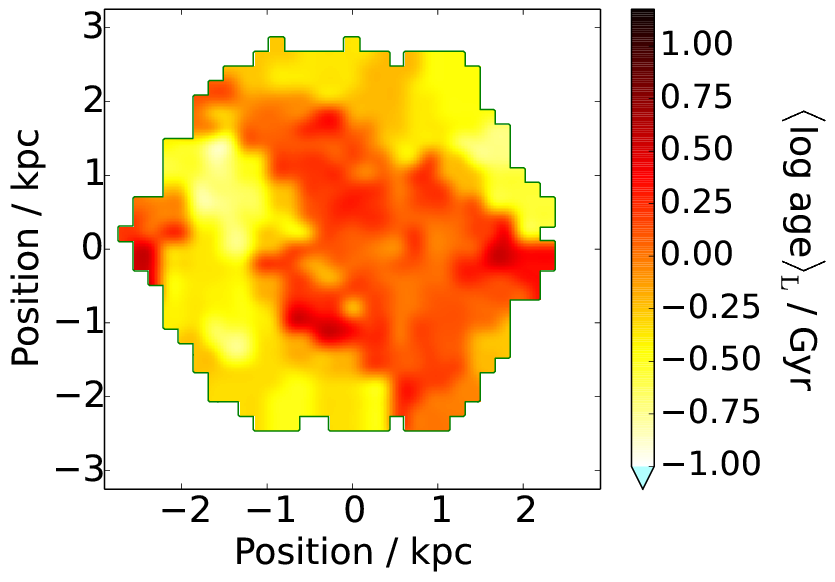}
	\caption{Luminosity-weighted stellar age.}
\end{subfigure}
\begin{subfigure}{0.32\linewidth}
	\includegraphics[width=\linewidth]{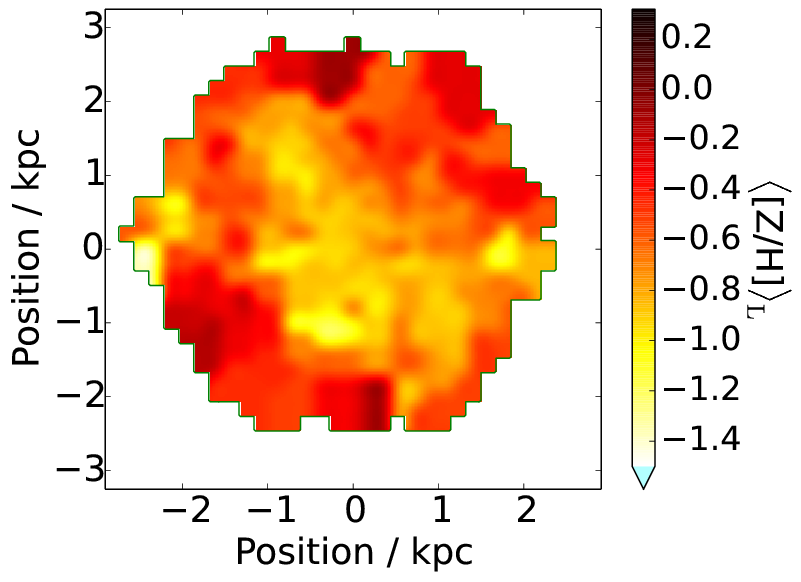}
	\caption{Luminosity-weighted metallicity.}
\end{subfigure}
\begin{subfigure}{0.32\linewidth}
	\includegraphics[width=\linewidth]{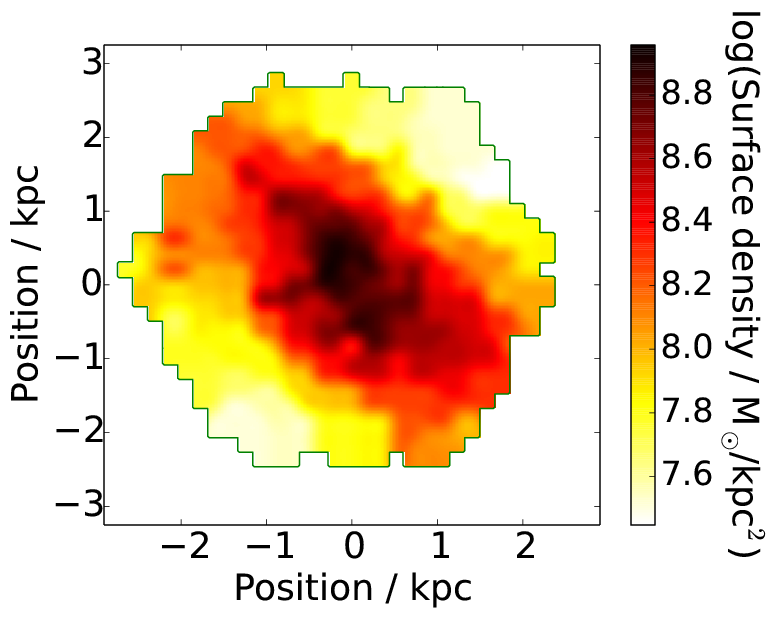}
	\caption{Stellar mass.}
\end{subfigure}
\begin{subfigure}{0.32\linewidth}
	\includegraphics[width=\linewidth]{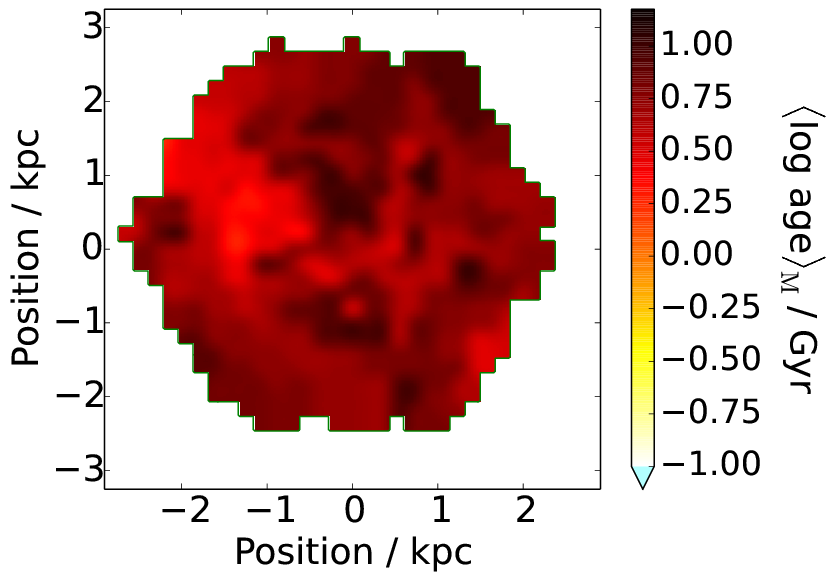}
	\caption{Mass-weighted stellar age.}
\end{subfigure}
\begin{subfigure}{0.32\linewidth}
	\includegraphics[width=\linewidth]{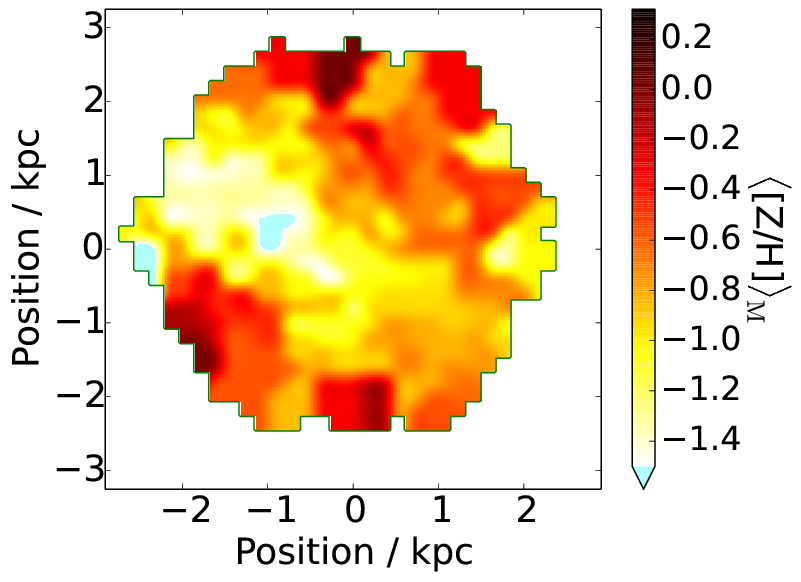}
	\caption{Mass-weighted metallicity.}
\end{subfigure}
\begin{subfigure}{0.32\linewidth}
	\includegraphics[width=\linewidth]{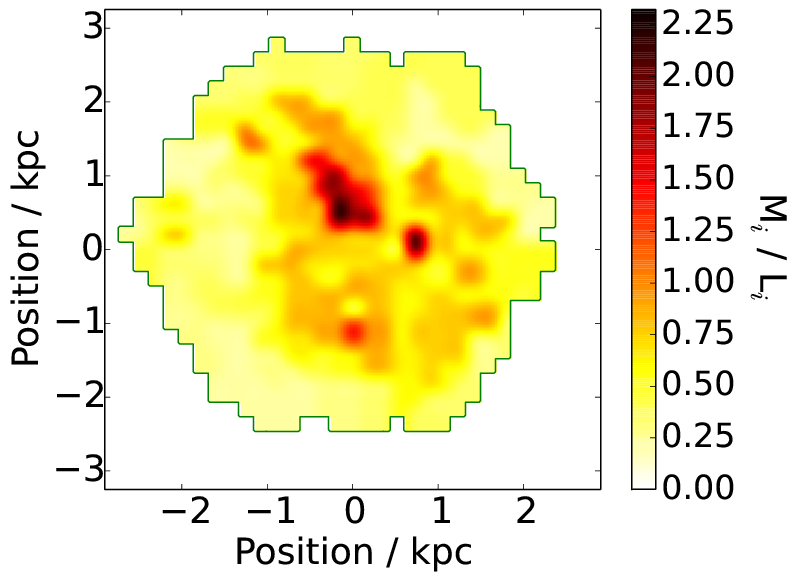}
	\caption{Stellar mass-to-light ratio in the SDSS $i$-band.}
\end{subfigure}
\begin{subfigure}{0.3\linewidth}
	\includegraphics[width=\linewidth]{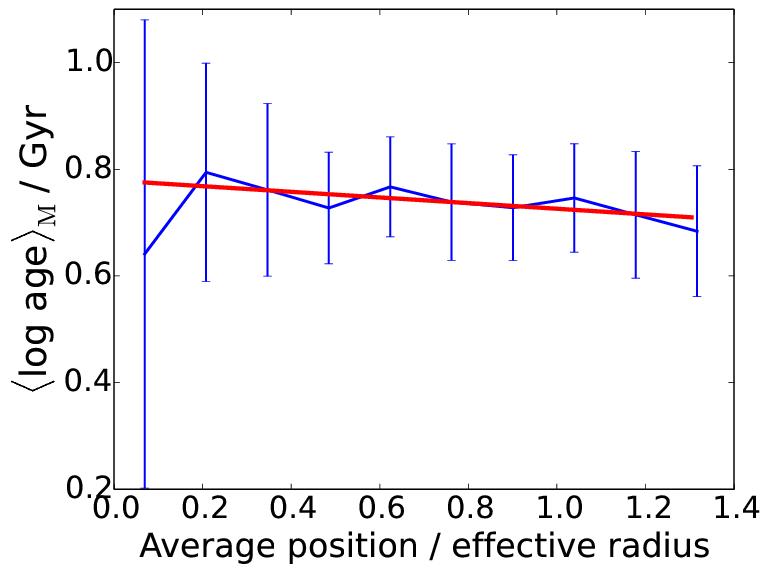}
	\caption{Radial age profile.}
\end{subfigure}\hspace{0.2cm}
\begin{subfigure}{0.3\linewidth}
	\includegraphics[width=\linewidth]{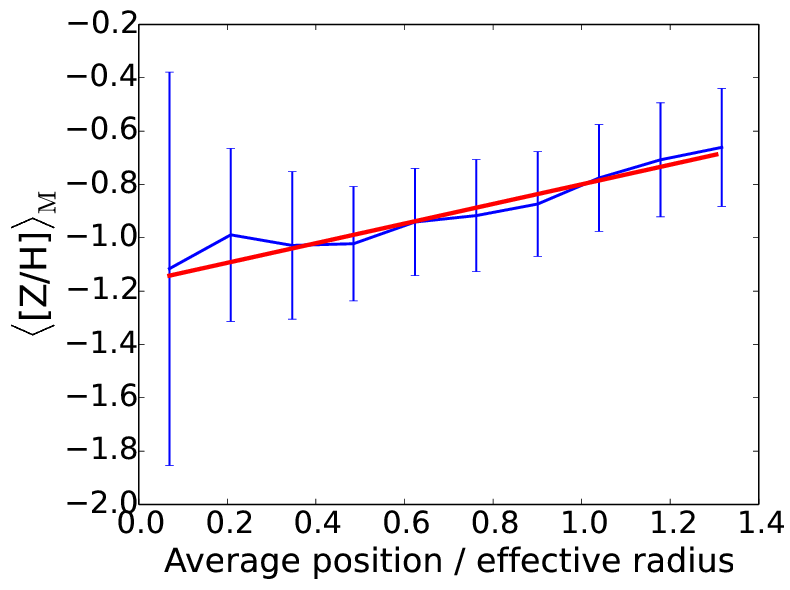}
	\caption{Radial metallicity profile.}
\end{subfigure}\hspace{0.4cm}
\begin{subfigure}{0.3\linewidth}
	\includegraphics[width=\linewidth]{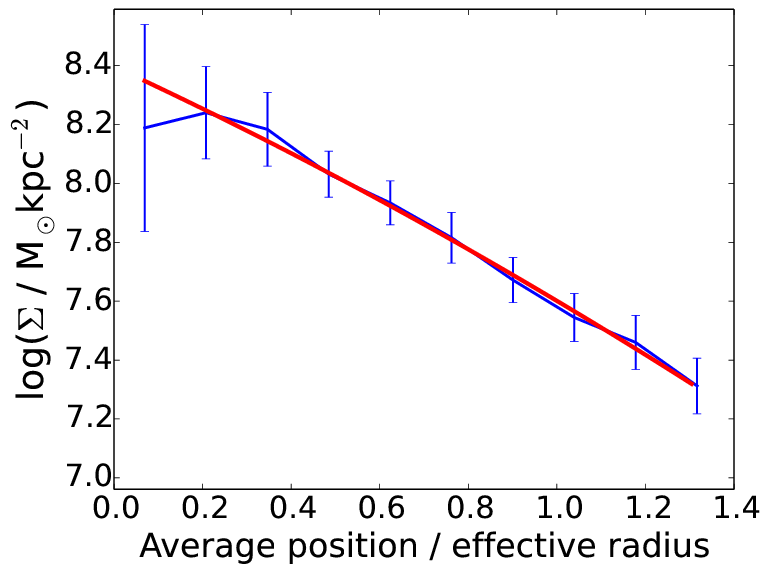}
	\caption{Stellar mass surface density gradient profile.}
\end{subfigure}\hspace{0.9cm}
\caption{{\bf Group $\gamma$, galaxy \galaxythirteen} as in table \protect\ref{tab:sample}. Stellar population maps and profiles analyzed using MILES-based models with their full parameter range, as described in detail in Figure \protect\ref{maps_eighteen}. This galaxy has been observed under poorer conditions and a lower exposure and dithering setup than expected for MaNGA. Unlike galaxy \galaxyfourteen (Figure \protect\ref{maps_fourteen}), we resolve a more complex stellar population structure that reflects the asymmetric colour distribution seen in the SDSS image.}
\label{maps_thirteen}
\end{figure*}

Galaxy p4-19B (Appendix Figure \ref{maps_thirteen}) is a late-type galaxy with a fairly complex structure in its stellar population maps. We see a band of \logageM = \logageL = 1 Gyr stretching across the maps from North-East to South-West. This band is visible in the other maps as having low \metalL and \metalM, but high \logageM compared to the background halo. This combination is suggestive an intermediate age disk population with a background old halo and/or disk population, with a high contribution of star formation in the outer disk regions of the galaxy giving a significant negative age gradient in \logageL or a flat gradient in \logageM.

\end{document}